\documentclass{article}
\usepackage{graphicx} 

\usepackage{hyperref}
\usepackage{amsmath}
\usepackage{amssymb}
\usepackage{xcolor}
\usepackage{arydshln} 
\usepackage{setspace}
\usepackage[top=20mm, bottom=20mm, left=25mm, right=25mm]{geometry}
\usepackage{color,soul} 
\usepackage{caption}
\usepackage{verbatim}

\newtheorem{prop}{Proposition}

\newcommand{\independenT}[2]{\vcenter{\hbox{$#1\perp\!\!\!\perp$}}}
\newcommand{\indep}{\mathrel{\mathpalette\independenT\perp}}
\newcommand{\nindep}{\not\!\indep}

\makeatletter
\newcommand*{\centernot}{
  \mathpalette\@centernot
}
\def\@centernot#1#2{
  \mathrel{
    \rlap{
      \settowidth\dimen@{$\m@th#1{#2}$}
      \kern.5\dimen@
      \settowidth\dimen@{$\m@th#1=$}
      \kern-.5\dimen@
      $\m@th#1\not$
    }
    {#2}
  }
}
\makeatother

\usepackage{enumitem}

\graphicspath{Images/}

\title{What is estimated in cluster randomized crossover trials with informative sizes? --- A survey of estimands and common estimators}
\author{Kenneth M. Lee$^{1*}$, Andrew B. Forbes$^{2}$, Jessica Kasza$^{2}$, Andrew Copas$^{3}$,\\Brennan C. Kahan$^{3}$, Paul J. Young$^{4,5,6,7}$, Michael O. Harhay$^{8,9}$, Fan Li$^{10,11}$}
\date{May 1, 2025}

\begin{document}

\maketitle

\noindent $^1$Department of Biostatistics, Epidemiology and Informatics, University of Pennsylvania, Philadelphia, PA, USA

\noindent $^2$School of Public Health and Preventive Medicine, Monash University, Melbourne, Australia

\noindent $^3$MRC Clinical Trials Unit at UCL, London, UK

\noindent $^4$Intensive Care Unit, Wellington Hospital, Wellington, New Zealand

\noindent $^5$Medical Research Institute of New Zealand, Wellington, New Zealand

\noindent $^6$Australian and New Zealand Intensive Care Research Centre, Monash University, Melbourne, Victoria, Australia

\noindent $^7$Department of Critical Care, University of Melbourne, Melbourne, Victoria, Australia

\noindent $^8$Clinical Trials Methods and Outcomes Lab, Palliative and Advanced Illness Research (PAIR) Center, Perelman School of Medicine, University of Pennsylvania, Philadelphia, PA, USA

\noindent $^9$Department of Biostatistics, Epidemiology, and Informatics, Perelman School of Medicine, University of Pennsylvania, Philadelphia, USA

\noindent $^{10}$Department of Biostatistics, Yale School of Public Health, New Haven, CT, USA 

\noindent $^{11}$Center for Methods in Implementation and Prevention Science, Yale School of Public Health, New Haven, CT, USA

\noindent * Corresponding Author. Center for Clinical Epidemiology and Biostatistics, University of Pennsylvania School of Medicine, Blockley Hall, 423 Guardian Drive, Philadelphia, PA 19104

\noindent E-mail: kenneth.lee@pennmedicine.upenn.edu

\newpage
\begin{abstract}
In the analysis of cluster randomized trials (CRTs), previous work has defined two meaningful estimands: the individual-average treatment effect (iATE) and cluster-average treatment effect (cATE) estimand, to address individual and cluster-level hypotheses.
In multi-period CRT designs, such as the cluster randomized crossover (CRXO) trial, additional weighted average treatment effect estimands help fully reflect the longitudinal nature of these trial designs, namely the cluster-period-average treatment effect (cpATE) and period-average treatment effect (pATE).
We define different forms of informative sizes, where the treatment effects vary according to cluster, period, and/or cluster-period sizes, which subsequently cause these estimands to differ in magnitude.
Under such conditions, we demonstrate which of the unweighted, inverse cluster-period size weighted, inverse cluster size weighted, and inverse period size weighted:
(i.) independence estimating equation, (ii.) fixed effects model, (iii.) exchangeable mixed effects model, and (iv.) nested exchangeable mixed effects model treatment effect estimators are consistent for the aforementioned estimands in 2-period cross-sectional CRXO designs with continuous outcomes.
We report a simulation study
and conclude with a reanalysis of a CRXO trial testing different treatments on hospital length of stay among patients receiving invasive mechanical ventilation.
Notably, with informative sizes, the unweighted and weighted nested exchangeable mixed effects model estimators are not consistent for any meaningful estimand and can yield biased results.
In contrast, the unweighted and weighted independence estimating equation, and under specific scenarios, the fixed effects model and exchangeable mixed effects model, can yield consistent and empirically unbiased estimators for meaningful estimands in 2-period CRXO trials.
\end{abstract}

{\raggedright \textbf{Keywords:} cluster randomized crossover trials, informative sizes, estimands, mixed effects, fixed effects, consistency \par}
\break

\section{Introduction}

Cluster randomized trials (CRTs) refer to the collection of study designs where randomization to different treatment arms or sequences is carried out at the cluster level (such as at a hospital, clinic, or worksite level), with outcome measurements typically collected at the individual-level \cite{turner_review_2017}.
The cluster randomized crossover (CRXO) trial design is a multi-period CRT design where clusters are randomized to initially receive the treatment or control, then crossover to the control or treatment condition in adjacent time periods, creating a ``checker-board" design \cite{arnup_understanding_2017}.
This CRXO trial design can be highly statistically efficient compared to other CRT designs and can have many logistical advantages \cite{hemming_use_2020, mckenzie_reporting_2025}.
Furthermore, a CRXO design can have multiple periods and crossovers, and under certain model assumptions, the efficiency of treatment effect estimators can increase as more crossovers are added \cite{grantham_how_2019, hemming_use_2020}.
In this article, we will primarily focus on the simplest 2-period cross-sectional CRXO trial design; an example is shown in Figure \ref{fig:CRXO_design}.

Multiple models with different correlation structures have been previously suggested for the analysis of CRXO trials that take into account the multi-period nature of the design \cite{turner_analysis_2007, li_power_2019,morgan_choosing_2017,mckenzie_reporting_2025}.
In this article, we will primarily survey the properties of treatment effect estimators derived from the independence estimating equation, exchangeable mixed effects model, nested exchangeable mixed effects model, and fixed effects model.
Unlike the majority of prior literature on CRXO trials, we are interested in which of these model-based treatment effect estimators generally produce consistent estimation for clearly defined estimands under the potential outcomes framework, regardless of whether other model assumptions hold. Therefore, it is critically important to define such transparent treatment effect estimands that accommodate the particular design features of the CRXO design. When a model-based treatment effect estimator targets a clear potential outcomes defined estimand of interest, it will be referred to as ``model-assisted" \cite{su_model-assisted_2021,chen_model-assisted_2025}.

In a standard parallel cluster randomized trial (P-CRT), where clusters are randomized to implement the treatment or control over the entire trial duration, recent studies have used the potential outcomes framework to define two target estimands with natural interpretations---the cluster-average treatment effect (cATE) and the individual-average treatment effect (iATE), corresponding to different target units of inference \cite{kahan_demystifying_2024,kahan_informative_2023}. Briefly, the cATE (sometimes also referred to as the “unit average treatment effect” (UATE) \cite{imai_essential_2009}) is the average treatment effect defined on the cluster-level, with all observed individuals pooled across their corresponding cluster (defined formally in Section \ref{sect:PO_Estimands}), and can be of interest when studying interventions designed for implementation at the cluster level.
The iATE (sometimes also referred to as the ``participant average treatment effect") is the average treatment effect defined on the individual-level, mimicking what would typically be targeted in an individually-randomized trial, and may often be of relevance when studying individual-level interventions that are cluster randomized due to logistical or administrative considerations.

In addition to the cATE and iATE, multi-period CRT designs can also use potential outcomes to define (across all time periods with treatment positivity) a period-average treatment effect (pATE)---the average treatment effect on the period-level, with all observed individuals pooled over all observed clusters in each period, and a cluster-period-average treatment effect (cpATE)---the average treatment effect on the cluster-period cell-level, which is often the implicit target estimand in the analysis of multi-period CRT designs using cluster-period cell summaries \cite{chen_model-assisted_2025}.
However, as we will formally demonstrate in Section \ref{sect:PO_Estimands}, these four estimands differ by definition and potentially in magnitude in multi-period CRTs, including the simplest 2-period CRXO trial.
The previously described model-based estimators are typically specified based on individual-level data and are implicitly intended to target the iATE. Accordingly, inverse cluster size, period size, or cluster-period size weights may be specified to conceptually ensure that all clusters, periods, or cluster-periods contribute equally for the estimators to ideally target the cATE, pATE, and cpATE estimands, respectively \cite{williamson_marginal_2003,kahan_demystifying_2024}. In this paper, we delineate conditions under which these estimators and their weighted counterparts are consistent for their corresponding weighted estimands in CRXO trials.

In the P-CRT and parallel cluster randomized trial with a baseline period (PB-CRT), the iATE and cATE can differ in magnitude in the presence of treatment effects that vary according to cluster size, also referred to as ``informative cluster sizes” \cite{bugni_inference_2024,kahan_informative_2023,wang_model-robust_2024,wang_two_2022,williamson_marginal_2003,kahan_demystifying_2024,lee_how_2025}.
In this article, we further define ``informative period sizes" and ``informative cluster-period sizes" for multi-period CRT designs, where there are treatment effects that vary according to period size or cluster-period size, respectively. Altogether, we will collectively refer to these ``informative sizes" as scenarios with treatment effects that vary according to either cluster, period, and/or cluster-period sizes, such that the aforementioned estimands (iATE, cATE, pATE, and cpATE) may differ in magnitude.

Notably, previous work in P-CRTs demonstrated that in the presence of informative cluster sizes, the unweighted and inverse cluster-size weighted exchangeable mixed effects model estimators converge to estimands that are notably neither the iATE nor the cATE and have no clear interpretation \cite{wang_two_2022}.
In contrast, the unweighted and inverse cluster-size weighted independence estimating equations will yield consistent estimators for the iATE and cATE estimands, respectively \cite{wang_two_2022}.
This work was recently extended to PB-CRTs, where in addition to the corresponding independence estimating equation estimators, the unweighted and weighted fixed effects model estimators were also shown to generally yield consistent estimators for the iATE and cATE, respectively \cite{lee_how_2025}.
Furthermore, the exchangeable mixed effects model is again shown to yield inconsistent estimators for these two natural estimands in the presence of informative cluster sizes \cite{lee_how_2025}. However, as a somewhat surprising result, estimators based on such a model were shown to be empirically robust to bias in a PB-CRT design \cite{lee_how_2025}.
In contrast, the unweighted and weighted nested-exchangeable mixed effects model estimators in a PB-CRT design with informative cluster sizes converge to estimands with data-dependent weights that are difficult to interpret and can greatly differ from the iATE and cATE estimands \cite{lee_how_2025}.
Importantly, while PB-CRT and CRXO designs permit similar model-based estimators, the additional complexity in estimand construction and definition under a CRXO trial introduces additional challenges when studying the performance of these common estimators (as we will further discuss in Section \ref{sect:PO_Estimands}); accordingly, the existing results from PB-CRTs are not directly applicable.

To address this critical knowledge gap, we will highlight important considerations regarding the appropriate use of different statistical methods in CRXO trials with informative sizes.
We start in Section \ref{sect:PO_Estimands} by formally presenting the four described estimands of interest with meaningful interpretations, targeting average treatment effects on the individual, cluster-period cell, cluster, and period-levels, before defining different forms of informative sizes under which these estimands may differ. To the best of our knowledge, this is the first article that introduces all 4 estimands under a CRXO design. We then review point estimators from standard statistical models used in CRXO trials in Section \ref{sect:analytic_models} and derive the probability limits of the associated treatment effect estimators in Section \ref{sect:Estimators} to determine which estimators are consistent for the previously defined estimands in the presence of informative sizes.
We build on the results with a simulation study in Section \ref{sect:simulation}, and use these estimators in a reanalysis of a CRXO trial exploring the effect of stress ulcer prophylaxis with proton pump inhibitors, as compared to histamine-2 receptor blockers, on hospital log-length of stay among patients receiving invasive mechanical ventilation.
Section \ref{sect:discussion} concludes with a discussion.

\section{Potential Outcomes \& Estimands}
\label{sect:PO_Estimands}

\begin{figure}[htp]
    \centering
    \includegraphics[width=12cm]{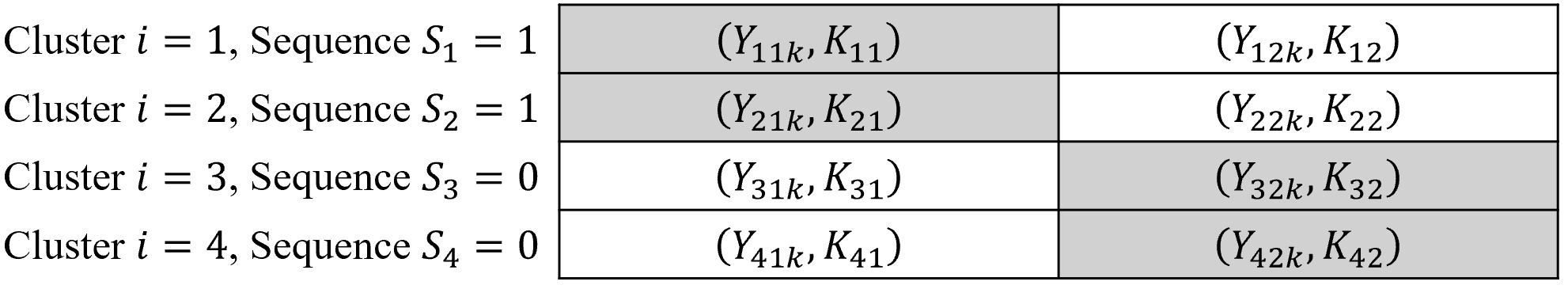}
    \caption{A standard 2-period, 2-sequence, 4 cluster cross-sectional CRXO trial design, with individuals $k \in (1,...,K_{ij})$, in period $j \in (1,2)$ of cluster $i \in (1,...,4)$ randomized to sequence $S_i$ with outcomes $Y_{ijk}$. Cluster-period cells assigned to receive the treatment (control) are shown in gray (white).}
   \label{fig:CRXO_design}
\end{figure}

We assume that data from each cluster, indexed by $i=1,...,I$, are collected in 2 discrete, equally-spaced periods, indexed by $j=1,2$.
We primarily focus on the basic cross-sectional CRXO trial design with 2 periods, 2 crossover sequences, $I/2$ clusters equally randomized to each sequence, and $K_{ij}$ individuals in each cluster-period cell (Figure \ref{fig:CRXO_design}).
Cluster $i$ randomized to sequence $S_{i}=1$ is assigned to receive the treatment in period $j=1$ and control in period $j=2$, whereas $S_{i}=0$ is assigned to receive the control in period $j=1$ and treatment in period $j=2$ (Figure \ref{fig:CRXO_design}).
Overall, the total number of individuals across all clusters and periods is $n=\sum_{i=1}^{I}\sum_{j=1}^{2}K_{ij}$.
In this article, we focus on continuous outcomes $Y_{ijk}$ for each individual $k$ in period $j$ of cluster $i$ (Figure \ref{fig:CRXO_design}).

We follow the potential outcomes framework and define the treatment effect estimands as a difference between potential outcomes under the treatment and control conditions in a CRXO trial. We denote $Y_{ijk}(z)$ as the potential outcome for individual $k$ in period $j$ of cluster $i$, assigned to receive treatment $z=1$ (treatment) or $0$ (control).
We can then define the individual treatment effect (ITE):
\[\text{ITE}_{ijk} = Y_{ijk}(1)-Y_{ijk}(0) 
\,.\]
The observed outcome $Y_{ijk}$ can be connected to the potential outcomes via the cluster-level Stable Unit Treatment Value Assumption:
\begin{equation}
\label{eq:PO}
Y_{i1k}=S_{i}Y_{i1k}(1)+(1-S_{i})Y_{i1k}(0) \,,~~~Y_{i2k}=S_{i}Y_{i2k}(0)+(1-S_{i})Y_{i2k}(1)
\end{equation}
in periods $j=1$ and $2$, with the sequence indicator $S_{i}$.

To define estimands, we consider a cluster superpopulation framework, where sampled clusters are independent and identically distributed draws from an infinite superpopulation of clusters. Accordingly, randomness is introduced in the sampling of clusters, and also by the subsequent randomization of half the sampled clusters to the different crossover sequences. 
All individuals within each cluster-period cell are observed with no further sub-sampling of individuals.

Under the 2-period, 2-sequence CRXO design, we adopt the general class of weighted average treatment effect (wATE) estimands described in Chen \& Li \cite{chen_model-assisted_2025} (originally studied for stepped wedge designs) as the finite population average, simplified to:
\begin{equation}
\label{eq:wATE}
    \text{wATE} = \frac{\sum_{i=1}^{I} \sum_{j=1}^{J} \sum_{k=1}^{K_{ij}}w_{ijk}[Y_{ijk}(1)-Y_{ijk}(0]}{\sum_{i=1}^I \sum_{j=1}^{J} \sum_{k=1}^{K_{ij}}w_{ijk}} \,,
\end{equation}
with the total number of periods $J=2$. In Table \ref{tab:estimands_weights}, we outline the corresponding individual-specific weight $w_{ijk} \geq 0$ to define the following estimands:
\begin{enumerate}
    \item individual-average treatment effect (iATE)
    \item cluster-period-average treatment effect (cpATE)
    \item cluster-average treatment effect (cATE)
    \item period-average treatment effect (pATE)
\end{enumerate}
as the average difference between the potential outcomes under the treatment and control conditions across (1.) individuals, (2.) cluster-period cells, (3.) clusters, and (4.) periods, respectively (Table \ref{tab:estimands_weights}). That is, due to the addition of a crossover period, there are two more distinct estimands, the cpATE and pATE, that can be defined and interpreted in contrast to a conventional P-CRT or PB-CRT.
In such multi-period CRT designs containing multiple periods with treatment positivity (where there is at least one cluster in either treatment condition), the cATE can be more accurately defined as the average treatment effect on the cluster-level, pooled over all observed time periods with the aforementioned treatment positivity.
Conversely, in a P-CRT or PB-CRT, the cATE and cpATE coincide by design, as do the iATE and pATE, since the treatment is typically only administered during a single period \cite{wang_two_2022,lee_how_2025}.
Importantly, while Chen \& Li \cite{chen_model-assisted_2025} have identified the iATE, cpATE, and pATE as special cases within the family of wATE estimands, we additionally identify the cATE as a member that carries a natural interpretation, akin to the cATE defined in a P-CRT and PB-CRT (Table \ref{tab:estimands_weights}).

\begin{table}
\caption{A summary of four interpretable causal estimands the individual-average treatment effect (iATE), cluster-period-average treatment effect (cpATE), cluster-average treatment effect (cATE), and period-average treatment effect (pATE), under the general family of wATE estimands (equation \ref{eq:wATE}). Estimands are defined for a finite population or superpopulation with potential outcomes $Y_{ijk}(z)$ for individuals $k \in (1,...,K_{ij})$ in period $j \in (1,2)$ of cluster $i \in (1,...,I)$ using estimand weights $w_{ijk}$.
}
\label{tab:estimands_weights}
\begin{center}
\bgroup
\def\arraystretch{1.5}
\begin{tabular}{|c l c c|} 
    \hline
    & \textbf{Weights ($w_{ijk}$)} & \textbf{Finite population} & \textbf{Superpopulation}\\
    \hline\hline
    \multicolumn{4}{|l|}{\textbf{iATE}: the average treatment effect defined on the individual-level.}
    \\
    & $w_{ijk}=1$ &
    \scalebox{0.8}{$ \displaystyle\frac{\sum_{i=1}^I \sum_{j=1}^{2} \sum_{k=1}^{K_{ij}} \left[Y_{ijk}(1)-Y_{ijk}(0)\right] }{\sum_{i=1}^I\sum_{j=1}^{2}K_{ij}}$} & \scalebox{0.8}{$\displaystyle\frac{E\left[\sum_{j=1}^{2} \sum_{k=1}^{K_{ij}} \left[Y_{ijk}(1)-Y_{ijk}(0)\right] \right]}{E\left[\sum_{j=1}^{2}K_{ij}\right]}$}\\
    \hline
    \multicolumn{4}{|l|}{\textbf{cpATE}: the average treatment effect defined across individuals on the cluster-period cell-level.}
    \\
    & $w_{ijk}=\frac{1}{K_{ij}}$ &
    \scalebox{0.8}{$\displaystyle\frac{1}{I} \sum_{i=1}^I \displaystyle\frac{1}{2} \sum_{j=1}^{2} \left[ \displaystyle\frac{ \sum_{k=1}^{K_{ij}} \left[Y_{ijk}(1)-Y_{ijk}(0)\right]}{K_{ij}} \right]$} & \scalebox{0.8}{$E\left[ \displaystyle\frac{1}{2} \sum_{j=1}^{2} \displaystyle\frac{ \sum_{k=1}^{K_{ij}} \left[Y_{ijk}(1)-Y_{ijk}(0)\right]}{K_{ij}} \right]$} \\
    \hline
    \multicolumn{4}{|l|}{\textbf{cATE}: the average treatment effect defined across individuals on the cluster-level.}
    \\
    & $w_{ijk}=\frac{1}{\sum_{j=1}^{2}K_{ij}}$ &
    \scalebox{0.8}{$\displaystyle\frac{1}{I} \sum_{i=1}^I \left[ \displaystyle\frac{\sum_{j=1}^{2} \sum_{k=1}^{K_{ij}} \left[Y_{ijk}(1)-Y_{ijk}(0)\right]}{\sum_{j=1}^{2}K_{ij}} \right]$} & \scalebox{0.8}{$E\left[ \displaystyle\frac{\sum_{j=1}^{2} \sum_{k=1}^{K_{ij}} \left[Y_{ijk}(1)-Y_{ijk}(0)\right]}{\sum_{j=1}^{2}K_{ij}} \right]$}\\
    \hline
    \multicolumn{4}{|l|}{\textbf{pATE}: the average treatment effect defined across individuals on the period level.}
    \\
    & $w_{ijk}=\frac{1}{\sum_{i=1}^{I}K_{ij}}$ &
    \scalebox{0.8}{$\displaystyle\frac{1}{2} \sum_{j=1}^{2} \left[ \displaystyle\frac{\sum_{i=1}^I\sum_{k=1}^{K_{ij}} \left[Y_{ijk}(1)-Y_{ijk}(0)\right]}{\sum_{i=1}^I K_{ij}} \right]$} & \scalebox{0.8}{$\displaystyle\frac{1}{2} \sum_{j=1}^{2} \left[ \displaystyle\frac{E\left[ \sum_{k=1}^{K_{ij}} \left[Y_{ijk}(1)-Y_{ijk}(0)\right] \right]}{E\left[K_{ij}\right]} \right]$}\\
    \hline
\end{tabular}
\egroup
\end{center}
\end{table}

The precise definitions of the four estimands are summarized in Table \ref{tab:estimands_weights}. Conceptually, these four estimands address different levels of hypothesis. The iATE addresses an individual-level effect hypothesis; cpATE, a cluster-period or cell-level effect hypothesis; cATE, a cluster-level effect hypothesis; pATE, a period-level effect hypothesis. We focus on these four estimands due to their natural and relevant interpretations, and by no means indicate that these are the only possible estimands of interest in CRXO trials.
A more detailed discussion of estimands in simpler P-CRT settings is included in Kahan et al. \cite{kahan_estimands_2023}. The cpATE and pATE have received comparatively less attention and, as we have highlighted, only differ in definition from the cATE and iATE, respectively, in multi-period CRT designs where multiple periods have within-period treatment positivity.
Notably, the cpATE is often the implicit target estimand in the analysis of multi-period CRT designs using cluster-period cell summaries.
Subsequently, the pATE can be interpreted as targeting the average treatment effect when treating each period in a multi-period CRT as a separate P-CRT then averaging the results across the different periods.

Similar to P-CRTs and PB-CRTs, the magnitude of these estimands may differ in the presence of ``informative sizes'' where there are treatment effects that may vary according to cluster, period, and/or cluster-period sizes.
Accordingly, these estimands will play a crucial role in distinguishing between different forms of informative sizes (Section \ref{sect:IS}).

Finally, when potential outcomes have identical marginal means over cluster-period cells and cluster-period sizes are independent of the potential outcomes $(K_{ij} \indep {Y_{ijk}(1), Y_{ijk}(0)})$, which is the case when there are non-informative sizes, the iATE, cpATE, cATE, and pATE estimands carry the same magnitude and all reduce to:
\[ \text{ATE}=E\left[Y_{ijk}(1)-Y_{ijk}(0)\right] \,,\]
which can be referred to as the average treatment effect (ATE). This non-informative size assumption has generally been implicitly assumed in previous literature on CRXO trials \cite{turner_analysis_2007, li_power_2019,morgan_choosing_2017,mckenzie_reporting_2025}. 

\subsection{Informative Sizes and illustrative examples}
\label{sect:IS}

We broadly define three types of informative sizes below, with more specifics regarding the sufficient conditions for them to occur detailed in Table \ref{tab:estimands_equiv}.2.
(1.) Informative cluster sizes (ICS) are scenarios where treatment effects vary by cluster size (aggregated over 2 periods), such that the iATE and cATE differ in magnitude $(\text{iATE} \neq \text{cATE})$. (2.) Informative period sizes (IPS) are scenarios where treatment effects vary by period size (aggregated over all clusters), such that the iATE and pATE differ $(\text{iATE} \neq \text{pATE})$. A more specific subset of scenarios with non-informative period sizes involve those where cluster-period cell sizes are equal between-periods, within-clusters $(K_{i1}=K_{i2} \, \forall \, i)$. Finally, (3.) informative cluster-period sizes (ICPS) are scenarios where treatment effects vary by cluster-period size, such that the cpATE differs from both the cATE and pATE $(\text{cpATE} \notin (\text{cATE}, \text{pATE}))$.
We can use $\text{ICPS}_c$ to refer to conditions where cpATE $\neq$ cATE and $\text{ICPS}_p$ when cpATE $\neq$ pATE. In other words, ICPS occurs when $\text{ICPS}_c \cap \text{ICPS}_p$ are true.

\begin{figure}[htp]
    \centering
    \includegraphics[width=10cm]{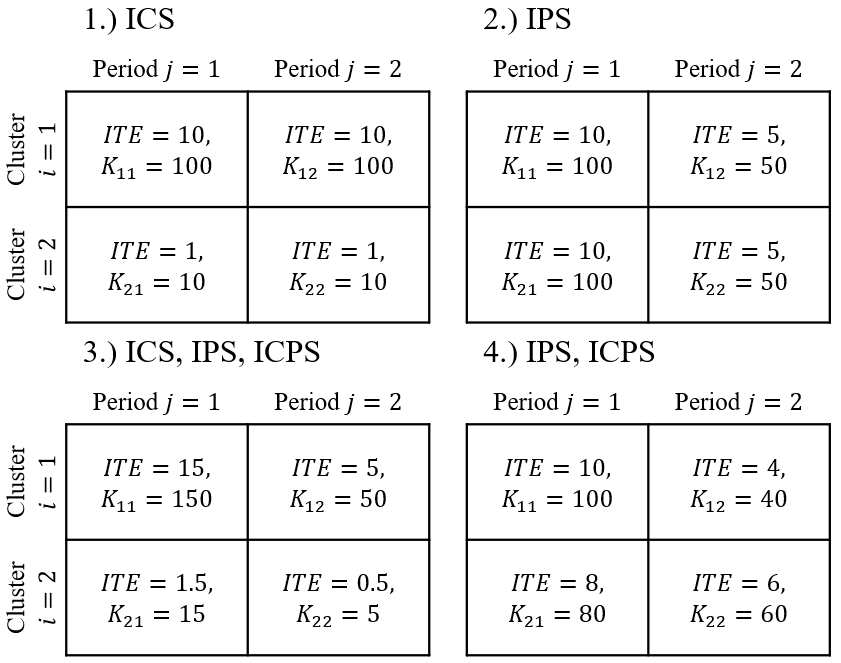}
    \caption{Different individual treatment effects (assuming a constant $ITE_{ijk}=Y_{ijk}(1)-Y_{ijk}(0)$ within each cluster-period for simplicity, simply referred to here as ``ITE") with potential outcomes and cluster sizes $(K_{ij})$ in a 2 cluster, 2-period cross-sectional CRXO trial design with 1.) informative cluster sizes, 2.) informative period sizes, 3.) informative cluster sizes, informative period sizes, and informative cluster-period sizes, and 4.) informative period sizes and informative cluster-period sizes.}
   \label{fig:CRXO_IS}
\end{figure}

Intuitively, ICS may occur when certain clusters with more effective treatment effects also recruit more individuals over the duration of the trial.
IPS may occur when there are time-varying treatment effects \cite{lee_analysis_2024,lee_cluster_2024} that coincide with changes in the sample size over time.
For example, treatments for certain diseases may change in effectiveness over time, corresponding to certain seasons, which also influence the number of patients affected by the disease.
Finally, ICPS may occur on top of ICS and/or IPS, with changes in treatment effects coinciding with changes in cluster-period cell sizes.
We include some different illustrative examples of scenarios with different informative sizes in Figure \ref{fig:CRXO_IS}.

Notably, informative sizes can also refer to scenarios where outcomes (not just treatment effects) vary according to cluster, period, or cluster-period sizes. However, we do not focus on this definition due to our estimands of interest being defined as differences on an absolute scale, therefore the described estimands only differ when treatment effects vary and not outcomes \cite{kahan_demystifying_2024}.

We can specify when the iATE, cATE, pATE, and cpATE estimands coincide under the following set of conditions:
\begin{enumerate}[label=(\alph*)]
    \item $\{Y_{i1k}(1)-Y_{i1k}(0),Y_{i2k}(1)-Y_{i2k}(0)\} \indep \left(K_{i1}, K_{i2}\right)$
    \item $E[Y_{i1k}(1)-Y_{i1k}(0)]=E[Y_{i2k}(1)-Y_{i2k}(0)]$
    \item $E[K_{i1}]=E[K_{i2}]$
    \item $\{Y_{i1k}(1)-Y_{i1k}(0)\} \indep K_{i1}$ and $\{Y_{i2k}(1)-Y_{i2k}(0)\} \indep K_{i2}$
    \item $K_{i1}+K_{i2}=\text{constant}$
    \item $K_{i1}=K_{i2}$
\end{enumerate}
for all clusters $i$. These conditions are interpreted as (a) all individual treatment effects are independent of all cluster-period cell sizes, (b) there is no between-period, within-cluster treatment effect heterogeneity, (c) there is no expected between-period, within-cluster sample size heterogeneity, (d) individual treatment effects are independent of their corresponding cluster-period cell-sizes, (e) there is a common cluster size across all clusters, and (f) cluster-period sizes are equivalent between-periods, within-clusters.
Notably (a) $\subseteq$ (d) and (f) $\subseteq$ (c).

Finally, the complementary of the above conditions, referred to as ($\text{a}^c$) through ($\text{f}^c$), are accordingly defined as the corresponding complementary inequalities and dependencies (and explicitly laid out in Table \ref{tab:estimands_equiv}). Notably, ($\text{d}^c$) $\subseteq$ ($\text{a}^c$) and ($\text{c}^c$) $\subseteq$ ($\text{f}^c$).

In Table \ref{tab:estimands_equiv}.1, we compare the different estimands to find under which of the above conditions that two different estimands may coincide. More details on establishing these results are included in the Appendix (\ref{sect:appendix_estimands_equiv}).
Subsequently, this allows us to specify some sufficient conditions for when informative sizes may occur, without any obvious practically relevant counterexamples (Table \ref{tab:estimands_equiv}.2).
Using these sufficient conditions, we make the following proposition:
\begin{prop}
\label{proposition:ICPS}
$\ $
\begin{enumerate}
    \item ICPS requires either ICS or IPS to occur.
    \item ICPS does not require both ICS and IPS to occur.
\end{enumerate}
\end{prop}
Support for Proposition \ref{proposition:ICPS} is included in the Appendix (\ref{sect:appendix_proposition}).

\begin{table}[htp]
\caption{Sufficient conditions under which the described estimands are i.) equal or ii.) unequal, respectively.
The sufficient conditions for i.) include: (a) $\{Y_{i1k}(1)-Y_{i1k}(0),Y_{i2k}(1)-Y_{i2k}(0)\} \indep \left(K_{i1}, K_{i2}\right)$,
(b) $E[Y_{i1k}(1)-Y_{i1k}(0)]=E[Y_{i2k}(1)-Y_{i2k}(0)]$,
(c) $E[K_{i1}]=E[K_{i2}]$,
(d) $\{Y_{i1k}(1)-Y_{i1k}(0)\} \indep K_{i1}$ and $\{Y_{i2k}(1)-Y_{i2k}(0)\} \indep K_{i2}$,
(e) $E[K_{i1}+K_{i2}]=\text{constant}$, and
(f) $K_{i1}=K_{i2}$.
The sufficient conditions for ii.) include:
($\text{a}^c$) $\{Y_{i1k}(1)-Y_{i1k}(0),Y_{i2k}(1)-Y_{i2k}(0)\} \nindep \left(K_{i1}, K_{i2}\right)$,
($\text{b}^c$) $E[Y_{i1k}(1)-Y_{i1k}(0)] \neq E[Y_{i2k}(1)-Y_{i2k}(0)]$,
($\text{c}^c$) $E[K_{i1}] \neq E[K_{i2}]$,
($\text{d}^c$) $\{Y_{i1k}(1)-Y_{i1k}(0)\} \nindep K_{i1}$ and $\{Y_{i2k}(1)-Y_{i2k}(0)\} \nindep K_{i2}$,
($\text{e}^c$) $K_{i1}+K_{i2} \neq \text{constant}$, and
($\text{f}^c$) $K_{i1} \neq K_{i2}$.}
\label{tab:estimands_equiv}
\begin{center}
\bgroup
\def\arraystretch{1.3}
{
\begin{tabular}{|c c c|} 
    \hline
    i.) & \textbf{Equalities} & \textbf{when the following conditions are \textit{TRUE}} \\
    \hline\hline
    & iATE = cATE & \{(a) $\cap$ (b)\} $\cup$ (e)\\
    \hdashline
    & iATE = pATE & \{(b) $\cap$ (d)\} $\cup$ (c)\\
    \hdashline
    & iATE = cpATE & \{(b) $\cup$ (c)\} $\cap$ (d)\\
    \hdashline
    & cATE = pATE & (a) $\cap$ (b) \\
    \hdashline
    & cATE = cpATE & \{(a) $\cap$ (b)\} $\cup$ (f)\\
    \hdashline
    & pATE = cpATE & (d) \\
    \hline \hline
    ii.) & \textbf{Informative sizes} & \textbf{when the following conditions are \textit{TRUE}} \\
    \hline \hline
    & ICS (iATE $\neq$ cATE) & \{($\text{a}^c$) $\cup$ ($\text{b}^c$)\} $\cap$  ($\text{e}^c$) \\
    \hdashline
    & IPS (iATE $\neq$ pATE) & \{($\text{b}^c$) $\cup$ ($\text{d}^c$)\} $\cap$ ($\text{c}^c$) \\
    \hdashline
    & ICPS (cpATE $\notin$ (cATE, pATE)) & ($\text{d}^c$) $\cap$ ($\text{f}^c$) \\
    \hline
\end{tabular}
}
\egroup
\end{center}
\end{table}

We use the conditions in Table \ref{tab:estimands_equiv} to illustrate how these estimands can differ in magnitude across four example scenarios with different combinations of informative sizes in Figure \ref{fig:CRXO_IS} and Table \ref{tab:CRXO_IS} where there are: 1.) ICS, but no IPS and ICPS, 2.) IPS, but no ICS and ICPS, 3.) ICS, IPS, and ICPS, 4.) IPS and ICPS, but no ICS.

\begin{table}[htp]
\caption{The iATE, cATE, pATE, and cpATE estimand values corresponding to the examples illustrated in Figure \ref{fig:CRXO_IS}.1 - \ref{fig:CRXO_IS}.4.}
\label{tab:CRXO_IS}
\begin{center}
\bgroup
\def\arraystretch{1.3}
\begin{tabular}{|l c c c c|} 
    \hline
    Example & iATE & cATE & pATE & cpATE \\
    \hline\hline
    1.) ICS & $\approx$ 9.1819 & 5.5 & $\approx$ 9.1819 & 5.5 \\
    \hdashline
    2.) IPS & $\approx$ 8.33 & $\approx$ 8.33 & 7.5 & 7.5\\
    \hdashline
    3.) ICS, IPS, ICPS & 11.47727 & 6.875 & $\approx$ 9.1819 & 5.5\\
    \hdashline
    4.) IPS, ICPS & 7.714286 & 7.714286 & $\approx$ 7.156 & 7 \\
    \hline
\end{tabular}
\egroup
\end{center}
\end{table}

In Figure \ref{fig:CRXO_IS}.1 and Table \ref{tab:CRXO_IS}.1, treatment effects and associated cluster-period sizes do not vary between-periods, within-clusters. However, they vary between clusters.
Here, ICS is present $(\text{cATE} \neq \text{iATE})$, however there are no IPS $(\text{pATE}=\text{iATE})$ and no ICPS $(\text{cpATE}=\text{cATE})$.
In this scenario, ICS occurs due to conditions ($\text{a}^c$) and ($\text{e}^c$) being true, but neither IPS nor ICPS occur due to conditions (c) and (f) being true, respectively (Table \ref{tab:estimands_equiv}).

In Figure \ref{fig:CRXO_IS}.2 and Table \ref{tab:CRXO_IS}.2, treatment effects and associated cluster-period sizes do not vary between clusters, within periods. However, they vary between-periods.
Here, IPS is present $(\text{pATE} \neq \text{iATE})$, however there are no ICS $(\text{cATE}=\text{iATE})$ and no IPS $(\text{cpATE}=\text{cATE})$.
In this scenario, IPS occurs due to conditions ($\text{b}^c$) and ($\text{c}^c$) being true, but neither ICS nor ICPS occur due to conditions (d) and (e) being true, respectively (Table \ref{tab:estimands_equiv}).

In Figure \ref{fig:CRXO_IS}.3 and Table \ref{tab:CRXO_IS}.3, treatment effects and associated cluster-period sizes vary between clusters and periods.
Here, ICS $(\text{cATE} \neq \text{iATE})$, IPS $(\text{pATE} \neq \text{iATE})$, and ICPS $(\text{cpATE} \notin (\text{cATE}, \text{pATE}))$ are all present.
In this scenario, ICS, IPS, and ICPS occur due to conditions ($\text{a}^c$) through ($\text{f}^c$) being true (Table \ref{tab:estimands_equiv}).

Finally, in Figure \ref{fig:CRXO_IS}.4 and Table \ref{tab:CRXO_IS}.4, treatment effects and associated cluster-period sizes vary between clusters and periods. However, the marginal cluster sizes and average treatment effects pooled across periods within-clusters do not vary between clusters.
Here, IPS $(\text{pATE} \neq \text{iATE})$ and ICPS $(\text{cpATE} \notin (\text{cATE}, \text{pATE}))$ are present, however there is no ICS $(\text{cATE}=\text{iATE})$.
In this scenario, ICPS and IPS occur due to conditions ($\text{a}^c$) through ($\text{d}^c$) and ($\text{f}^c$) being true, but ICS does not occur due to condition (e) being true (Table \ref{tab:estimands_equiv}).
As in Proposition \ref{proposition:ICPS}.2, we illustrate here that ICPS does not require the presence of both ICS and IPS.

\section{Analytic Models for CRXO Trials}
\label{sect:analytic_models}

There are several common analytic models using individual-level data for CRXO trials \cite{turner_analysis_2007,arnup_appropriate_2016}. First, we describe an analysis with an ``independence estimating equation" (IEE) specified with treatment and period indicator variables. With an identity link, the IEE accordingly yields an equivalent estimator to an ordinary least squares (OLS) estimator. Such an estimator strictly makes ``vertical" within-period comparisons \cite{matthews_stepped_2017}.

CRXO trials can also be analyzed with mixed effects models that account for the within-cluster correlation. In this article, we consider an ``exchangeable mixed effects" (EME) model which resembles the IEE model but additionally specifies a cluster random intercept to induce an exchangeable correlation structure between outcomes. We also consider a ``nested exchangeable mixed effects" (NEME) model, which specifies both a cluster random intercept and a cluster-period random interaction term to induce a nested exchangeable correlation structure between outcomes. 

Finally, we will also analyze CRXO trials with a ``two-way fixed effects models" which includes fixed effects for both clusters and periods to make both between and within-cluster comparisons. Throughout this article, we will simply refer to this as the ``fixed effects" (FE) model. The FE model resembles the EME model, but instead specifies cluster intercepts as fixed rather than random effects. Interestingly, the FE model has been observed to yield identical results to the EME model in the analysis of CRXO trials with equal cluster-period sizes across all clusters and periods \cite{turner_analysis_2007}.
Additional discussion of different considerations when choosing between mixed effects and fixed effects models can be generally found in Mundlak \cite{mundlak_pooling_1978} and Allison \cite{allison_fixed_2009}, and specifically regarding CRTs in previous work \cite{lee_fixed-effects_2024,lee_inclusion_2022,lee_inclusion_2024}. However, the main purpose of our article is not to discuss the philosophical differences between specifying cluster or cluster-period effects via random or fixed effects. Instead, we are interested in the induced point estimators under these different models, without requiring some possibly stringent model assumptions to be correct.

\subsection{IEE model}

The independence estimating equation (IEE) model has an independence correlation structure for outcomes within and between-clusters, and is specified as:
\begin{equation}
\label{eq:IEE_model}
    Y_{ijk} = X_{ij}\delta + \Phi_{j} + e_{ijk},~~~~e_{ijk} \mathop \sim^{\text{iid}} N(0,\sigma^{2}_{w}),
\end{equation}
where $X_{ij}$ and $\delta$ are the indicator and coefficient for the treatment, $\Phi_j$ are the period indicators for periods $j=1,2$, and $e_{ijk}$ is the residual error. The treatment effect in the IEE model can be estimated using ordinary least squares (OLS), with equation \ref{eq:IEE_model} rewritten as:
\begin{equation}
\label{eq:IEE_model2}
Y = Z\theta_{IEE} + \dot{\epsilon},~~~~\dot{\epsilon} \mathop \sim^{\text{iid}} N(0,\dot{V}),
\end{equation}
where $Y$ is the $n \times 1$ vector of individual-level outcomes $Y_{ijk}$, $Z$ is the conventional $n \times (J+1)$ design matrix and $\theta_{IEE}=(\delta, \Phi_1, \Phi_2)'$ is the $(J+1) \times 1$ vector of parameters (with $J=2$), $\dot{V}=\sigma^2_w \textbf{I}_n$ (where $\textbf{I}_n$ is an $n\times n$ identity matrix) denotes the variance-covariance matrix of $Y$, and $n=\sum_{i=1}^{I}\sum_{j=1}^{2}K_{ij}$ being the total trial sample size.

The resulting IEE point estimator is then:
\begin{equation}
\label{eq:IEE_model_matrix}
\hat{\theta}_{IEE}=(Z'Z)^{-1}Z'Y \,.
\end{equation}

\subsection{EME model}
\label{sect:EME_model}
We can define the exchangeable mixed effects (EME) model by adding a cluster random intercept to the IEE model (equation \ref{eq:IEE_model}) to produce:
\begin{equation}
\label{eq:EME_model}
    Y_{ijk} = X_{ij}\delta + \Phi_{j} + \alpha_i + e_{ijk},~~~~\alpha_{i} \overset{\text{iid}}{\sim} N(0,\tau^{2}_{\alpha}),~~~~e_{ijk} \overset{\text{iid}}{\sim} N(0,\sigma^{2}_{w}),
\end{equation}
where $\alpha_i$ is a normally distributed cluster random intercept.

The treatment effect in the EME model can then be rewritten as:
\begin{equation}
\label{eq:EME_model2}
Y = Z\theta_{EME} + \ddot{\epsilon},~~~~\ddot{\epsilon} \mathop \sim^{\text{iid}} N(0,\ddot{V}),
\end{equation}
\sloppy which resembles the IEE model (equation \ref{eq:IEE_model2}) but is instead defined with $\ddot{V} = \oplus_{i=1}^I R_{i}^{EME} = \text{diag}\left(R_{1}^{EME},R_{2}^{EME},\ldots,R_{I}^{EME}\right)$ denoting the variance-covariance matrix of $Y$, and each block for cluster $i$ $R_{i}^{EME} = \textbf{I}_{\sum_{j=1}^{2} K_{ij}} (\sigma^2_w) + \textbf{J}_{\sum_{j=1}^{2} K_{ij}} (\tau^2_\alpha)$ is a $\sum_{j=1}^{2} K_{ij}\times \sum_{j=1}^{2} K_{ij}$ symmetric matrix (where $\textbf{I}_{\sum_{j=1}^{2} K_{ij}}$ and $\textbf{J}_{\sum_{j=1}^{2} K_{ij}}$ are a $\sum_{j=1}^{2} K_{ij}\times \sum_{j=1}^{2} K_{ij}$ dimension identity matrix and matrix of ones, respectively). The resulting exchangeable mixed effects point estimator, estimated using generalized least squares (GLS), is then:
\begin{equation}
\label{eq:EME_model_matrix}
\hat{\theta}_{EME}=(Z'\ddot{V}^{-1}Z)^{-1}Z'\ddot{V}^{-1}Y \,.
\end{equation}

\subsection{NEME model}
\label{sect:NEME_model}
We can further define the nested exchangeable mixed effects (NEME) model by adding a cluster-period random interaction term to the EME model (equation \ref{eq:EME_model}) to produce:
\begin{equation}
\label{eq:NEME_model}
    Y_{ijk} = X_{ij}\delta + \Phi_{j} + \alpha_i + \gamma_{ij} + e_{ijk},~~~~\alpha_{i} \mathop \sim^{\text{iid}} N(0,\tau^{2}_{\alpha}),~~~~\gamma_{ij} \mathop \sim^{\text{iid}} N(0,\tau^{2}_{\gamma}),~~~~e_{ijk} \mathop \sim^{\text{iid}} N(0,\sigma^{2}_{w}),
\end{equation}
where $\gamma_{ij}$ are the normally distributed cluster-period random interaction terms.

The treatment effect in the NEME model can then be rewritten as:
\begin{equation}
\label{eq:NEME_model2}
Y = Z\theta_{NEME} + \dddot{\epsilon},~~~\dddot{\epsilon} \mathop \sim^{\text{iid}} N(0,\dddot{V}),
\end{equation}
which resembles the IEE and EME models (equations \ref{eq:IEE_model2} \& \ref{eq:EME_model2}) but is instead defined with $\dddot{V}=\oplus_{i=1}^I R_{i}^{NEME}=\text{diag}\left(R_{1}^{NEME},R_{2}^{NEME},\ldots,R_{I}^{NEME}\right)$ denoting the variance-covariance matrix of $Y$, and each block $R_{i}^{NEME}$ is a $\sum_{j=1}^{2} K_{ij}$ by $\sum_{j=1}^{2} K_{ij}$ symmetric matrix that can be written as the following block matrices:
\[
    R_{i}^{NEME} =
    \begin{pmatrix}
        R_{i1}^{NEME} & R_{i2}^{NEME}\\
        R_{i3}^{NEME} & R_{i4}^{NEME}
    \end{pmatrix} \,.
\]
Assuming equal cluster-period cell sizes between-periods, within-clusters: $K_{i1}=K_{i2}=K_{i-} \, \forall \, i$, for simplicity, the components of the correlation matrix are then: $R_{i1}^{NEME} = R_{i4}^{NEME} = \textbf{I}_{K_{i-}} (\sigma^2_w) + \textbf{J}_{K_{i-}} (\tau^2_\alpha + \tau^2_\gamma)$ and $R_{i2}^{NEME} = R_{i3}^{NEME} = \textbf{J}_{K_{i-}} (\tau^2_\alpha)$ (where $\textbf{I}_{K_{i-}}$ and $\textbf{J}_{K_{i-}}$ are a $K_{i-}$ by $K_{i-}$ dimension identity matrix and matrix of ones, respectively). The resulting nested exchangeable mixed effects point estimator, estimated using generalized least squares (GLS), is then:
\begin{equation}
\hat{\theta}_{NEME}=(Z'\dddot{V}^{-1}Z)^{-1}Z'\dddot{V}^{-1}Y \,.
\end{equation}

\subsection{FE model}
We can define the fixed effects (FE) model in the analysis of a CRXO trial with treatment, period, and cluster indicators, shown below:
\begin{equation}
\label{eq:FE_model}
    Y_{ijk}=X_{ij}\delta + \Phi_2 + \alpha_i + e_{ijk},~~~~e_{ijk} \mathop \sim^{\text{iid}} N(0,\sigma^{2}_{w}),
\end{equation}
where $\Phi_1=0$ for identifiability, and $\alpha_i$ are the $I$ fixed cluster intercepts. The treatment effect in the FE model can be estimated using OLS with equation \ref{eq:FE_model} rewritten as:
\begin{equation}
\label{eq:FE_model2}
    Y = \Tilde{Z}\theta_{FE} + \dot{\epsilon},~~~~\dot{\epsilon} \mathop \sim^{\text{iid}} N(0,\dot{V}).
\end{equation}
This resembles the IEE model (equation \ref{eq:IEE_model2}), but is instead defined with $\Tilde{Z}$ as the $n$ by $(I+2)$ design matrix and $\theta_{FE}=(\delta,\Phi_2, \alpha_1 ,..., \alpha_I)'$ is the $(I+2)$ by $1$ vector of coefficients in the described 2-period CRXO design.

The resulting fixed effects point estimator is then:
\begin{equation}
\hat{\theta}_{FE}=(\Tilde{Z}'\Tilde{Z})^{-1}\Tilde{Z}'Y \,.
\end{equation}

\section{Convergence Limits and Implied Estimands of Different Estimators in CRXO Trials}
\label{sect:Estimators}

In this section, we discuss the different estimators and their convergence probability limits; full derivations of the estimators are included in the Appendix (\ref{sect:appendix_IEE_IEEcpw_IEEcw} - \ref{sect:appendix_FE_FEcpw_FEcw}). The sufficient conditions under which each estimator converges to the iATE, cpATE, cATE, or pATE estimands are summarized in Table \ref{tab:summary}.

The individual-level estimators described in Section \ref{sect:analytic_models} are expected to target the iATE estimand.
Intuitively, the cpATE, cATE, and pATE estimands may then be targeted by utilizing inverse cluster-period size weights, inverse cluster size weights, and inverse period size weights, respectively.
Such weighted treatment effect estimators are commonly implemented as described in Williamson et al. \cite{williamson_marginal_2003}. Here, we will initially assume for simplicity that cluster-period sizes vary between-clusters but not between-periods, within-clusters, $K_{i1} = K_{i2} = K_{i-} \, \forall \, i$.
Accordingly, the weighted treatment effect estimators can be obtained by solving for the weighted estimating equations:
\[
\sum_{i=1}^{I} \frac{D'_{i}V_{i}^{-1}(Y_i-\mu_i)}{w_i}=0
\]
where $Y_i = (Y_{i11},...,Y_{i1K_{i-}},Y_{i21},...,Y_{i2K_{i-}})'$ and $\mu_i = (\mu_{i1},...,\mu_{i1},\mu_{i2},...,\mu_{i2})'$ are the $(2K_{i-})$ by $1$ vector of individual-level outcomes and marginal means $\mu_i = E[Y_{ijk}|Z_i] = Z_i\theta$ ($\Tilde{Z}$ in a fixed effects model) in cluster $i$. Additionally, $D_i = \frac{d\mu_i}{d\theta}$. Here, the quantity $w_i$ is defined as the cluster-specific weight equal to 1 to equally weigh all individuals, $K_{i-}$ to equally weigh all cluster-periods, or $\sum_{j=1}^2 K_{ij} = 2K_{i-}$ to equally weigh all clusters. Period-specific weights can be difficult to define under such a specification, due to multiple periods being nested within-clusters. We will discuss this in more detail near the end of this section.
Accordingly, we can rewrite the above equation as:
\[
\sum_{i=1}^{I} D'_{i} \textit{W}_{i}^{-1} (Y_i-\mu_i)=0
\]
and solve for $\theta$ in the above estimating equations with an identity link:
\begin{equation}
    \hat{\theta} = \left(\sum_{i=1}^{I} Z'_{i} \textit{W}_{i}^{-1} Z_i \right)^{-1} \left(\sum_{i=1}^{I} Z'_{i} \textit{W}_{i}^{-1} Y_i \right) = \left(Z' \textit{W}^{-1} Z \right)^{-1} \left(Z' \textit{W}^{-1} Y \right)
\end{equation}
($\Tilde{Z}$ in a fixed effects model).
We define $\textit{W} = \oplus_{i=1}^I Q_i$ for $I$ total clusters. With IEE, EME, NEME, and FE estimators, we define the corresponding model and cluster-specific values of $Q_i$ as: $Q_i^{IEE} = Q_i^{FE} = w_i \sigma^2_e{\textbf{I}_{\sum_{j=1}^{2}K_{ij}}}$, $Q_i^{EME} = w_i R_i^{EME}$, and $Q_i^{NEME} = w_i R_i^{NEME}$ (with $R_i^{EME}$ and $R_i^{NEME}$ previously defined in Sections \ref{sect:EME_model} and \ref{sect:NEME_model}, respectively).

Notably, when cluster-period cell sizes are equal between-periods, within-clusters ($K_{i1} = K_{i2} = K_{i-} \, \forall \, i$), the cpATE and cATE estimands will coincide and implementing inverse cluster-period or inverse cluster-size weights in a given treatment effect estimator will yield equivalent estimators.
We can extend this described weighting, when appropriate, to scenarios where cluster-period sizes vary between-periods, within-clusters, $K_{i1} \neq K_{i2}$. This is easy to extend in analyses with uncorrelated errors (IEE and FE) and can also be simply implemented by performing the corresponding analyses with cluster-period cell means or cluster means \cite{kahan_demystifying_2024}. However, to our knowledge, this extension to analyses with correlated errors within-clusters (EME and NEME) is not clear in the existing literature, nor is it obvious if such weighted mixed effects analyses are even desirable, as we will discuss.

Similarly, inverse-period size weights cannot be implemented in the mixed effects analyses with correlated errors within-clusters (EME and NEME) unless cluster-period sizes are equivalent between-periods, within-clusters. Otherwise, only the analyses with uncorrelated errors (IEE and FE) can be specified by implementing inverse period-size weights: $Q_i^{IEE} = Q_i^{FE} = \sigma^2_e diag(w_1 1_{K_{i1}}, w_2 1_{K_{i2}})$, where $1_{K_{ij}}$ is a vector of ones with $K_{ij}$ entries and $w_{j}=\sum_{i=1}^{I}K_{ij}$ is the weight corresponding to period $j$. 

Next, we will discuss the convergence limit and implied estimand under each weighted model approach in the subsequent sections. For ease of reference, we have summarized in Table \ref{tab:summary} the minimum sufficient conditions under which each treatment effect estimator converges to the iATE, cpATE, cATE, or pATE estimands in a CRXO design.

\begin{table}
\caption{The minimum sufficient conditions under which each treatment effect estimator converges to the iATE, cpATE, cATE, or pATE estimands in a CRXO design.
``Always" denotes an estimator that is always consistent in the presence of arbitrary informative sizes.
Conditions with non-informative cluster sizes, non-informative period sizes, and non-informative cluster-period sizes are labeled with ``No ICS" (cATE=iATE), ``No IPS" (pATE=iATE), ``No $\text{ICPS}_c$" (cpATE=cATE), and ``No $\text{ICPS}_p$" (cpATE=pATE), respectively. Scenarios where cluster-period cell sizes $K_{ij}$ are equivalent, $K_{i1}=K_{i2} \,\forall \, i$, are considered a subset of scenarios with no IPS. 
Furthermore, we denote scenarios with no informative sizes as ``$\star$" (no ICS, no IPS, no ICPS, \& iATE=cpATE=cATE=pATE).
The within-period and between-period ICC are denoted as $\rho_{wp}=(\tau^2_\alpha+\tau^2_\gamma)/(\tau^2_\alpha+\tau^2_\gamma+\sigma^2_w)$ and $\rho_{bp}=\tau^2_\alpha/(\tau^2_\alpha+\tau^2_\gamma+\sigma^2_w)$, respectively, and $\lambda_i=K_{i2}/K_{i1}$.}
\label{tab:summary}
\begin{center}
\bgroup
\def\arraystretch{1.2}
\begin{tabular}{|c c c c c|} 
    \hline
    $\hat{\delta}$ & $\xrightarrow{P}$ iATE & $\xrightarrow{P}$ cpATE & $\xrightarrow{P}$ cATE & $\xrightarrow{P}$ pATE \\ [0.5ex] 
    
    \hline\hline
    IEE & \textbf{Always} & $\star$ & No ICS & No IPS \\ \hdashline 
    IEEcpw & $\star$ & \textbf{Always} & No $\text{ICPS}_c$ & No $\text{ICPS}_p$\\ \hdashline
    IEEcw & No ICS & No $\text{ICPS}_c$ & \textbf{Always} & $\star$\\
    \hdashline
    IEEpw & No IPS & No $\text{ICPS}_p$ & $\star$ & \textbf{Always}\\
    
    \hline
    EME & $K_{i1}=K_{i2}$ &
    $\star$ &
    \begin{tabular}{@{}c@{}}$K_{i1}=K_{i2}$ \\ $\cap$ No ICS \end{tabular} & 
    $K_{i1}=K_{i2}$\\ \hdashline
    EMEcpw & $\star$ & $K_{i1}=K_{i2}$ & $K_{i1}=K_{i2}$ & \begin{tabular}{@{}c@{}} $K_{i1}=K_{i2}$ \\ $\cap$ No $\text{ICPS}_p$ \end{tabular} \\ \hdashline
    EMEcw & 
    \begin{tabular}{@{}c@{}} $K_{i1}=K_{i2}$ \\ $\cap$ No ICS \end{tabular} & 
    $K_{i1}=K_{i2}$ & 
    $K_{i1}=K_{i2}$ & 
    $\star$\\ 
    
    \hline
    NEME & 
    \begin{tabular}{@{}c@{}} $\rho_{wp}=\rho_{bp}$ \\ $\cap$ $K_{i1}=K_{i2}$ \end{tabular} & 
    $\star$ & 
    \begin{tabular}{@{}c@{}} $\rho_{wp}=\rho_{bp}$ \\ $\cap$ $K_{i1}=K_{i2}$ \\ $\cap$ No ICS \end{tabular} & 
    \begin{tabular}{@{}c@{}}$\rho_{wp}=\rho_{bp}$ \\ $\cap$ $K_{i1}=K_{i2}$ \end{tabular}\\ \hdashline
    
    NEMEcpw & 
    $\star$ & 
    \begin{tabular}{@{}c@{}}$\rho_{wp}=\rho_{bp}$ \\ $\cap$ $K_{i1}=K_{i2}$ \end{tabular} &
    \begin{tabular}{@{}c@{}}$\rho_{wp}=\rho_{bp}$ \\ $\cap$ $K_{i1}=K_{i2}$ \end{tabular} & 
    \begin{tabular}{@{}c@{}} $\rho_{wp}=\rho_{bp}$ \\ $\cap$ $K_{i1}=K_{i2}$ \\ $\cap$ No $\text{ICPS}_p$ \end{tabular} \\ \hdashline 
    
    NEMEcw & 
    \begin{tabular}{@{}c@{}} $\rho_{wp}=\rho_{bp}$ \\ $\cap$ $K_{i1}=K_{i2}$ \\ $\cap$ No ICS \end{tabular} & 
    \begin{tabular}{@{}c@{}} $\rho_{wp}=\rho_{bp}$ \\ $\cap$ $K_{i1}=K_{i2}$ \end{tabular} & 
    \begin{tabular}{@{}c@{}} $\rho_{wp}=\rho_{bp}$ \\ $\cap$ $K_{i1}=K_{i2}$ \end{tabular} & 
    $\star$\\ 
    
    \hline
    FE & 
    \begin{tabular}{@{}c@{}} $\lambda_i = \lambda \, \forall \, i$ \\ $\cap$ No IPS \end{tabular} & 
    \begin{tabular}{@{}c@{}} $\lambda_i = \lambda \, \forall \, i$ \\ $\cap$ No $\text{ICPS}_p$ \end{tabular} & 
    $\star$ & 
    $\lambda_i = \lambda \, \forall \, i$ \\ \hdashline
    
    FEcpw & $\star$ & \textbf{Always} & No $\text{ICPS}_c$ & No $\text{ICPS}_p$ \\ \hdashline
    
    FEcw & 
    $\star$ & 
    $\lambda_i = \lambda \, \forall \, i$ 
    & \begin{tabular}{@{}c@{}} $\lambda_i = \lambda \, \forall \, i$ \\ $\cap$ No $\text{ICPS}_c$ \end{tabular} 
    & \begin{tabular}{@{}c@{}} $\lambda_i = \lambda \, \forall \, i$ \\ $\cap$ No $\text{ICPS}_p$ \end{tabular}
    \\ \hdashline
    
    FEpw & 
    \begin{tabular}{@{}c@{}} $\lambda_i = \lambda \, \forall \, i$ \\ $\cap$ No IPS \end{tabular} & 
    \begin{tabular}{@{}c@{}} $\lambda_i = \lambda \, \forall \, i$ \\ $\cap$ No $\text{ICPS}_p$ \end{tabular} & 
    $\star$ & 
    $\lambda_i = \lambda \, \forall \, i$ \\
    \hline
\end{tabular}
\egroup
\end{center}
\end{table}

\subsection{Unweighted \& weighted independence estimating equation estimators}
\subsubsection{IEE estimator}

The independence estimating equation (IEE) treatment effect estimator can be written as:
\small
\begin{equation}
    \begin{split}
        &\hat{\delta}_{IEE} = \\
        &\left(\frac{ (\sum_{i=1}^{I}S_{i}\sum_{k=1}^{K_{i1}}Y_{i1k}(1)) (\sum_{i=1}^{I}(1-S_{i})K_{i1}) (\sum_{i=1}^{I}K_{i2}) + (\sum_{i=1}^{I}S_{i}K_{i2}) (\sum_{i=1}^{I}(1-S_{i})\sum_{k=1}^{K_{i2}}Y_{i2k}(1)) (\sum_{i=1}^{I}K_{i1}) }
        { (\sum_{i=1}^{I}S_{i}K_{i1}) (\sum_{i=1}^{I}(1-S_{i})K_{i1}) (\sum_{i=1}^{I}K_{i2}) + (\sum_{i=1}^{I}S_{i}K_{i2}) (\sum_{i=1}^{I}(1-S_{i})K_{i2}) (\sum_{i=1}^{I}K_{i1}) }\right) \\
        &-\left(\frac{ (\sum_{i=1}^{I}S_{i}K_{i1}) (\sum_{i=1}^{I}(1-S_{i})\sum_{k=1}^{K_{i1}}Y_{i1k}(0)) (\sum_{i=1}^{I}K_{i2}) + (\sum_{i=1}^{I}S_{i}\sum_{k=1}^{K_{i2}}Y_{i2k}(0)) (\sum_{i=1}^{I}(1-S_{i})K_{i2}) (\sum_{i=1}^{I}K_{i1}) }
        { (\sum_{i=1}^{I}S_{i}K_{i1}) (\sum_{i=1}^{I}(1-S_{i})K_{i1}) (\sum_{i=1}^{I}K_{i2}) + (\sum_{i=1}^{I}S_{i}K_{i2}) (\sum_{i=1}^{I}(1-S_{i})K_{i2}) (\sum_{i=1}^{I}K_{i1}) }\right) \,.
    \end{split}
\end{equation}
\normalsize
Under cluster randomization, the sequence variable $S_{i}$ is independent of the potential outcomes and cluster-period sizes, $S_{i} \indep \Omega$, with $\Omega=\{Y_{ijk}(0), Y_{ijk}(1), K_{ij}\}_{i=1, k=1}^{I, K_{ij}}$.
Subsequently, we can demonstrate that this estimator is consistent and asymptotically unbiased for the iATE:
\begin{equation}
  \hat{\delta}_{IEE} \xrightarrow{P} \frac{E\left[\sum_{j=1}^{2} \sum_{k=1}^{K_{ij}} \left[Y_{ijk}(1)-Y_{ijk}(0)\right] \right]}{E\left[\sum_{j=1}^{2}K_{ij}\right]}
\end{equation}
with more information included in the Appendix (\ref{sect:appendix_IEE}).

\subsubsection{IEEcpw estimator}
\label{sect:IEEcpw}

The independence estimating equation with inverse cluster-period size weights (IEEcpw) treatment effect estimator can be written as:
\begin{equation}
    \hat{\delta}_{IEEcpw} =\frac{1}{I}\sum_{i=1}^{I} 
    \left(
        S_{i}\left( \frac{\sum_{k=1}^{K_{i1}}Y_{i1k}(1)}{K_{i1}} - \frac{\sum_{k=1}^{K_{i2}}Y_{i2k}(0)}{K_{i2}} \right) + 
        (1-S_{i}) \left(\frac{\sum_{k=1}^{K_{i2}}Y_{i2k}(1)}{K_{i2}} - \frac{\sum_{k=1}^{K_{i1}}Y_{i1k}(0)}{K_{i1}} \right) 
    \right) \,.
\end{equation}
Under cluster randomization, not only can we demonstrate that the IEEcpw estimator is consistent for the cpATE estimand:
\begin{equation}
    \hat{\delta}_{IEEcpw} \xrightarrow{P} E\left[ \frac{1}{2} \sum_{j=1}^{2} \frac{ \sum_{k=1}^{K_{ij}} \left[Y_{ijk}(1)-Y_{ijk}(0)\right]}{K_{ij}} \right] \,,
\end{equation}
but we can further demonstrate that the IEEcpw estimator is unbiased in expectation for the cpATE:
\begin{equation}
    E[E[\hat{\delta}_{IEEcpw}|\Omega]] = E[\hat{\delta}_{IEEcpw}] = E\left[ \frac{1}{2} \sum_{j=1}^{2} \frac{ \sum_{k=1}^{K_{ij}} \left[Y_{ijk}(1)-Y_{ijk}(0)\right]}{K_{ij}} \right]
\end{equation}
with more information included in the Appendix (\ref{sect:appendix_IEEcpw}).

\subsubsection{IEEcw estimator}

The independence estimating equation with inverse cluster size weights (IEEcw) treatment effect estimator can be written as:
\begin{equation}
    \begin{split}
        &\hat{\delta}_{IEEcw} =\\
        &\left(\frac{
        \begin{tabular}{l}
            $\left(\sum_{i=1}^{I}S_{i}\frac{\sum_{k=1}^{K_{i1}}Y_{i1k}(1)}{\sum_{j=1}^{2}K_{ij}}\right) \left(\sum_{i=1}^{I}(1-S_{i})\frac{K_{i1}}{\sum_{j=1}^{2}K_{ij}}\right) \left(\sum_{i=1}^{I}\frac{K_{i2}}{\sum_{j=1}^{2}K_{ij}}\right)$ \\
            \indent $+ \left(\sum_{i=1}^{I}S_{i}\frac{K_{i2}}{\sum_{j=1}^{2}K_{ij}}\right) \left(\sum_{i=1}^{I}(1-S_{i})\frac{\sum_{k=1}^{K_{i2}}Y_{i2k}(1)}{\sum_{j=1}^{2}K_{ij}}\right) \left(\sum_{i=1}^{I}\frac{K_{i1}}{\sum_{j=1}^{2}K_{ij}}\right)$
        \end{tabular}
        }{ 
            \begin{tabular}{l}
                $\left(\sum_{i=1}^{I}S_{i}\frac{K_{i1}}{\sum_{j=1}^{2}K_{ij}}\right) \left(\sum_{i=1}^{I}(1-S_{i})\frac{K_{i1}}{\sum_{j=1}^{2}K_{ij}}\right) \left(\sum_{i=1}^{I}\frac{K_{i2}}{\sum_{j=1}^{2}K_{ij}}\right)$ \\
                \indent $+ \left(\sum_{i=1}^{I}S_{i}\frac{K_{i2}}{\sum_{j=1}^{2}K_{ij}}\right) \left(\sum_{i=1}^{I}(1-S_{i})\frac{K_{i2}}{\sum_{j=1}^{2}K_{ij}}\right) \left(\sum_{i=1}^{I}\frac{K_{i1}}{\sum_{j=1}^{2}K_{ij}}\right)$
            \end{tabular} 
        }\right)\\
        &\indent -\left(\frac{ 
        \begin{tabular}{l}
        $\left(\sum_{i=1}^{I}S_{i}\frac{K_{i1}}{\sum_{j=1}^{2}K_{ij}}\right) \left(\sum_{i=1}^{I}(1-S_{i})\frac{\sum_{k=1}^{K_{i1}}Y_{i1k}(0)}{\sum_{j=1}^{2}K_{ij}}\right) \left(\sum_{i=1}^{I}\frac{K_{i2}}{\sum_{j=1}^{2}K_{ij}}\right)$ \\
        \indent $+ \left(\sum_{i=1}^{I}S_{i}\frac{\sum_{k=1}^{K_{i2}}Y_{i2k}(0)}{\sum_{j=1}^{2}K_{ij}}\right) \left(\sum_{i=1}^{I}(1-S_{i})\frac{K_{i2}}{\sum_{j=1}^{2}K_{ij}}\right) \left(\sum_{i=1}^{I}\frac{K_{i1}}{\sum_{j=1}^{2}K_{ij}}\right)$
        \end{tabular}
        }{ 
        \begin{tabular}{l}
            $\left(\sum_{i=1}^{I}S_{i}\frac{K_{i1}}{\sum_{j=1}^{2}K_{ij}}\right) \left(\sum_{i=1}^{I}(1-S_{i})\frac{K_{i1}}{\sum_{j=1}^{2}K_{ij}}\right) \left(\sum_{i=1}^{I}\frac{K_{i2}}{\sum_{j=1}^{2}K_{ij}}\right)$ \\
            \indent $+ \left(\sum_{i=1}^{I}S_{i}\frac{K_{i2}}{\sum_{j=1}^{2}K_{ij}}\right) \left(\sum_{i=1}^{I}(1-S_{i})\frac{K_{i2}}{\sum_{j=1}^{2}K_{ij}}\right) \left(\sum_{i=1}^{I}\frac{K_{i1}}{\sum_{j=1}^{2}K_{ij}}\right)$ 
            \end{tabular}
        }\right) \,.
    \end{split}
\end{equation}
Under cluster randomization, we can demonstrate that this estimator is consistent and asymptotically unbiased for the cATE:
\begin{equation}
    \hat{\delta}_{IEEcw} \xrightarrow{P} E\left[ \frac{\sum_{j=1}^{2} \sum_{k=1}^{K_{ij}} \left[Y_{ijk}(1)-Y_{ijk}(0)\right]}{\sum_{j=1}^{2}K_{ij}} \right]
\end{equation}
with more information included in the Appendix (\ref{sect:appendix_IEEcw}).

\subsubsection{IEEpw estimator}

The independence estimating equation with inverse period size weights (IEEpw) treatment effect estimator can be written as:

\begin{equation}
    \begin{split}
    &\hat{\delta}_{IEEpw} =\\
    &\left(\frac{
        \left(\sum_{i=1}^{I}S_{i}\frac{\sum_{k=1}^{K_{i1}}Y_{i1k}(1)}{\sum_{i=1}^{I}K_{i1}}\right) \left(\sum_{i=1}^{I}(1-S_{i})\frac{K_{i1}}{\sum_{i=1}^{I}K_{i1}}\right)
        + \left(\sum_{i=1}^{I}S_{i}\frac{K_{i2}}{\sum_{i=1}^{I}K_{i2}}\right) \left(\sum_{i=1}^{I}(1-S_{i})\frac{\sum_{k=1}^{K_{i2}}Y_{i2k}(1)}{\sum_{i=1}^{I}K_{i2}}\right)
    }{ 
        \left(\sum_{i=1}^{I}S_{i}\frac{K_{i1}}{\sum_{i=1}^{I}K_{i1}}\right) \left(\sum_{i=1}^{I}(1-S_{i})\frac{K_{i1}}{\sum_{i=1}^{I}K_{i1}}\right)
        + \left(\sum_{i=1}^{I}S_{i}\frac{K_{i2}}{\sum_{i=1}^{I}K_{i2}}\right) \left(\sum_{i=1}^{I}(1-S_{i})\frac{K_{i2}}{\sum_{i=1}^{I}K_{i2}}\right)
    }\right) \\
    & \indent -\left(\frac{ 
        \left(\sum_{i=1}^{I}S_{i}\frac{K_{i1}}{\sum_{i=1}^{I}K_{i1}}\right) \left(\sum_{i=1}^{I}(1-S_{i})\frac{\sum_{k=1}^{K_{i1}}Y_{i1k}(0)}{\sum_{i=1}^{I}K_{i1}}\right)
        + \left(\sum_{i=1}^{I}S_{i}\frac{\sum_{k=1}^{K_{i2}}Y_{i2k}(0)}{\sum_{i=1}^{I}K_{i2}}\right) \left(\sum_{i=1}^{I}(1-S_{i})\frac{K_{i2}}{\sum_{i=1}^{I}K_{i2}}\right)
    }{ 
        \left(\sum_{i=1}^{I}S_{i}\frac{K_{i1}}{\sum_{i=1}^{I}K_{i1}}\right) \left(\sum_{i=1}^{I}(1-S_{i})\frac{K_{i1}}{\sum_{i=1}^{I}K_{i1}}\right)
        + \left(\sum_{i=1}^{I}S_{i}\frac{K_{i2}}{\sum_{i=1}^{I}K_{i2}}\right) \left(\sum_{i=1}^{I}(1-S_{i})\frac{K_{i2}}{\sum_{i=1}^{I}K_{i2}}\right)
    }\right) \,.
    \end{split}
\end{equation}

Under cluster randomization, we can demonstrate that this estimator is consistent and asymptotically unbiased for the pATE:

\begin{equation}
    \hat{\delta}_{IEEpw} \xrightarrow{P} \frac{1}{2} \sum_{j=1}^{2} \left[ \frac{E\left[ \sum_{k=1}^{K_{ij}} \left[Y_{ijk}(1)-Y_{ijk}(0)\right] \right]}{E\left[K_{ij}\right]} \right] \,.
\end{equation}
with more information included in the Appendix (\ref{sect:appendix_IEEpw}).

\subsection{Unweighted \& weighted mixed effects model estimators}

In general, we can demonstrate that the unweighted and weighted EME and NEME treatment effect estimators all asymptotically converge to the same general form of:
\begin{equation}
\label{eq:ME_general}
    \hat{\delta} \xrightarrow{P} \frac{ E\left[ \frac{1}{K_{i-}}(A_{i}-C_{i})\left(\sum_{j=1}^{2}\sum_{k=1}^{K_{i-}}\left[Y_{ijk}(1)-Y_{ijk}(0)\right]\right)\right] }
    { E\left[ 2(A_{i}-C_{i}) \right] }
\end{equation}
with model and weight-specific values of $A_{i}$ and $C_{i}$ (defined below) when cluster-period sizes are equal between-periods and within-clusters ($K_{i1} = K_{i2} = K_{i-} \, \forall \, i$). Recall that in such conditions, the cpATE and cATE are equivalent, as are the iATE and pATE.
Unless $A_i-C_i$ is constant over clusters $i$, $(A_i - C_i) \propto 1$, the mixed effects model treatment effect estimators will generally converge to weighted average treatment effect estimands with data-dependent and model-specific weights that can be difficult to interpret.

\subsubsection{NEME \& EME estimators}

With a nested exchangeable mixed effects (NEME) model treatment effect estimator $\hat{\delta}_{NEME}$, the model-specific values of $A_{i}$ and $C_{i}$ are:
\[A_{i}=(K_{i-})\left(\frac{\sigma^{2}_{w}+(K_{i-})(\tau^{2}_{\alpha}+\tau^{2}_{\gamma})}{(\sigma^{2}_{w}+(K_{i-})(\tau^{2}_{\alpha}+\tau^{2}_{\gamma}))^{2}-((K_{i-})(\tau^{2}_{\alpha}))^{2}}\right)\]
and:
\[C_{i}=-(K_{i-})\left(\frac{(K_{i-})(\tau^{2}_{\alpha})}{(\sigma^{2}_{w}+(K_{i-})(\tau^{2}_{\alpha}+\tau^{2}_{\gamma}))^{2}-((K_{i-})(\tau^{2}_{\alpha}))^{2}}\right)\]
when cluster-period sizes are equal between-periods, within-clusters ($K_{i1} = K_{i2} = K_{i-} \, \forall \, i$).
Recall that $\sigma^{2}_{w}$, $\tau^{2}_{\alpha}$, and $\tau^{2}_{\gamma}$ are the variances of the residual errors, cluster random intercepts, and cluster-period random interaction terms, respectively.
In the NEME estimator:
\[A_{i}-C_{i}=(K_{i-})(\sigma^{2}_{w}+\tau^{2}_{\alpha}+\tau^{2}_{\gamma})\left(\frac{1}{1+(K_{i-}-1)\rho_{wp}-(K_{i-})\rho_{bp}}\right)\]
where $\rho_{wp}=\frac{\tau^{2}_{\alpha}+\tau^{2}_{\gamma}}{\sigma^{2}_{w}+\tau^{2}_{\alpha}+\tau^{2}_{\gamma}}$ and $\rho_{bp}=\frac{\tau^{2}_{\alpha}}{\sigma^{2}_{w}+\tau^{2}_{\alpha}+\tau^{2}_{\gamma}}$ are the within-period (wp-ICC) and between-period (bp-ICC) intracluster correlation, respectively.

Altogether, the NEME estimator converges in probability to:
\begin{equation}
\label{eq:NEME_estimator}
    \hat{\delta}_{NEME} \xrightarrow{P} \frac{ E\left[ \left(\displaystyle\frac{1}{1+(K_{i-}-1)\rho_{wp}-(K_{i-})\rho_{bp}}\right)\left(\sum_{j=1}^{2}\sum_{k=1}^{K_{i-}}\left[Y_{ijk}(1)-Y_{ijk}(0)\right]\right)\right] }
    { E\left[ \left(\displaystyle\frac{1}{1+(K_{i-}-1)\rho_{wp}-(K_{i-})\rho_{bp}}\right)2K_{i-} \right] } \,,
\end{equation}
given $K_{i1} = K_{i2} = K_{i-} \, \forall \, i$. More information is included in the Appendix (\ref{sect:appendix_NEME}).

Notably, the EME estimator is a special case of the NEME estimator where $\rho_{wp}=\rho_{bp}$. The EME estimator can then be derived from Equation \ref{eq:NEME_estimator}: 
\begin{equation}
    \hat{\delta}_{EME} \xrightarrow{P} \frac{ E\left[ \sum_{j=1}^{2}\sum_{k=1}^{K_{i-}}\left[Y_{ijk}(1)-Y_{ijk}(0)\right]\right] }
    { E\left[2K_{i-}\right] } \,,
\end{equation}
demonstrating that the EME estimator is consistent for iATE = pATE estimands when cluster-period sizes are equal between-periods, within-clusters ($K_{i1} = K_{i2} = K_{i-} \, \forall \, i$).
More information is included in the Appendix (\ref{sect:appendix_EME}).

\subsubsection{NEMEcpw, NEMEcw, EMEcpw, \& EMEcw estimators}

Subsequently, the nested exchangeable mixed effects with inverse cluster-period size weights (NEMEcpw) and with inverse cluster size weights (NEMEcw) treatment effect estimators converge to:
\begin{equation}
    \hat{\delta}_{NEMEcpw} = \hat{\delta}_{NEMEcw} \xrightarrow{P} E\left[ \left(\frac{\left(\displaystyle\frac{1}{1+(K_{i-}-1)\rho_{wp}-(K_{i-})\rho_{bp}}\right)}{E\left[\displaystyle\frac{1}{1+(K_{i-}-1)\rho_{wp}-(K_{i-})\rho_{bp}}\right]}\right) \frac{ \sum_{j=1}^{2} \sum_{k=1}^{K_{i-}} \left[Y_{ijk}(1)-Y_{ijk}(0)\right]}{2K_{i-}} \right]
\end{equation}
given $K_{i1} = K_{i2} = K_{i-} \, \forall \, i$. More information is included in the Appendix (\ref{sect:appendix_NEME}).

Altogether, with a nested exchangeable correlation structure, $A_{i}-C_{i}$ remains cluster-specific such that $\hat{\delta}_{NEME}$, $\hat{\delta}_{NEMEcpw}$, and $\hat{\delta}_{NEMEcw}$ converge to weighted average treatment effect estimands with data-dependent weights that depend on both the cluster-period size as well as the probability limit of the model-based wp-ICC and bp-ICC estimators. These weights are generally difficult to interpret, and the resulting estimators are typically not consistent for the iATE, cpATE, nor cATE estimands.

As in the previous section, it is then straightforward to extend the results above to the EMEcpw and EMEcw estimators, when $\rho_{wp}=\rho_{bp}$, and demonstrate that the exchangeable mixed effects with inverse cluster-period size weights (EMEcpw) and with inverse cluster size weights (EMEcw) treatment effect estimators are consistent for the cpATE and cATE estimands:
\begin{equation}
    \hat{\delta}_{EMEcpw} = \hat{\delta}_{EMEcw} \xrightarrow{P} E\left[ \frac{1}{2} \sum_{j=1}^{2} \frac{ \sum_{k=1}^{K_{i-}} \left[Y_{ijk}(1)-Y_{ijk}(0)\right]}{K_{i-}} \right] = E\left[ \frac{\sum_{j=1}^{2} \sum_{k=1}^{K_{i-}} \left[Y_{ijk}(1)-Y_{ijk}(0)\right]}{2K_{i-}} \right]
\end{equation}
given $K_{i1} = K_{i2} = K_{i-} \, \forall \, i$.
More information is included in the Appendix (\ref{sect:appendix_EMEcpw_EMEcw}).

\subsection{Unweighted \& weighted fixed effects model estimators}

\subsubsection{FE estimator}
\label{sect:FE}

The fixed effects (FE) treatment effect estimator can be written as:
\begin{equation}
    \hat{\delta}_{FE} = \left(\frac{1}{4}\right) \left(
        \begin{tabular}{l}
            $\frac{1}{\left( \frac{1}{I/2} \right) \left(\sum_{i=1}^{I} S_i \frac{K_{i1}K_{i2}}{K_{i1}+K_{i2}} \right)} \left( \left( \frac{1}{I/2} \right) \sum_{i=1}^{I} S_{i} \left(\sum_{k=1}^{K_{i1}} Y_{i1k}(1) - \sum_{k=1}^{K_{i2}} Y_{i2k}(0) \right) \right)$
            \\
            \indent $+ \frac{1}{\left( \frac{1}{I/2} \right) \left( \sum_{i=1}^{I} (1-S_i) \frac{K_{i1}K_{i2}}{K_{i1}+K_{i2}}\right)}
            \left( \left( \frac{1}{I/2} \right) \sum_{i=1}^{I} (1-S_i) \sum_{k=1}^{K_{i2}} Y_{i2k}(1) \right)$
            \\ 
            \indent $- \frac{1}{\left( \frac{1}{I/2} \right) \left( \sum_{i=1}^{I} (1-S_i) \frac{K_{i1}K_{i2}}{K_{i1}+K_{i2}}\right)} \left(2 \left( \frac{1}{I/2} \right) \sum_{i=1}^{I} (1-S_i) \left( \frac{K_{i2}}{K_{i1}+K_{i2}} \right) \sum_{k=1}^{K_{i1}} Y_{i1k}(0) \right)$ 
            \\
            \indent $- \frac{1}{\left( \frac{1}{I/2} \right) \left(\sum_{i=1}^{I} S_i \frac{K_{i1}K_{i2}}{K_{i1}+K_{i2}} \right)} \left( \left( \frac{1}{I/2} \right) \sum_{i=1}^{I} S_i \left( \frac{K_{i1}-K_{i2}}{K_{i1}+K_{i2}} \right) \left(\sum_{k=1}^{K_{i1}} Y_{i1k}(1) + \sum_{k=1}^{K_{i2}} Y_{i2k}(0) \right) \right)$
            \\
            \indent $+ \frac{1}{ \left( \frac{1}{I/2} \right) \left( \sum_{i=1}^{I} (1-S_i) \frac{K_{i1}K_{i2}}{K_{i1}+K_{i2}}\right)} \left( \left( \frac{1}{I/2} \right) \sum_{i=1}^{I} (1-S_i) \left( \frac{K_{i1}-K_{i2}}{K_{i1}+K_{i2}} \right) \sum_{k=1}^{K_{i2}} Y_{i2k}(1) \right)$
        \end{tabular} 
        \right) \,.
\end{equation}

Under cluster randomization, the sequence variable $S_{i}$ is independent of the potential outcomes and cluster-period sizes, $S_{i} \indep \Omega$, with $\Omega=\{Y_{ijk}(0), Y_{ijk}(1), K_{ij}\}_{i=1, k=1}^{I, K_{ij}}$. We can then demonstrate that this estimator converges in probability to:
\small
\begin{equation}
\hat{\delta}_{FE} \xrightarrow{P}
\left(
    \frac{
    \left(
        E\left[\left(\frac{K_{i2}}{K_{i1}+K_{i2}}\right)\left(\sum_{k=1}^{K_{i1}} Y_{i1k}(1) - \sum_{k=1}^{K_{i1}} Y_{i1k}(0)\right)\right]
        +
        E\left[\left(\frac{K_{i1}}{K_{i1}+K_{i2}}\right)\left(\sum_{k=1}^{K_{i2}} Y_{i2k}(1) - \sum_{k=1}^{K_{i2}} Y_{i2k}(0)\right)\right]
    \right)
    }
    {2E\left[\frac{K_{i1}K_{i2}}{K_{i1}+K_{i2}}\right]}
\right) \,.
\end{equation}
\normalsize

In a CRXO design, suppose we define $\lambda_{i} = K_{i2}/K_{i1}$ and then assume $\lambda_i=\lambda \, \forall \, i$, then interestingly the FE estimator is consistent for the pATE estimand, where:
\begin{equation}
    \hat{\delta}_{FE} \xrightarrow{P} \frac{1}{2} \sum_{j=1}^{2} \left[ \frac{E\left[ \sum_{k=1}^{K_{ij}} \left[Y_{ijk}(1)-Y_{ijk}(0)\right] \right]}{E\left[K_{ij}\right]} \right] \,.
\end{equation}
That is, if the sample sizes in period $j=2$, relative to period $j=1$, are inflated by a fixed ratio of $\lambda$ for all clusters $i$, the FE estimator is surprisingly a consistent estimator for the pATE estimand, instead of the iATE estimand.
Only when additionally assuming no IPS will the FE estimator be consistent for the iATE estimand. More information is included in the Appendix (\ref{sect:appendix_FE}).

\subsubsection{FEcpw estimator}
\label{sect:FEcpw}

The fixed effects model with inverse cluster-period size weights (FEcpw) treatment effect estimator can be written as:
\begin{equation}
     \hat{\delta}_{FEcpw} = \frac{1}{I}\sum_{i=1}^{I} 
    \left(
        S_{i}\left( \frac{\sum_{k=1}^{K_{i1}}Y_{i1k}(1)}{K_{i1}} - \frac{\sum_{k=1}^{K_{i2}}Y_{i2k}(0)}{K_{i2}} \right) + 
        (1-S_{i}) \left(\frac{\sum_{k=1}^{K_{i2}}Y_{i2k}(1)}{K_{i2}} - \frac{\sum_{k=1}^{K_{i1}}Y_{i1k}(0)}{K_{i1}} \right) 
    \right) \,.
\end{equation}
Under cluster randomization, not only can we demonstrate that the FEcpw estimator is consistent for the cpATE estimand:
\begin{equation}
    \hat{\delta}_{FEcpw} \xrightarrow{P} E\left[ \frac{1}{2} \sum_{j=1}^{2} \frac{ \sum_{k=1}^{K_{ij}} \left[Y_{ijk}(1)-Y_{ijk}(0)\right]}{K_{ij}} \right] \,,
\end{equation}
but we can further demonstrate that the FEcpw estimator is unbiased in expectation for the cpATE:
\begin{equation}
    E[E[\hat{\delta}_{FEcpw}|\Omega]] = E[\hat{\delta}_{FEcpw}] = E\left[ \frac{1}{2} \sum_{j=1}^{2} \frac{ \sum_{k=1}^{K_{ij}} \left[Y_{ijk}(1)-Y_{ijk}(0)\right]}{K_{ij}} \right]
\end{equation}
with more information included in the Appendix (\ref{sect:appendix_FEcpw}).

\subsubsection{FEcw estimator}
\label{sect:FEcw}

The fixed effects model with inverse cluster size weights (FEcw) treatment effect estimator can be written as:
\small
\begin{equation}
    \begin{split}
        &\hat{\delta}_{FEcw} = \\
        &\left(\frac{1}{4}\right) \left(
        \begin{tabular}{l}
            $\frac{1}{\left(\sum_{i=1}^{I} S_i \frac{K_{i1}K_{i2}}{(K_{i1}+K_{i2})^2} \right)} \left(\sum_{i=1}^{I} S_{i} \left( \frac{\sum_{k=1}^{K_{i1}} Y_{i1k}}{\sum_{j=1}^2 K_{ij}} - \frac{\sum_{k=1}^{K_{i2}} Y_{i2k}}{\sum_{j=1}^2 K_{ij}} \right) \right)$
            \\
            \indent $+ \frac{1}{\left( \sum_{i=1}^{I} (1-S_i) \frac{K_{i1}K_{i2}}{(K_{i1}+K_{i2})^2}\right)}
            \left( \sum_{i=1}^{I} (1-S_i) \left(\frac{\sum_{k=1}^{K_{i2}} Y_{i2k}}{\sum_{j=1}^2 K_{ij}}\right) 
            - 2 \sum_{i=1}^{I} (1-S_i) \left( \frac{K_{i2}}{K_{i1}+K_{i2}} \right) \left(\frac{\sum_{k=1}^{K_{i1}} Y_{i1k}}{\sum_{j=1}^2 K_{ij}}\right) \right)$ 
            \\
            \indent $- \frac{1}{\left(\sum_{i=1}^{I} S_i \frac{K_{i1}K_{i2}}{(K_{i1}+K_{i2})^2} \right)} \left( \sum_{i=1}^{I} S_i \left( \frac{K_{i1}-K_{i2}}{K_{i1}+K_{i2}} \right) \sum_{j=1}^{2} \left(\frac{\sum_{k=1}^{K_{ij}} Y_{ijk}}{\sum_{j=1}^2 K_{ij}}\right) \right)$
            \\
            \indent $+ \frac{1}{\left( \sum_{i=1}^{I} (1-S_i) \frac{K_{i1}K_{i2}}{(K_{i1}+K_{i2})^2}\right)} \left( \sum_{i=1}^{I} (1-S_i) \left( \frac{K_{i1}-K_{i2}}{K_{i1}+K_{i2}} \right) \left(\frac{\sum_{k=1}^{K_{i2}} Y_{i2k}}{\sum_{j=1}^2 K_{ij}}\right) \right)$
        \end{tabular}
        \right)
    \end{split}
\end{equation}
\normalsize

Under cluster randomization and defining $\lambda_i=K_{i2}/K_{i1}$, we can demonstrate that the FEcw treatment effect estimator converges in probability to:
\begin{equation}
\begin{split}
\hat{\delta}_{FEcw} \xrightarrow{P} 
 \left(\frac{1}{2E\left[ \frac{\lambda_i}{(1+\lambda_i)^2} \right]}\right)
    E\left[ 
    \begin{tabular}{l}
        $\frac{\lambda_i}{1+\lambda_i}
        \left( \frac{\sum_{k=1}^{K_{i1}} Y_{i1k}(1)}{\sum_{j=1}^2 K_{ij}} - \frac{\sum_{k=1}^{K_{i1}} Y_{i1k}(0)}{\sum_{j=1}^2 K_{ij}} \right)$ \\
        \indent $+ \frac{1}{1+\lambda_i}
        \left( \frac{\sum_{k=1}^{K_{i2}} Y_{i2k}(1)}{\sum_{j=1}^2 K_{ij}} - \frac{\sum_{k=1}^{K_{i2}} Y_{i2k}(0)}{\sum_{j=1}^2 K_{ij}} \right)$
    \end{tabular}
    \right]
 \,.
\end{split}
\end{equation}
Assuming $\lambda_i = \lambda \, \forall \, i$, we can show that the FEcw estimator is surprisingly consistent for the cpATE estimand in a CRXO design:
\begin{equation}
    \hat{\delta}_{FEcw} \xrightarrow{P} E\left[ \frac{1}{2} \sum_{j=1}^{2} \frac{ \sum_{k=1}^{K_{ij}} \left[Y_{ijk}(1)-Y_{ijk}(0)\right]}{K_{ij}} \right] \,.
\end{equation}
In other words, if the sample sizes in period $j=2$, relative to period $j=1$, are inflated by a fixed ratio of $\lambda$ for all clusters $i$, the FEcw estimator will be a consistent estimator for the cpATE estimand, instead of the cATE estimand.
Only when additionally assuming no $\text{ICPS}_c$ (cpATE = cATE) will the FEcw estimator be consistent for the cATE estimand.
More information is included in the Appendix (\ref{sect:appendix_FEcw}).

\subsubsection{FEpw estimator}

Finally, the fixed effects model with inverse period size weights (FEpw) treatment effect estimator can be written as:
\small
\begin{equation}
    \begin{split}
    & \hat{\delta}_{FEpw} = \\ 
    & \left(\frac{1}{4}\right) \left(
        \begin{tabular}{l}
            $\frac{1}{\left(\sum_{i=1}^{I} S_i \textbf{A}_i \right)} \left(\sum_{i=1}^{I} S_{i} \left( \frac{\sum_{k=1}^{K_{i1}} Y_{i1k}}{\sum_{i=1}^I K_{i1}} - \frac{\sum_{k=1}^{K_{i2}} Y_{i2k}}{\sum_{i=1}^I K_{i2}} \right) \right)$
            \\
            \indent $+ \frac{1}{\left( \sum_{i=1}^{I} (1-S_i) \textbf{A}_i\right)} \left( \sum_{i=1}^{I} (1-S_i) \left(\frac{\sum_{k=1}^{K_{i2}} Y_{i2k}}{\sum_{i=1}^I K_{i2}}\right) 
            - 2 \sum_{i=1}^{I} (1-S_i) \left(\textbf{C}_i\right) \left(\frac{\sum_{k=1}^{K_{i1}} Y_{i1k}}{\sum_{i=1}^I K_{i1}}\right) \right)$ 
            \\
            \indent $- \frac{1}{\left(\sum_{i=1}^{I} S_i \textbf{A}_i \right)} \left( \sum_{i=1}^{I} S_i \left( \textbf{B}_i-\textbf{C}_i \right) \sum_{j=1}^{2} \left(\frac{\sum_{k=1}^{K_{ij}} Y_{ijk}}{\sum_{i=1}^I K_{ij}}\right) \right)$
            \\
            \indent $+ \frac{1}{\left( \sum_{i=1}^{I} (1-S_i) \textbf{A}_i\right)} \left( \sum_{i=1}^{I} (1-S_i) \left( \textbf{B}_i-\textbf{C}_i \right) \left(\frac{\sum_{k=1}^{K_{i2}} Y_{i2k}}{\sum_{i=1}^I K_{i2}}\right) \right)$
        \end{tabular}
        \right)
    \end{split}
\end{equation}
\normalsize
where
\[
    \textbf{A}_i = \frac{\left(\frac{K_{i1}}{\sum_{i=1}^I K_{i1}}\right)\left(\frac{K_{i2}}{\sum_{i=1}^I K_{i2}}\right)}{\left(\frac{K_{i1}}{\sum_{i=1}^I K_{i1}}+\frac{K_{i2}}{\sum_{i=1}^I K_{i2}}\right)} = \frac{K_{i1}K_{i2}}{(\sum_{i=1}^I K_{i2})K_{i1}+(\sum_{i=1}^I K_{i1})K_{i2}} \,,
\]
\[
    \textbf{B}_i = \frac{\left(\frac{K_{i1}}{\sum_{i=1}^I K_{i1}}\right)}{\left(\frac{K_{i1}}{\sum_{i=1}^I K_{i1}}+\frac{K_{i2}}{\sum_{i=1}^I K_{i2}}\right)} = \frac{(\sum_{i=1}^I K_{i2})K_{i1}}{(\sum_{i=1}^I K_{i2})K_{i1}+(\sum_{i=1}^I K_{i1})K_{i2}} \,,
\]
\[
    \textbf{C}_i = \frac{\left(\frac{K_{i2}}{\sum_{i=1}^I K_{i2}}\right)}{\left(\frac{K_{i1}}{\sum_{i=1}^I K_{i1}}+\frac{K_{i2}}{\sum_{i=1}^I K_{i2}}\right)} = \frac{(\sum_{i=1}^I K_{i1})K_{i2}}{(\sum_{i=1}^I K_{i2})K_{i1}+(\sum_{i=1}^I K_{i1})K_{i2}} \,.
\]

Under cluster randomization and
defining $\lambda_i=K_{i2}/K_{i1}$, we can then demonstrate that the FEpw treatment effect estimator converges in probability to:
\begin{equation}
    \begin{split}
    \hat{\delta}_{FEpw} \xrightarrow{P} \\
    & \left(\frac{1}{E\left[\frac{2K_{i1}\lambda_i}{ (E[K_{i1}\lambda_i]+E[K_{i1}]\lambda_i)}\right]}\right) \left(
    \begin{tabular}{l}
        $\frac{E[K_{i1}]\lambda_i}{E[K_{i1}\lambda_i]+E[K_{i1}]\lambda_i}
        E\left[ \frac{\sum_{k=1}^{K_{i1}} Y_{i1k}(1)}{\sum_{j=1}^2 K_{ij}} - \frac{\sum_{k=1}^{K_{i1}} Y_{i1k}(0)}{\sum_{j=1}^2 K_{ij}} \right]$ \\
        \indent $+ \frac{E[K_{i1}\lambda_i]}{E[K_{i1}\lambda_i]+E[K_{i1}]\lambda_i}
        E\left[ \frac{\sum_{k=1}^{K_{i2}} Y_{i2k}(1)}{\sum_{j=1}^2 K_{ij}} - \frac{\sum_{k=1}^{K_{i2}} Y_{i2k}(0)}{\sum_{j=1}^2 K_{ij}} \right]$
    \end{tabular}
    \right) \,.
 \end{split}
\end{equation}
Assuming $\lambda_i=\lambda \, \forall \, i$ we can then demonstrate that the FEpw estimator can be consistent for the pATE estimand in a CRXO design:
\begin{equation}
    \hat{\delta}_{FEpw} \xrightarrow{P} \frac{1}{2} \sum_{j=1}^{2} \left[ \frac{E\left[ \sum_{k=1}^{K_{ij}} \left[Y_{ijk}(1)-Y_{ijk}(0)\right] \right]}{E\left[K_{ij}\right]} \right]
\end{equation}
with more information included in the Appendix (\ref{sect:appendix_FEpw}).

\section{A Simulation Study}
\label{sect:simulation}

In this section, we simulated scenarios of CRXO trial data with continuous outcomes and non-informative or informative sizes to empirically study the operating characteristics of each estimator and to demonstrate the results derived in Section \ref{sect:Estimators}. Individual-level potential outcomes nested within clusters arising from different cluster-level subpopulations arose from the following data generating process (DGP):
\begin{align*}
    Y_{i1k}(0)&= \phi_1 + \alpha_i + \gamma_{ij} + e_{ijk} \,, \\
    Y_{i1k}(1)&=\delta_{u1} + \phi_1 + \alpha_i + \gamma_{ij} + e_{ijk} \,, \\
    Y_{i2k}(0)&=\phi_2 + \alpha_i + \gamma_{ij} + e_{ijk} \,, \\
    Y_{i2k}(1)&=\delta_{u2} + \phi_2 + \alpha_i + \gamma_{ij} + e_{ijk} \,, \\
    \alpha_{i} \mathop \sim^{\text{iid}} N(0,\tau^{2}_{\alpha}=0.053) \,,~~~
    \gamma_{ij} &\mathop \sim^{\text{iid}} N(0,\tau^{2}_{\gamma}=0.013) \,,~~~
    e_{ijk} \mathop \sim^{\text{iid}} N(0,\sigma^{2}_{w}=1) \,,
\end{align*}
with $\delta_{uj}$ being the subpopulation $u$, period $j$-specific heterogeneous treatment effects (specific parameter values are described in the following subsections). We set the two period effects $\phi_1 = 1$ (which can alternatively be re-paramaterized as the grand-mean parameter $\mu$) and $\phi_2 = 0.5$. 
Cluster random intercepts $\alpha_i$ and cluster-period random interaction terms $\gamma_{ij}$ are drawn from independent normal distributions with variances $\tau^2_\alpha = 0.053$ and $\tau^2_\gamma=0.013$, respectively, to yield a within-period intracluster correlation coefficient of $\rho_{wp}\approx 0.06$, between-period intracluster correlation coefficient of $\rho_{bp} \approx 0.05$, and a cluster auto-correlation of $CAC = \rho_{bp}/\rho_{wp} = 0.8$.
For illustration, the results included here are simulated for a 10-cluster (5-clusters/sequence), 2-period CRXO trial. To further confirm our theoretical asymptotic results, we included additional simulations for a 50-cluster (25-clusters/sequence), 2-period CRXO trial, with those results included in the Appendix (\ref{sect:appendix_sim50}).

In simulation scenarios without (Section \ref{sect:simulation_no_ics}) and with (Sections \ref{sect:simulation_ics} and \ref{sect:simulation_ips}) informative sizes, half of the clusters arose from subpopulation $u=1$ with the other half from subpopulation $u=2$, with cluster sampling probabilities of $P(u=1)=P(u=2)=0.5$. Half of all clusters were randomized to receive the treatment in period $j=1$, with the other half receiving the treatment in period $j=2$, corresponding with the 2 $\times$ 2 CRXO design illustrated in Figure \ref{fig:CRXO_design}.

In scenarios with noninformative sizes (Section \ref{sect:simulation_no_ics}), we simulated a homogeneous treatment effect across subpopulations $u$ and period $j$, whereas scenarios with ICS (Section \ref{sect:simulation_ics}) were simulated with a heterogeneous treatment effect that varies by subpopulation $u$.
In Sections \ref{sect:simulation_no_ics} \& \ref{sect:simulation_ics}, we fixed the cluster-period sizes to be the same between-periods, within-clusters, $K_{i1} = K_{i2} = K_{i-} \, \forall \, i$.
Accordingly, the inverse cluster-period size weighted and inverse cluster size weighted estimators will be equivalent.
Cluster-period sizes from subpopulations $u=1$ and $2$ were generated with $K_{i-,1} \sim Poisson(20)$ and $K_{i-,2} \sim Poisson(100)$, respectively.
We additionally include results from an additional set of simulations with ICS in the Appendix (\ref{sect:appendix_ICPS}) where we allowed cluster-period sizes to randomly vary between-periods, within-clusters, as is common in practice. In simulation settings with IPS (Section \ref{sect:simulation_ips}), we simulated a heterogeneous treatment effect that varies by period $j$ and set cluster-periods sizes from periods $j=1$ and $2$ to be generated with $K_{i1} \sim Poisson(20)$ and $K_{i2} \sim Poisson(100)$, respectively.

While the theoretical results described in Section \ref{sect:Estimators} primarily focus on consistency of the estimators, in this section, we use the simulations to test the potential empirical unbiasedness of the previously described estimators.
We simulated 1000 CRXO datasets for each scenario (no informative sizes, ICS, IPS). We present the results in terms of percent relative bias $\left(=\frac{\overline{\hat{\delta}}-ATE}{ATE} \times 100\right)$, with the bar denoting the average over the 1000 simulated datasets. Furthermore, we explored the accuracy of different variance estimators described in the next paragraph $\left(=\overline{Var(\hat{\delta})}\right)$, presented alongside the empirical variance (the variance of the 1000 simulated point estimates, also commonly referred to as the “observed” or “sampling” variances of the point estimates over the simulation replicates \cite{morris_using_2019}). We then used the model-based and jackknife variance estimators to obtain the corresponding normality-based 95\% confidence intervals and measure the coverage probability (CP) and power.

The weighted mixed effects models were run using the WeMix package in R \cite{bailey_wemix_2018}. The leave-one-cluster-out jackknife variance estimator was manually programmed in R version 4.3.2 following the description by Bell \& McCaffrey \cite{bell_bias_2002}. Notably, a similar jackknife variance estimator has been previously demonstrated to yield robust inference with arbitrary mixed effects model misspecification in SW-CRTs \cite{ouyang_maintaining_2024}. We additionally evaluated the sandwich variance estimator \cite{liang_longitudinal_1986} and the bias-reduced linearization robust variance \cite{pustejovsky_small-sample_2018} with the clubSandwich package in R . However, clubSandwich is not compatible with WeMix and is not implemented with the weighted EME and NEME estimators.

\subsection{Simulation results with non-informative sizes}
\label{sect:simulation_no_ics}

\begin{figure}[htp]
    \centering
    \includegraphics[width=12cm]{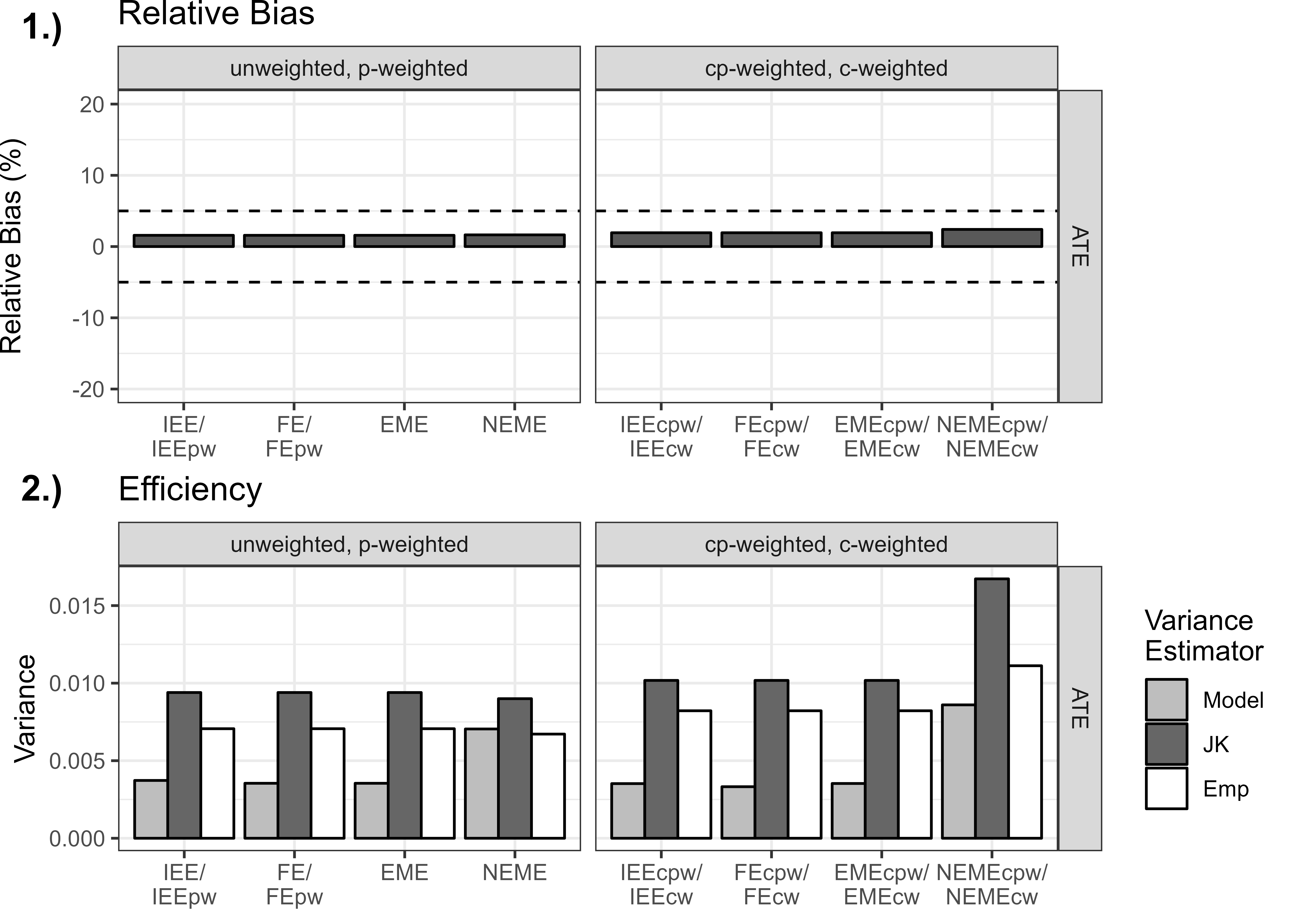}
    \caption{1.) Simulation relative bias (\%) results in scenarios with homogeneous treatment effects (non-informative sizes). Dashed lines show a relative bias of $5\%$ and $-5\%$. 2.) The efficiency of the different unweighted and weighted models as captured by the average of the model-based (``Model") and leave-one-cluster jackknife (``JK") variance estimates over the 1000 simulation replicates, graphed alongside the corresponding empirical (``Emp") variances.}
   \label{fig:noICS_bias_var}
\end{figure}

In scenarios with noninformative sizes, we simulated a homogeneous treatment effect across subpopulations $u$ and periods $j$, $\delta_{uj} = iATE= cpATE = cATE = pATE = ATE = 0.4 \, \forall \, u,j$. With fixed cluster-period cell sizes, $K_{i1}=K_{i2}=K_{i-} \, \forall \, i$, the IEE and IEEpw estimators are identical, as are the FE and FEpw estimators. Furthermore, the IEEcpw and IEEcw estimators are identical, as are the FEcpw and FEcw estimators.
As expected, all the unweighted and weighted treatment effect estimators were unbiased for the true average treatment effect estimand (Figure \ref{fig:noICS_bias_var}.1). 

When comparing the weighted models against their unweighted counterparts, we observe that modelling with inverse cluster-period or cluster-size weights may lead to worse efficiency in terms of empirical variances (Figure \ref{fig:noICS_bias_var}.2).
Across the unweighted analyses, the empirical variances were all roughly equivalent (Figure \ref{fig:noICS_bias_var}.2) and accordingly are all similarly efficient in the analysis of a CRXO trial.
Across the weighted estimators, the NEMEcpw and NEMEcw estimators had the largest empirical variances (Figure \ref{fig:noICS_bias_var}.2) and is observed to be an empirically less efficient estimator than the other similarly weighted estimators, despite the true underlying DGP having a nested-exchangeable correlation structure.

The averages of the model-based and leave-one-cluster-out jackknife variance estimates are included in Figure \ref{fig:noICS_bias_var}.2, alongside the empirical variance. The model-based and jackknife variance estimators explicitly target the empirical variance, with systematic deviations representing a bias in the estimation of the variance \cite{morris_using_2019}. Comparing the variance estimates in Figure \ref{fig:noICS_bias_var}.2 where the true underlying DGP has a nested exchangeable correlation structure, we observe that the jackknife variance estimator typically overestimates the empirical variances and tends to be conservative in our simulations with $10$ clusters.
Overall, these results hold in simulations with 50 total clusters, where the jackknife variance estimator closely approximates the empirical variance with larger samples of clusters (Appendix \ref{sect:appendix_sim50}).

\begin{figure}[htp]
    \centering
    \includegraphics[width=12cm]{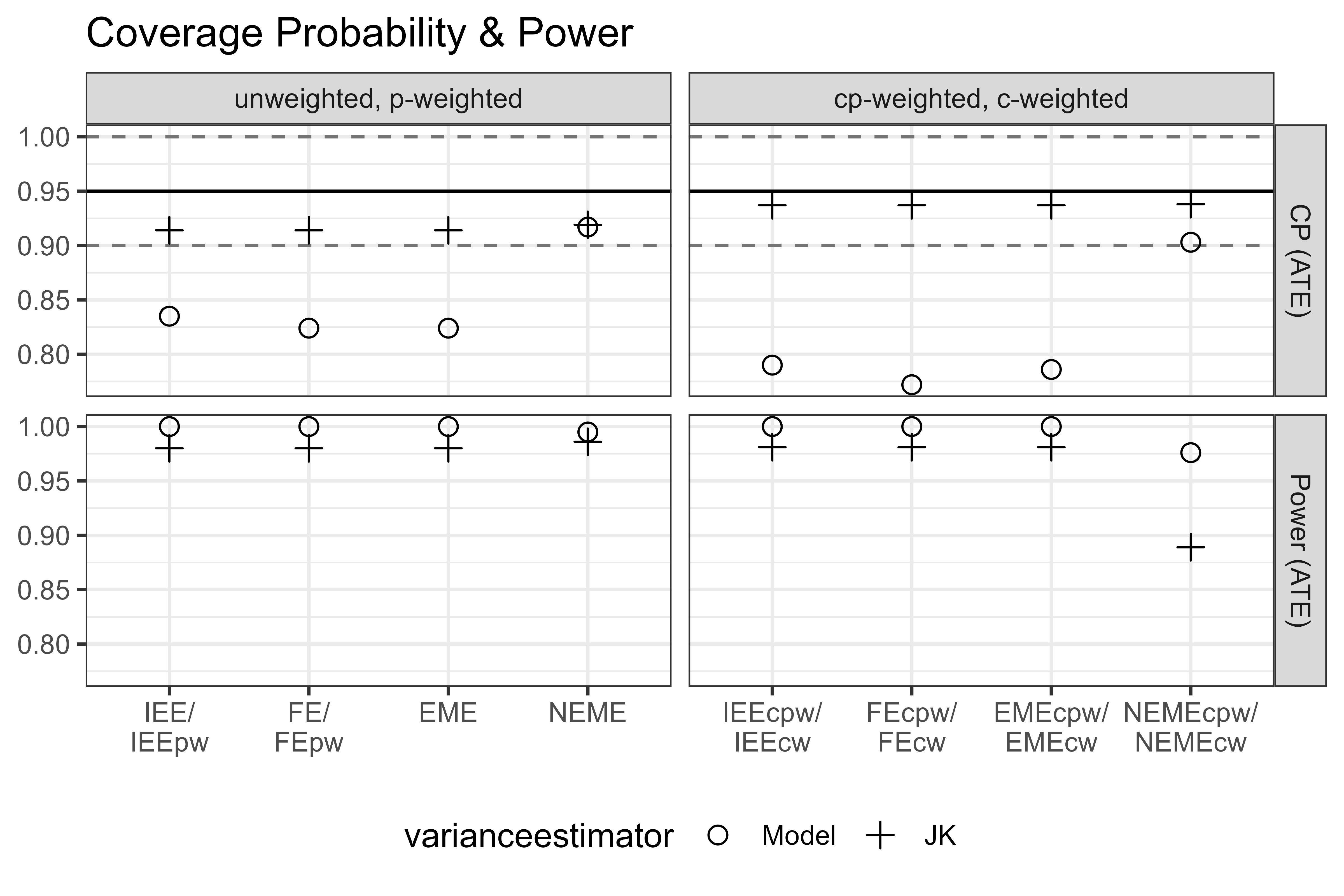}
    \caption{The coverage probability of the 95\% confidence interval and the power of the different unweighted and weighted models using the model-based (``Model") and leave-one-cluster jackknife (``JK") variance estimators in scenarios with a homogeneous treatment effect (non-informative sizes). The solid lines lines show a coverage probability of 0.95, with the dashed lines denoting the range from 0.90 to 1.}
   \label{fig:noICS_CP_Power}
\end{figure}

With the true underlying DGP having a nested exchangeable correlation, all models had close to proper coverage of the 95\% confidence intervals with the jackknife variance estimator (Figure \ref{fig:noICS_CP_Power}, Appendix \ref{sect:appendix_sim50}).
This largely corresponds with previous work that demonstrated the robust variance estimators can help yield robust inference with correlation structure misspecification in SW-CRTs \cite{ouyang_maintaining_2024}. As expected, the power across the different analyses are then slightly reduced in these scenarios when using the jackknife variance estimator.

Notably, the WeMix package \cite{bailey_wemix_2018} automatically returns the sandwich variance estimator \cite{liang_longitudinal_1986}. For demonstration purposes, we manually programmed the model-based variance estimators for the EMEcpw/EMEcw and NEMEcpw/NEMEcw estimators that were programmed using WeMix.
In Appendices (\ref{sect:appendix_noICS_efficiency}) and (\ref{sect:appendix_sim50}), we also compared the efficiency, coverage probability, and power results in Figures \ref{fig:noICS_bias_var}.2 and \ref{fig:noICS_CP_Power} to the corresponding results when using the ``CR0" sandwich variance estimator \cite{liang_longitudinal_1986}, and the ``CR2" bias-reduced linearization robust variance estimator \cite{pustejovsky_small-sample_2018} (as implemented with the ``clubSandwich" package in R \cite{pustejovsky_clubsandwich_2016}).
To reiterate, the output from the WeMix package is incompatible with clubSandwich, and the bias-reduced linearization robust variance estimates and corresponding coverage probability and power results are excluded for EMEcpw/EMEcw and NEMEcow/NEMEcw estimators. We observe that the sandwich variance estimator underestimates the empirical variance and yields undercoverage of the 95\% confidence interval (Appendices \ref{sect:appendix_noICS_efficiency} \& \ref{sect:appendix_sim50}); this is not unexpected because the uncorrected sandwich variance estimator often has downward bias when applied to CRXO trials with a small number of clusters \cite{li_sample_2018}. In contrast, we observe that the bias-reduced linearization robust variance estimator closely approximated the empirical variances;
however, it has slight under-coverage of the 95\% confidence interval, especially in unweighted estimators (Appendices \ref{sect:appendix_noICS_efficiency} \& \ref{sect:appendix_sim50}).

\subsection{Simulation results with informative cluster sizes}
\label{sect:simulation_ics}

In scenarios with ICS between clusters arising from subpopulations $u=1$ and $2$, we simulated heterogeneous treatment effects $\delta_{1j} = \delta_1 = 0.2$ and $\delta_{2j} = \delta_2 = 0.6$ $\forall \, j$ that correspond with average cluster-period sizes of $E[K_{i-,1}]=20$ or $E[K_{i-,2}]=100$, respectively. Cluster-period sizes were fixed between-periods, within-clusters ($K_{i1} = K_{i2} = K_{i-} \, \forall \, i$).

With the described ICS in the underlying DGP, the true iATE and pATE estimands are equal and given by:
\begin{align*}
   iATE &= pATE = E\left[ \frac{\sum_{j=1}^{2} \sum_{k=1}^{K_{i-}} [Y_{ijk}(1)-Y_{ijk}(0)]}{E[2K_{i-}]} \right]
    = \frac{E[E[K_{i-}\delta_u|u]]}{E[E[K_{i-}|u]]} \\
    &= \frac{P(u=1)E[K_{i-}\delta_u|u=1] + P(u=2)E[K_{i-}\delta_u|u=2]}{P(u=1)E[K_{i-}|u=1] + P(u=2)E[K_{i-}|u=2]}
 =\frac{0.5(20\delta_1) + 0.5(100\delta_2)}{0.5(20) + 0.5(100)} = \frac{10\delta_1 + 50\delta_2}{60} = 0.5\overline{3}
\end{align*}
(where $\delta_1 = 0.2$ and $\delta_2 = 0.5$). Subsequently, the true cpATE and cATE are equal and given by:
\begin{align*}
    cpATE = cATE &= E\left[ \frac{\sum_{j=1}^{2}\sum_{k=1}^{K_{i-}}[Y_{ijk}(1)-Y_{ijk}(0)]}{2K_{i-}} \right]
    = E[E[\delta_u|u]] \\
    &= P(u=1)E[\delta_u|u=1] + P(u=2)E[\delta_u|u=2]
    = 0.5\delta_1 + 0.5\delta_2 = 0.4 \,.
\end{align*}

With cluster-period sizes fixed between-periods, within-clusters, the IEE and IEEpw estimators are identical, as are the FE and FEpw estimators. Furthermore the IEEcpw and IEEcw estimators are identical, as are the FEcpw and FEcw estimators.

\begin{figure}[htp]
    \centering
    \includegraphics[width=12cm]{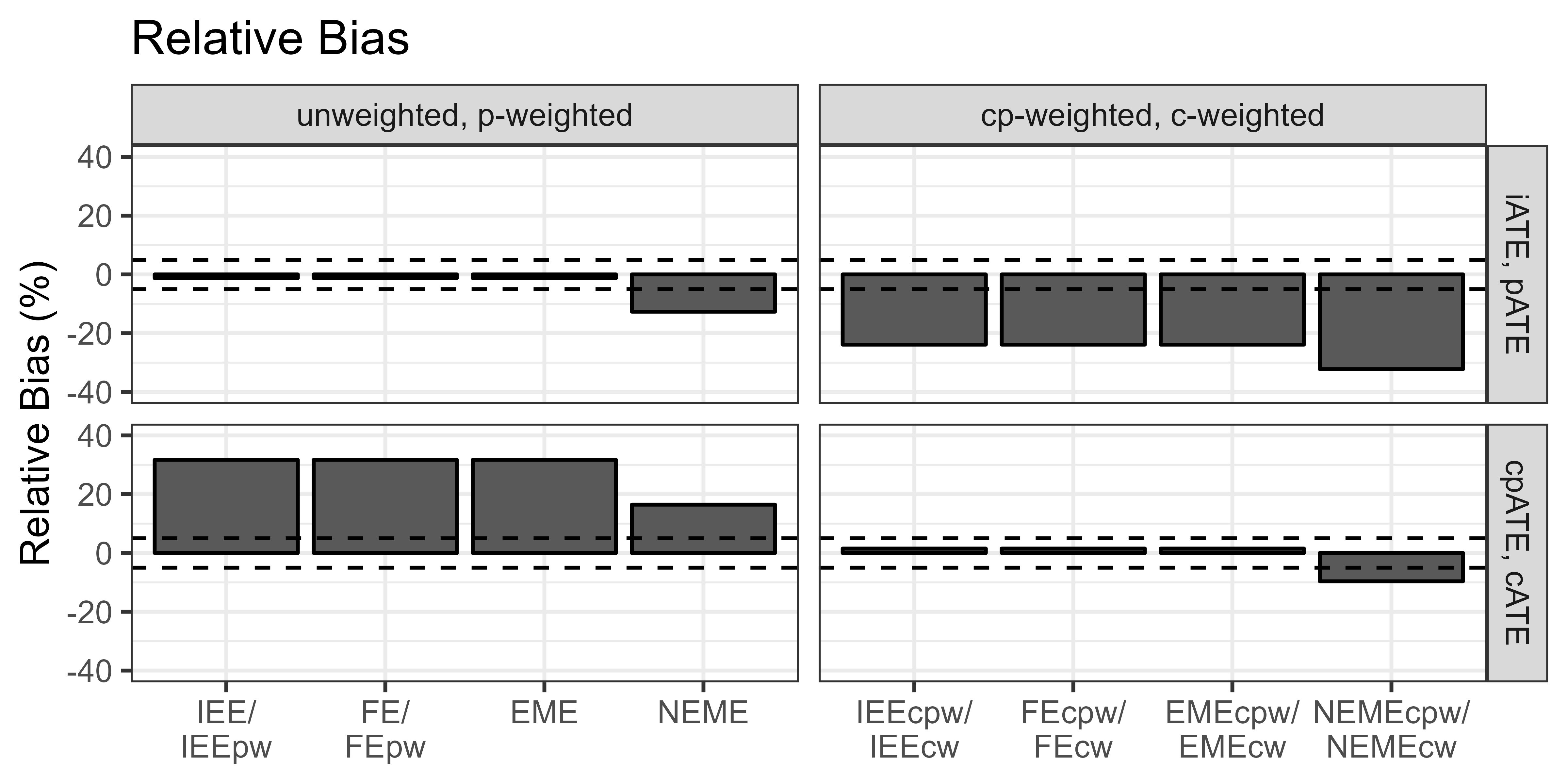}
    \caption{Simulation relative bias (\%) results in scenarios with informative cluster sizes. Dashed lines show a relative bias of $5\%$ and $-5\%$.}
   \label{fig:ICS_bias}
\end{figure}

We observe in scenarios with informative cluster sizes that the NEME estimator can yield biased results for the iATE and pATE, and the NEMEcpw and NEMEcw estimators can yield biased results for the cpATE and cATE (Figure \ref{fig:ICS_bias}, Appendix \ref{sect:appendix_sim50}). In contrast, the IEE/IEEpw, FE/FEpw, and EME yielded empirically unbiased results for the iATE and pATE, and the IEEcpw/IEEcw, FEcpw/FEcw, and EMEcpw/EMEcw for the cpATE and cATE (Figure \ref{fig:ICS_bias}, Appendix \ref{sect:appendix_sim50}). 

In Appendix (\ref{sect:appendix_ICPS}), we additionally simulated data with informative cluster sizes while allowing cluster-period sizes to vary between-periods, within-clusters, with results presented. Generally, we observe that the results in such a condition correspond with results when cluster-period sizes are fixed between-periods, within-clusters as in Figure \ref{fig:ICS_bias}.
The IEE, EME, FE, IEEcpw, FEcpw, IEEcw, EMEcw, FEcw, IEEpw, and FEpw estimators all yield empirically unbiased results for their corresponding weighted estimands (Appendix \ref{sect:appendix_ICPS}). It is not clear how to specify inverse cluster-period size weights for the EMEcpw estimator when cluster-period sizes vary between-periods, within-clusters. However, in the absence of informative cluster-period sizes, the EMEcw estimator will yield comparable results to the EMEcpw estimator. Furthermore, in such a setting, the IEEcpw (Section \ref{sect:IEEcpw}) and FEcpw (Section \ref{sect:FEcpw}) may be preferable given that they're consistent estimators and unbiased in expectation for the cpATE and cATE (Table \ref{tab:summary}), while being similarly efficient (Figure \ref{fig:noICS_bias_var}.2).

\subsection{Simulation results with informative period sizes}
\label{sect:simulation_ips}

In scenarios with informative period sizes between periods $j=1$ and $j=2$, we simulated heterogeneous treatment effects $\delta_{u1}=\delta_1=0.2$ and $\delta_{u2}=\delta_2=0.6 \, \forall \, u$ that corresponds with average cluster-period sizes of $E[K_{i1,u}] = E[K_{i1}] = 20$ and $E[K_{i2,u}] = E[K_{i2}] = 100$.

While the cATE and iATE are not explicitly equal here due to Jensen's inequality, we can observe via simulation that $E\left[\frac{K_{i1}}{K_{i1}+K_{i2}}\right]$ and $E\left[\frac{K_{i2}}{K_{i1}+K_{i2}}\right]$ approach $\frac{E[K_{i1}]}{E[K_{i1} + K_{i2}]}$ and $\frac{E[K_{i2}]}{E[K_{i1} + K_{i2}]}$, respectively. Accordingly, with the described IPS in the underlying DGP, the true iATE and cATE estimand are approximately equal:

\begin{align*}
    &cATE = E\left[\frac{\sum_{j=1}^2 \sum_{k=1}^{K_{ij}}[Y_{ijk}(1)-Y_{ijk}(0)]}{\sum_{j=1}^2K_{ij}}\right]
    = E\left[\frac{K_{i1}}{K_{i1}+K_{i2}}\right]\delta_1 + E\left[\frac{K_{i2}}{K_{i2}+K_{i2}}\right]\delta_2 \\
    &\approx iATE = E\left[\frac{\sum_{j=1}^2 \sum_{k=1}^{K_{ij}}[Y_{ijk}(1)-Y_{ijk}(0)]}{E[\sum_{j=1}^2K_{ij}]}\right] = \frac{E[K_{i1}]\delta_1 + E[K_{i2}]\delta_2}{E[K_{i1}] + E[K_{i2}]} 
    =\frac{20\delta_{1} + 100\delta_{2}}{20+100} = 0.5\bar{3}
\end{align*}
(where $\delta_{1}=0.2$ and $\delta_{2}=0.5$). Subsequently, the true cpATE and pATE are equivalent:
\[
\begin{split}
    cpATE = pATE &= E\left[ \frac{1}{2} \sum_{j=1}^{2} \frac{ \sum_{k=1}^{K_{ij}} \left[Y_{ijk}(1)-Y_{ijk}(0)\right]}{K_{ij}} \right]
    =\frac{1}{2} \sum_{j=1}^{2} \left[ \frac{E\left[ \sum_{k=1}^{K_{ij}} \left[Y_{ijk}(1)-Y_{ijk}(0)\right] \right]}{E\left[K_{ij}\right]} \right] \\
&= \frac{\delta_{u1} + \delta_{u2}}{2} = 0.4 \,.
\end{split}
\]
\begin{figure}[htp]
    \centering
    \includegraphics[width=15cm]{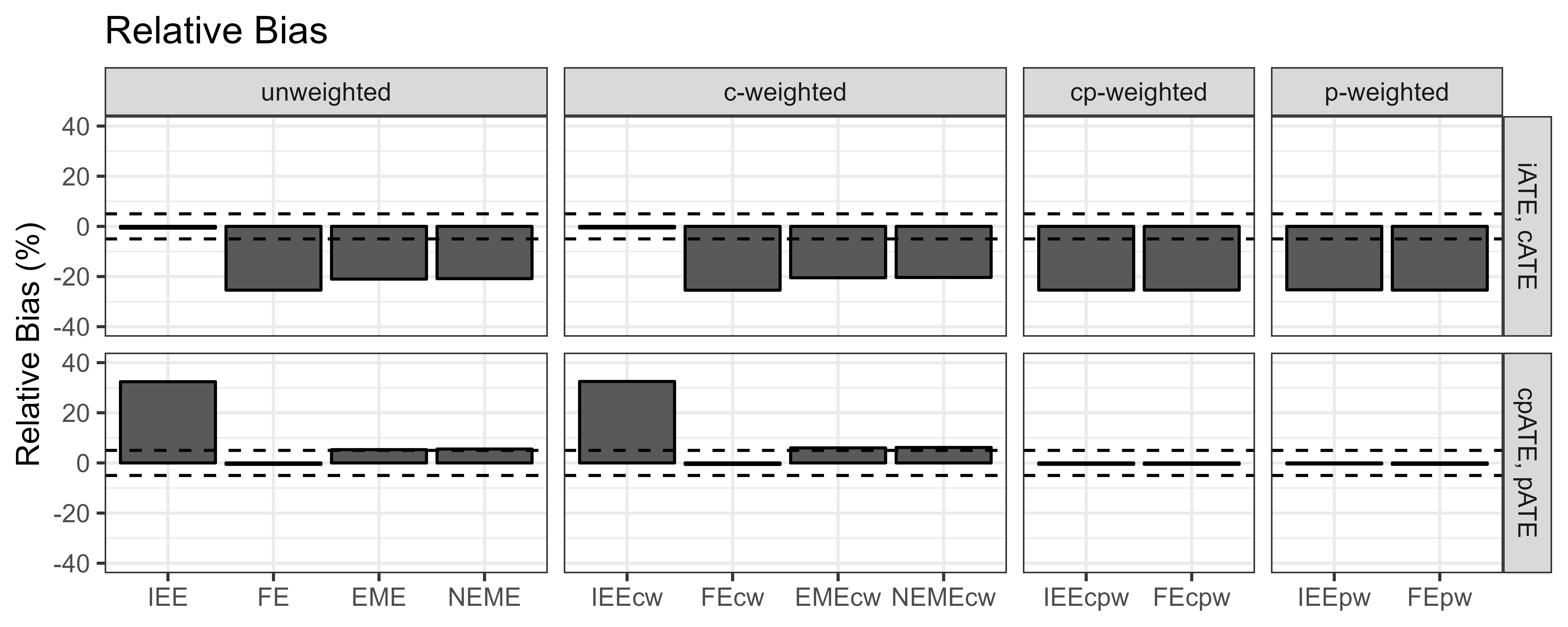}
    \caption{Simulation relative bias (\%) results in scenarios with informative period sizes. Dashed lines show a relative bias of $5\%$ and $-5\%$.}
   \label{fig:IPS_bias}
\end{figure}

The relative bias results in these simulations with IPS are shown in Figure \ref{fig:IPS_bias}.
Only the IEE and IEEcw estimators were empirically unbiased for the approximately equal iATE and cATE estimands (Figure \ref{fig:IPS_bias}, Appendix \ref{sect:appendix_sim50}).
As previously demonstrated, with minimal assumptions, the FE (Section \ref{sect:FE}) and FEcw (Section \ref{sect:FEcw}) estimators are generally consistent for the pATE and cpATE estimands (instead of the iATE and cATE estimands, Table \ref{tab:summary}), respectively (Figure \ref{fig:IPS_bias}, Appendix \ref{sect:appendix_sim50}).
These two estimators, alongside the IEEcpw, FEcpw, IEEpw, and FEpw estimators were empirically unbiased for the equivalent cpATE and pATE estimands (Figure \ref{fig:IPS_bias}, Appendix \ref{sect:appendix_sim50}).

In the presence of IPS, we observe that the unweighted and weighted mixed effects models (EME, NEME, EMEcw, NEMEcw) yielded biased results for all of the described estimands (Figure \ref{fig:IPS_bias}, Appendix \ref{sect:appendix_sim50}). Overall, the EME, NEME, EMEcw, and NEMEcw estimators yield results with around 5\% relative bias and are not recommended when IPS is suspected. Furthermore, it is not clear how to specify inverse cluster-period or inverse period size weights with these mixed effects models when cluster-period sizes vary between-period, within-cluster, as occurs in the presence of IPS.

\section{A Case Study: Reanalysis of a CRXO Trial}
\label{sect:case_study}

\begin{figure}[htp]
    \centering
    \includegraphics[width=10cm]{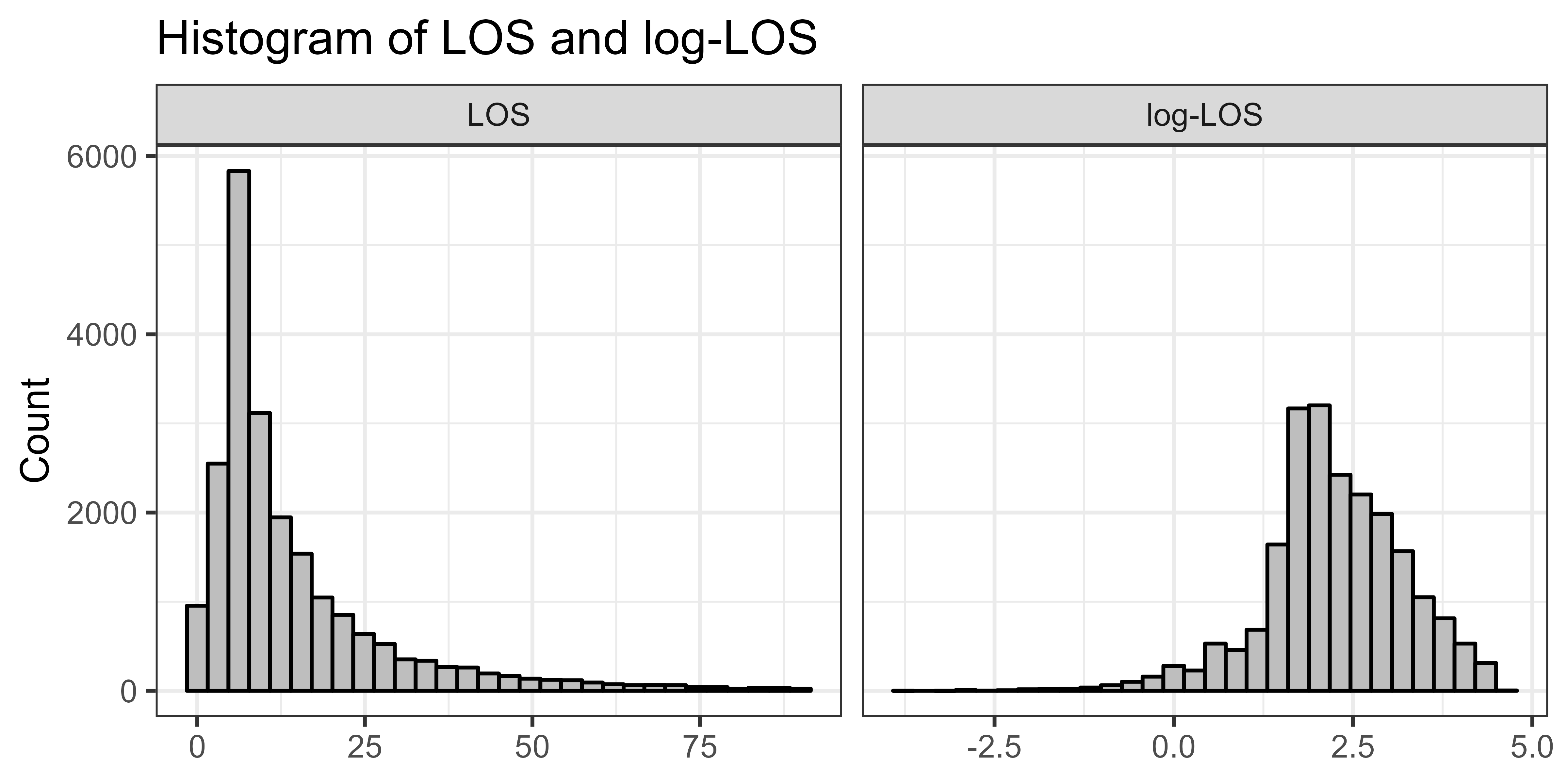}
    \caption{Histograms showing the distribution of hospital LOS and log-LOS among patients in a cross-sectional 2-period, 49 cluster CRXO trial.}
   \label{fig:LOS}
\end{figure}

In this section, we reanalyzed a CRXO trial dataset, exploring the effect of stress ulcer prophylaxis with proton pump inhibitors (PPIs; treatment) versus histamine-2 receptor blockers (H$_{2}$RBs; control) on hospital log-length of stay (log-LOS) among patients receiving invasive mechanical ventilation \cite{the_peptic_investigators_for_the_australian_and_new_zealand_intensive_care_society_clinical_trials_group_alberta_health_services_critical_care_strategic_clinical_network_and_the_irish_critical_care_trials_group_effect_2020}.
This trial had a 2-period cross-sectional CRXO design (corresponding with the design described in Figure \ref{fig:CRXO_design}) with 49 hospital ICU's serving as clusters contributing individual patient observations in both periods (Appendix \ref{sect:appendix_casestudy}).
The distribution of hospital LOS and log-LOS are shown in Figure \ref{fig:LOS}.

Primary analyses of a cross-sectional, 2-period CRXO design should be based on pre-specified analyses, as is standard.
When selecting an appropriate estimand and evaluating the potential for different informative sizes, researchers should carefully consider \textit{a priori} information about the trial's intervention, design, and implementation.
Below, we suggest a few practical steps to guide analyses and evaluate the robustness of results while accounting for informative sizes.

\setlist{nolistsep} 
\begin{enumerate}[noitemsep]
    \item Pre-specify the estimand of interest (iATE, cATE, pATE, cpATE) for each main comparison.
    \item Consult \textit{a priori} information to determine the potential for different informative sizes.
    \item Pre-specify an appropriate primary analysis estimator for the target estimand, and specify the required assumptions.
    \begin{enumerate}
        \item If informative sizes are anticipated, we generally recommend the IEE or its weighted counterparts with an approriate robust standard error estimator (i.e., leave-one-cluster-out jackknife variance estimator) to target the corresponding pre-specified estimand of interest.
    \end{enumerate}
    \item After trial completion, check the robustness of the pre-specified primary analysis estimator. If a ``no informative sizes" assumption was required, evaluate the plausibility of this assumption through empirical evaluation.
    \begin{enumerate}
        \item Examine the cluster, period, and cluster-period sizes, which may inform the initial risk of different informative sizes.
        \item Perform sensitivity analyses by comparing the pre-specified primary analysis estimator results against an accordingly weighted consistent estimator. Observe if there are any discrepancies between the primary analysis and consistent estimator results, which may be attributed to departures from the ``no informative sizes" assumption.
        \item Perform sensitivity analyses using estimators that are always consistent for the target and other weighted estimands, even in the presence of informative sizes, to evaluate if estimates from differently weighted consistent estimators differ in magnitude. 
        \item Perform sensitivity analyses using methods that are known to be inconsistent for the target estimand in the presence of informative sizes (i.e., NEME with appropriate weighting) to evaluate to what extent the produced estimates are noticeably different in magnitude from those of known consistent estimators (i.e., IEE with appropriate weighting).
    \end{enumerate}
\end{enumerate}

In this present trial \cite{the_peptic_investigators_for_the_australian_and_new_zealand_intensive_care_society_clinical_trials_group_alberta_health_services_critical_care_strategic_clinical_network_and_the_irish_critical_care_trials_group_effect_2020}, we anticipate investigators will be more interested in targeting the iATE, with interest being primarily on the effectiveness of the treatment among patients receiving invasive mechanical ventilation. Furthermore, treatment effects on log-LOS may conceivably change by cluster size due to the capacity of large versus small ICU's, leading to ICS.

We approach the following reanalysis following the above recommendations (step 4).
We observe that cluster-period cell sizes and period sizes do not appear to vary much between-periods (Appendix \ref{sect:appendix_casestudy}), which lessens the risk of informative period sizes. However, cluster-period cell sizes and cluster sizes do vary considerably between clusters (Appendix \ref{sect:appendix_casestudy}), which may heighten the risk of informative cluster sizes.

We reanalyzed  the effect of PPIs (compared to H$_{2}$RBs as the control) on log-LOS with the unweighted and weighted IEE, EME, NEME, and FE models in Table \ref{tab:case_study_PEPTIC}.
In the reanalysis of this CRXO trial, the EMEcpw, NEMEcpw, EMEpw, and NEMEpw are not well defined due to cluster-period sizes differing between-periods, within-clusters (Appendix \ref{sect:appendix_casestudy}) and are accordingly excluded from the reanalyses (Table \ref{tab:case_study_PEPTIC}).

\begin{table}
\caption{Case study reanalysis results of the effect of proton pump inhibitors (treatment) versus histamine-2 receptor blockers (control) on hospital log-length of stay. 95\% confidence intervals are formed with the leave-one-cluster-out jackknife variance estimators.}
\label{tab:case_study_PEPTIC}
\begin{center}
\bgroup
\def\arraystretch{1.2}
\begin{tabular}{|c c|} 
    \hline
    Estimator & $\hat{\delta}$ (95 \% CI) \\
    \hline\hline
    IEE & 0.018 (-0.017,0.053) \\
    \hdashline
    FE & 0.015 (-0.020, 0.050) \\
    \hdashline
    EME & 0.015 (-0.020, 0.050) \\
    \hdashline
    NEME & 0.024 (-0.015, 0.063) \\
    \hline
    IEEcpw & 0.033 (-0.022, 0.088) \\
    \hdashline
    FEcpw & 0.033 (-0.022, 0.088) \\
    \hline
    IEEcw & 0.040 (-0.018, 0.098) \\
    \hdashline
    FEcw & 0.033 (-0.022, 0.087) \\
    \hdashline
    EMEcw & 0.034 (-0.021, 0.089) \\
    \hdashline
    NEMEcw & 0.055 (-0.090, 0.200) \\
    \hline
    IEEpw & 0.018 (-0.017, 0.053) \\
    \hdashline
    FEpw & 0.015 (-0.020, 0.050) \\
    \hline
\end{tabular}
\egroup
\end{center}
\end{table}

We observe that the unweighted and inverse period size weighted estimators generally produced point estimates ranging between 0.015 to 0.018 (corresponding to geometric mean ratio estimates of 1.015 to 1.018); whereas the inverse cluster size and inverse cluster-period size weighted estimators generally produced larger point estimates ranging between 0.033 to 0.040 (corresponding to geometric mean ratio estimates of 1.033 to 1.040) (Table \ref{tab:case_study_PEPTIC}).
However, the differences between the unweighted and inverse cluster-size weighted estimates are small in comparison to the 95\% confidence intervals (formed using the leave-one-cluster-out jackknife variance estimators) which overlap across the differently weighted estimates (Table \ref{tab:case_study_PEPTIC}).
In contrast to the other estimators, the NEME and NEMEcw estimators yielded inflated estimates of 0.024 and 0.055 (corresponding to geometric mean ratio estimates of 1.024 to 1.056), respectively (Table \ref{tab:case_study_PEPTIC}). Although, this inflation is again small in comparison to the 95\% confidence intervals, which largely coincide with the other estimators (Table \ref{tab:case_study_PEPTIC}).
Still, the observed discrepancy between the unweighted and weighted estimators, alongside the deviation of the NEME and NEMEcw estimates, may imply the presence of ICS, but neither IPS nor ICPS.

To clarify, the above recommendations for evaluating the potential for informative sizes after trial completion are purely qualitative. We have not proposed a statistical test for detecting the presence of informative sizes, which can be the focus of future work.

\section{Discussion}
\label{sect:discussion}

As a multi-period cluster randomized trial design, we have shown that the 2-period cross-sectional CRXO trial can yield four natural estimands of interest, including the individual-average treatment effect (iATE), cluster-period-average treatment effect (cpATE), cluster-average treatment effect (cATE), and period-average treatment effect (pATE). This additional complexity is owing to the fact that potential outcomes collected from more than one period can be used to define multiple different marginal estimands, and represents a major distinction from the previous discussions regarding estimands in P-CRTs \cite{wang_two_2022} or PB-CRTs \cite{lee_how_2025}. We formally define these four estimands under a unified general class of weighted-average treatment effect estimands \cite{chen_model-assisted_2025} and clarify the conditions under which they differ in magnitude. Notably, when there are informative cluster sizes (ICS, iATE $\neq$ cATE), informative period sizes (IPS, iATE $\neq$ pATE), or informative cluster-period sizes (ICPS, cpATE $\notin$ (cATE, pATE)), then common estimators in the analysis of CRXO designs can converge to distinctly different estimands.

Overall, we demonstrate that the independence estimating equation (IEE) estimator is always consistent for the iATE estimand, regardless of the presence of informative sizes. This corresponds to the results for P-CRTs \cite{wang_two_2022} and PB-CRTs \cite{lee_how_2025} in the presence of ICS. Furthermore, its inverse cluster-period size (IEEcpw), inverse cluster size (IEEcw), and inverse period size (IEEpw) weighted counterparts are similarly always consistent for the cpATE, cATE, and pATE estimands, respectively.

Among the different weighted fixed effects model estimators, only the inverse cluster-period size weighted fixed effects model (FEcpw) is always consistent for its corresponding weighted estimand (the cpATE).
Surprisingly, we demonstrate that the inverse cluster size weighted fixed effects model (FEcw) estimator is consistent for the cpATE, and the fixed effects model (FE) estimator and its inverse period size weighted counterpart (FEpw) are both consistent for the pATE, when the proportion of individuals across periods is fixed for all clusters ($\lambda_i = \lambda \, \forall \, i$). A more specific example of this condition includes scenarios where cluster-period sizes are fixed between-periods, within-clusters ($K_{i1} = K_{i2} \, \forall \, i$).
Accordingly, careful consideration is recommended when employing the FE, FEcw, or FEpw estimators, especially when IPS is suspected.

In contrast to previous work in P-CRTs \cite{wang_two_2022} and PB-CRTs \cite{lee_how_2025}, we demonstrate that, under a CRXO design, the exchangeable mixed effects model (EME) estimator and its inverse cluster-period size (EMEcpw) and inverse cluster size (EMEcw) weighted counterparts can be consistent for the iATE, cpATE, and cATE estimands when cluster-period sizes are fixed between-periods, within-clusters ($K_{i1} = K_{i2} \, \forall \, i$). This result is particularly important given the widespread use of linear mixed models in the analysis of CRXO trials and has direct implications for study planning. 

However, we demonstrate that the unweighted and weighted nested exchangeable model estimators (NEME, NEMEcpw, NEMEcw) converge to weighted average treatment effect estimands with data-dependent weights that are difficult to interpret and are typically not consistent for the iATE, cpATE, cATE, nor pATE estimands in the presence of informative sizes. This corresponds with results previously observed in PB-CRT designs \cite{lee_how_2025}. In the absence of informative cluster sizes, previous literature has recommended using NEME for study planning and data analysis, as the nested exchangeable correlation structure is considered a more realistic representation of the underlying correlation structure in a 2-period, 2-sequence CRXO trial \cite{giraudeau_sample_2008,turner_analysis_2007, li_power_2019,morgan_choosing_2017,mckenzie_reporting_2025}. We still support this recommendation. However, when informative sizes are suspected, the use of such models should be reconsidered when the interest lies in marginal estimands, as their data-adaptive weighting scheme can inadvertently target marginal estimands that are more challenging to interpret. As we have shown in our derivations, the implied estimands targeted by unweighted and weighted NEME estimators depend on unknown intracluster correlation coefficients, cluster-period sizes, and may vary by the outcome of interest, complicating their interpretation in practical settings.

We use Table \ref{tab:advantages_disadvantages} to summarize some general advantages and disadvantages of these different estimators in the presence of informative sizes.
Notably, trials with balanced cluster-period sizes within clusters, such that $K_{i1} = K_{i2} \, \forall \, i$, ensures that the routine application of EME, FE, and their weighted counterparts can reliably target well-defined potential outcomes estimands in CRXO trials. Furthermore, this reduces the pool of potential estimands by constraining the following estimands to be equivalent: iATE = pATE \& cATE = cpATE.
Altogether, we highlight the need to concurrently consider the trial design and intended analysis method during the CRXO study design phase, especially if researchers are determined to use the EME or FE model in their primary analysis.

When the specified correlation structure in a ``model-assisted" estimator is considered inadequate for capturing the true underlying correlation structure, a bias-corrected sandwich variance estimator or the leave-one-cluster-out jackknife variance estimator should be considered for valid inference, as demonstrated in our simulation study. The jackknife variance estimator is easy to manually program and implement across different R packages, including WeMix.

Across our simulation scenarios with ICS, the IEE/IEEpw (IEEcpw/IEEcw), EME (EMEcpw/EMEcw), and FE/FEpw (FEcpw/FEcw) estimators were all empirically unbiased for the iATE/pATE (cpATE/cATE) estimands, even when cluster-period sizes varied between-periods, within-clusters, despite the unweighted and weighted EME and FE estimators not being consistent in such conditions. 
However, the NEME (NEMEcpw, NEMEcw) estimator could potentially yield very biased results.
In contrast, across our simulation scenarios with IPS, only the IEE and IEEcw estimators yielded empirically unbiased estimates for the iATE and cATE estimands, respectively. Whereas the IEEcpw, IEEpw, FE, FEcpw, FEcw, and FEpw estimators yielded empirically unbiased estimates for the cpATE and pATE estimands.
Overall, we observe that the unweighted and weighted IEE, FE, and EME estimators can operate similarly in terms of efficiency.
Accordingly, researchers may preferably opt to use the unweighted and weighted IEE estimators which, in addition to their general consistency results, does not dramatically compromise efficiency in a CRXO and allows the easy application of many cluster robust variance estimators via clubSandwich in R, including the bias-reduced linearization robust variance estimator \cite{pustejovsky_clubsandwich_2016}, which is also observed to perform well (Appendix \ref{sect:appendix_noICS_efficiency}).

We suggest a few practical steps for considering the risk of informative sizes in the analysis of cross-sectional 2-period CRXO designs and apply these recommendations in the reanalysis of a CRXO trial exploring the effect of stress ulcer prophylaxis with proton pump inhibitors (PPIs; treatment) versus histamine-2 receptor blockers (H$_{2}$RBs; control) on hospital log-length of stay (log-LOS) among patients receiving invasive mechanical ventilation \cite{the_peptic_investigators_for_the_australian_and_new_zealand_intensive_care_society_clinical_trials_group_alberta_health_services_critical_care_strategic_clinical_network_and_the_irish_critical_care_trials_group_effect_2020}.
When pre-specifying primary analyses, we emphasize the importance of consulting \textit{a priori} information regarding the trial intervention and design to determine the estimand of interest and the potential for different informative sizes to occur. In the described case study, we anticipate researchers to be more interested in the iATE, with interest being primarily on the effectiveness of the treatment among patients receiving invasive mechanical ventilation. Furthermore, treatment effects on log-LOS may change by cluster size due to the capacity of large versus small ICU's, leading to ICS.
When evaluating the potential for informative sizes after trial completion, we highlight the importance of evaluating the cluster, period, and cluster-period sizes. This trial had considerable variation in cluster sizes, but not much variation in period sizes, which may risk ICS but not IPS (Appendix \ref{sect:appendix_casestudy}).
Furthermore, we evaluate if the unweighted and weighted estimates from known consistent estimators, such as the IEE, IEEcw, IEEpw, IEEcpw, and FEcpw differ in magnitude. In this trial, the IEE and IEEpw estimators yielded similar estimates, as did the IEEcpw, FEcpw, and IEEcw estimators.
We also compare the NEME and NEMEcw, generally inconsistent estimators for the iATE and cATE, and compare the results against the IEE, IEEcw and other potentially consistent estimators. In this trial, the NEME and NEMEcw produced estimates that strayed from the IEE, FE, EME, and IEEcw, FEcw, EMEcw estimates, respectively.
Altogether, the case study reanalysis may hint at the presence of ICS.
Still, we reiterate that the recommendations described here for deducing the presence of informative sizes are purely qualitative. Future work can focus on producing statistical tests for formally detecting the presence of meaningful informative sizes.

The awareness regarding the risks of informative sizes in cluster randomized trials has recently gathered momentum \cite{kahan_estimands_2023,kahan_demystifying_2024,wang_two_2022,lee_how_2025}. While the prospect of informative sizes in clustered designs appears theoretically realistic, given the relatively nascent interest in this issue, it has not yet been practically and systematically explored across different studies. A comprehensive examination of past CRTs with the practical steps highlighted in our case study reanalysis and formal statistical tests (to be developed) will be crucial to identify the prevalence of such informative sizes.

We would like to emphasize that there are also other important considerations to note when choosing between these different estimators. For example, the fixed effects model has been illustrated to be effective in controlling for chance covariate imbalance, especially when the number of clusters are low in multi-period CRT designs \cite{lee_fixed-effects_2024}. In contrast, the mixed effects models have the benefit of automatically estimating the intracluster correlations.
While we continue to endorse such considerations, our primary focus in this article is to describe modeling considerations when analyzing CRXO trials with suspected informative sizes, and demystify their implied target estimands under the potential outcomes framework for more clear and transparent use of these models in practice.

\subsection{Conclusion}

Previous work in P-CRTs \cite{wang_two_2022} and PB-CRTs \cite{lee_how_2025} demonstrated that common estimators in the analysis of these CRT designs may not converge to interpretable estimands in the presence of informative cluster sizes.
Notably, the unweighted and weighted EME was demonstrated to be inconsistent for the iATE and cATE in P-CRTs \cite{wang_two_2022} and PB-CRTs \cite{lee_how_2025}, potentially yielding biased results in P-CRT designs. 
However, these estimators were previously shown to yield surprisingly minimal bias in PB-CRT designs \cite{lee_how_2025}.
In this work, we describe additional weighted average treatment effect estimands and clarify more forms of informative sizes. We further observe that when cluster-period sizes are fixed between-periods, within-clusters ($K_{i1} = K_{i2} \, \forall \, i$), the unweighted and weighted EME estimators can be consistent for the iATE, cpATE, and cATE estimands in CRXO trial designs, and generally yield empirically unbiased estimates.
This highlights the need to concurrently consider the trial design and intended analysis method during the CRXO study design phase.
Overall, we reveal that whether an estimator is consistent for its corresponding weighted estimand in the presence of informative sizes depends on the intrinsic properties of both the estimator and the study design.

Our simulation results indicated that the unweighted and weighted NEME estimators can yield unacceptably biased estimates for the iATE, cpATE, cATE, and pATE estimands in the presence of informative sizes.
In contrast, the unweighted and weighted IEE treatment effect estimators are generally trustworthy estimators for the iATE, cpATE, cATE, and pATE estimands in CRXO trials with informative sizes. The unweighted and weighted FE and EME estimators can also be consistent and reliable, however they often require additional assumptions, oftentimes requiring cluster-period sizes to be fixed between-periods, within-clusters ($K_{i1} = K_{i2} \, \forall \, i$) (Tables \ref{tab:summary} \& \ref{tab:advantages_disadvantages}).
However, the FE and FEcw estimators generally target the pATE and cpATE estimands, respectively, which can be very misleading in CRXO trials with IPS (Tables \ref{tab:summary} \& \ref{tab:advantages_disadvantages}).

Altogether, the unweighted and weighted IEE estimators are always consistent in these settings and are easily implemented in standard statistical software when specifying different weights and robust standard errors, making them preferable in the analysis of CRXO trials with informative sizes.

\begin{table}[p]
\fontsize{8.75}{10.3}\selectfont 
\caption{General advantages and disadvantages of commonly used models for estimating the individual-average treatment effect (iATE), cluster-period-average treatment effect (cpATE), cluster-average treatment effect (cATE), and period-average treatment effect (pATE) estimands in CRXO trials when informative sizes are present.}
\label{tab:advantages_disadvantages}
\begin{center}
\bgroup
\def\arraystretch{1.3}
\begin{tabular}{|p{0.1\textwidth} | p{0.4\textwidth} p{0.4\textwidth}|} 
    \hline
    \multicolumn{1}{|c|}{\textbf{Methods}} & \multicolumn{1}{c}{\textbf{Advantages}} & \multicolumn{1}{c|}{\textbf{Disadvantages}} \\
    \hline\hline
    \multicolumn{3}{|c|}{\textbf{iATE}} \\
    \hline
    IEE & + Theoretically consistent & - Doesn't automatically estimate the wp-ICC and bp-ICC \\
    \hdashline
    EME & + Theoretically consistent (when cluster-period sizes don't vary) \newline + Automatically estimates the ICC & - Doesn't automatically estimate the wp-ICC and bp-ICC \\
    \hdashline
    NEME & + Automatically estimates the wp-ICC and bp-ICC & - Not theoretically consistent \\
    \hdashline
    FE & + Theoretically consistent (when cluster-period sizes don't vary) & - Doesn't automatically estimate the wp-ICC and bp-ICC \newline - Generally targets the pATE\\
    \hline
    \multicolumn{3}{|c|}{\textbf{cpATE}} \\
    \hline
    IEEcpw & + Theoretically consistent \newline + Unbiased in expectation & - Doesn't automatically estimate the wp-ICC and bp-ICC \\
    \hdashline
    EMEcpw & + Theoretically consistent (when cluster-period sizes don't vary) \newline + Automatically estimates the ICC & - Doesn't automatically estimate the wp-ICC and bp-ICC \newline - Can only be implemented when cluster-period sizes don't vary \\
    \hdashline
    NEMEcpw & + Automatically estimates the wp-ICC and bp-ICC & - Not theoretically consistent \newline - Can only be implemented when cluster-period sizes don't vary \newline - Can be inefficient \\
    \hdashline
    FEcpw & + Theoretically consistent \newline + Unbiased in expectation & - Doesn't automatically estimate the wp-ICC and bp-ICC \\
    \hline
    \multicolumn{3}{|c|}{\textbf{cATE}} \\
    \hline
    IEEcw & + Theoretically consistent & - Doesn't automatically estimate the wp-ICC and bp-ICC \\
    \hdashline
    EMEcw & + Theoretically consistent (when cluster-period sizes don't vary) \newline + Automatically estimates the ICC & - Doesn't automatically estimate the wp-ICC and bp-ICC \\
    \hdashline
    NEMEcw & + Automatically estimates the wp-ICC and bp-ICC & - Not theoretically consistent \newline - Can be inefficient\\
    \hdashline
    FEcw & + Theoretically consistent (when cluster-period sizes don't vary) & - Doesn't automatically estimate the wp-ICC and bp-ICC \newline - Generally targets the cpATE \\
    \hline
    \multicolumn{3}{|c|}{\textbf{pATE}} \\
    \hline
    IEEpw & + Theoretically consistent & - Doesn't automatically estimate the wp-ICC and bp-ICC \\
    \hdashline
    EMEpw & + Theoretically consistent (when cluster-period sizes don't vary) \newline + Automatically estimates the ICC & - Doesn't automatically estimate the wp-ICC and bp-ICC \newline - Can only be implemented when cluster-period sizes don't vary (equivalent to EME) \\
    \hdashline
    NEMEpw & + Automatically estimates the wp-ICC and bp-ICC & - Not theoretically consistent \newline - Can only be implemented when cluster-period sizes don't vary (equivalent to NEME) \\
    \hdashline
    FEpw & + Theoretically consistent (when cluster-period sizes don't vary) & - Doesn't automatically estimate the wp-ICC and bp-ICC \\
    \hline\end{tabular}
\egroup
\end{center}
\end{table}

\break

\section*{Acknowledgment}
Research in this article was supported by a Patient-Centered Outcomes Research Institute Award\textsuperscript{\textregistered} (PCORI\textsuperscript{\textregistered} Award ME-2022C2-27676). The statements presented in this article are solely the responsibility of the authors and do not necessarily represent the official views of the National Institutes of Health or PCORI\textsuperscript{\textregistered}, its Board of Governors, or the Methodology Committee. B.C.K. and A.C. are funded by the UK MRC, grants MC\_UU\_00004/07 and MC\_UU\_00004/09.

\section*{Data Availability Statement}

\noindent Data sharing is not applicable to this article as no new data were created or analyzed in this study. 
\break

\bibliographystyle{wileyNJD-AMA}
\bibliography{references}

\clearpage

\appendix

\section{General conditions under which the described estimands are equivalent}
\label{sect:appendix_estimands_equiv}

We specify when the iATE, cATE, pATE, and cpATE estimands coincide under the following set of conditions in the data generating process:
\begin{enumerate}[label=(\alph*)]
    \item $\{Y_{i1k}(1)-Y_{i1k}(0),Y_{i2k}(1)-Y_{i2k}(0)\} \indep \left(K_{i1}, K_{i2}\right)$
    \item $E[Y_{i1k}(1)-Y_{i1k}(0)]=E[Y_{i2k}(1)-Y_{i2k}(0)]$
    \item $E[K_{i1}]=E[K_{i2}]$
    \item $\{Y_{i1k}(1)-Y_{i1k}(0)\} \indep K_{i1}$ and $\{Y_{i2k}(1)-Y_{i2k}(0)\} \indep K_{i2}$
    \item $K_{i1}+K_{i2}=\text{constant}$
    \item $K_{i1}=K_{i2}$
\end{enumerate}
for all clusters $i$. These conditions are interpreted as (a) all individual treatment effects are independent of all cluster-period cell sizes, (b) there is no between-period, within-cluster treatment effect heterogeneity, (c) there is no expected between-period, within-cluster sample size heterogeneity, (d) individual treatment effects are independent of their corresponding cluster-period cell-sizes, (e) there is a common cluster size across all clusters, (f) cluster-period sizes are equivalent between-periods, within-clusters.

Notably (a) is a more specific subset of (d), since joint independence (a) implies marginal independence (d), but not vice versa. Similarly, (f) is a more specific subset of (c).

\subsection{iATE versus cATE}
We can compare the iATE and cATE estimands to find under which general conditions the two estimands coincide.
\[
iATE-cATE=
\frac{E\left[\sum_{j=1}^{2} \sum_{k=1}^{K_{ij}} \left[Y_{ijk}(1)-Y_{ijk}(0)\right] \right]}{E\left[\sum_{j=1}^{2}K_{ij}\right]} 
- E\left[ \frac{\sum_{j=1}^{2} \sum_{k=1}^{K_{ij}} \left[Y_{ijk}(1)-Y_{ijk}(0)\right]}{\sum_{j=1}^{2}K_{ij}} \right]
\]
\[
=E\left[
    \left(
        \sum_{k=1}^{K_{i1}}E[Y_{i1k}(1)-Y_{i1k}(0)|K_{i1},K_{i2}] + \sum_{k=1}^{K_{i2}}E[Y_{i2k}(1)-Y_{i2k}(0)|K_{i1},K_{i2}]
    \right)
    \left(\frac{1}{E[K_{i1}+K_{i2}]}-\frac{1}{K_{i1}+K_{i2}}\right)
\right]
\]
For the above equation to equal zero, then generally (a) and (b) need to be true, or (e) needs to be true.

\subsection{iATE versus pATE}
We can compare the iATE and pATE estimands to determine under which general conditions the two estimands coincide.
\[
iATE-pATE=
\frac{E\left[\sum_{j=1}^{2} \sum_{k=1}^{K_{ij}} \left[Y_{ijk}(1)-Y_{ijk}(0)\right] \right]}{E\left[\sum_{j=1}^{2}K_{ij}\right]} 
- \frac{1}{2} \sum_{j=1}^{2} \left[ \frac{E\left[ \sum_{k=1}^{K_{ij}} \left[Y_{ijk}(1)-Y_{ijk}(0)\right] \right]}{E\left[K_{ij}\right]} \right]
\]
\[
\begin{split}
&= \frac{\left(E\left[ \sum_{k=1}^{K_{i1}}[Y_{i1k}(1)-Y_{i1k}(0)]E[K_{i2}] - \sum_{k=1}^{K_{i2}}[Y_{i2k}(1)-Y_{i2k}(0)]E[K_{i1}] \right]\right) E\left[K_{i1}-K_{i2}\right]}{E\left[K_{i1}+K_{i2}\right]E\left[K_{i1}\right]E\left[K_{i2}\right]} \\
&\propto \left(E\left[ \sum_{k=1}^{K_{i1}}[Y_{i1k}(1)-Y_{i1k}(0)]E[K_{i2}] - \sum_{k=1}^{K_{i2}}[Y_{i2k}(1)-Y_{i2k}(0)]E[K_{i1}] \right]\right) E\left[K_{i1}-K_{i2}\right]
\end{split}
\]
For the above equation to equal zero, then either (c) is true or (d) and (b) are true.

\subsection{iATE versus cpATE}
We can compare the iATE and cpATE estimands to determine under which general conditions the two estimands coincide.
\[
iATE-cpATE=
\frac{E\left[\sum_{j=1}^{2} \sum_{k=1}^{K_{ij}} \left[Y_{ijk}(1)-Y_{ijk}(0)\right] \right]}{E\left[\sum_{j=1}^{2}K_{ij}\right]} 
- E\left[ \frac{1}{2} \sum_{j=1}^{2} \frac{ \sum_{k=1}^{K_{ij}} \left[Y_{ijk}(1)-Y_{ijk}(0)\right]}{K_{ij}} \right]
\]
\[
=E\left[ \sum_{k=1}^{K_{i1}} [Y_{i1k}(1)-Y_{i1k}(0)]\left(\frac{1}{E[K_{i1}+K_{i2}]/2}-\frac{1}{K_{i1}}\right) \right] + E\left[ \sum_{k=1}^{K_{i2}} [Y_{i2k}(1)-Y_{i2k}(0)]\left(\frac{1}{E[K_{i1}+K_{i2}]/2}-\frac{1}{K_{i2}}\right) \right]
\]
\begin{align*}
=E&\left[ \sum_{k=1}^{K_{i1}} E[Y_{i1k}(1)-Y_{i1k}(0)|K_{i1}]\left(\frac{1}{E[K_{i1}+K_{i2}]/2}-\frac{1}{K_{i1}}\right) \right] + \\
&E\left[ \sum_{k=1}^{K_{i2}} E[Y_{i2k}(1)-Y_{i2k}(0)|K_{i2}]\left(\frac{1}{E[K_{i1}+K_{i2}]/2}-\frac{1}{K_{i2}}\right) \right]
\end{align*}
For the above equation to equal zero, then either (d) and (b), or (d) and (c), are true.

\subsection{cATE versus pATE}
We can compare the cATE and pATE estimands to determine under which general conditions the two estimands coincide.
\[
cATE-pATE = E\left[ \frac{\sum_{j=1}^{2} \sum_{k=1}^{K_{ij}} \left[Y_{ijk}(1)-Y_{ijk}(0)\right]}{\sum_{j=1}^{2}K_{ij}} \right] - \frac{1}{2} \sum_{j=1}^{2} \left[ \frac{E\left[ \sum_{k=1}^{K_{ij}} \left[Y_{ijk}(1)-Y_{ijk}(0)\right] \right]}{E\left[K_{ij}\right]} \right]
\]
\[
= E\left[\sum_{k=1}^{K_{i1}}[Y_{i1k}(1)-Y_{i1k}(0)]\left(\frac{2}{K_{i1}+K_{i2}}-\frac{1}{E[K_{i1}]}\right)\right] + E\left[\sum_{k=1}^{K_{i2}}[Y_{i2k}(1)-Y_{i2k}(0)]\left(\frac{2}{K_{i1}+K_{i2}}-\frac{1}{E[K_{i2}]}\right)\right]
\]
\begin{align*}
= E\left[\sum_{k=1}^{K_{i1}}E[Y_{i1k}(1)-Y_{i1k}(0)|K_{i1},K_{i2}]\left(\frac{2}{K_{i1}+K_{i2}}-\frac{1}{E[K_{i1}]}\right)\right] + \\ E\left[\sum_{k=1}^{K_{i2}}E[Y_{i2k}(1)-Y_{i2k}(0)|K_{i1},K_{i2}]\left(\frac{2}{K_{i1}+K_{i2}}-\frac{1}{E[K_{i2}]}\right)\right]
\end{align*}
For the above equation to equal zero, then (a) and (b) need to be true.

\subsection{cATE versus cpATE}
We can compare the cATE and cpATE estimands to determine under which general conditions the two estimands coincide.
\[
cATE-cpATE = E\left[ \frac{\sum_{j=1}^{2} \sum_{k=1}^{K_{ij}} \left[Y_{ijk}(1)-Y_{ijk}(0)\right]}{\sum_{j=1}^{2}K_{ij}} \right] - E\left[ \frac{1}{2} \sum_{j=1}^{2} \frac{ \sum_{k=1}^{K_{ij}} \left[Y_{ijk}(1)-Y_{ijk}(0)\right]}{K_{ij}} \right]
\]
\[
=E\left[\frac{\left(\sum_{k=1}^{K_{i1}}\left[Y_{i1k}(1)-Y_{i1k}(0)\right]K_{i2}-\sum_{k=1}^{K_{i2}}\left[Y_{i2k}(1)-Y_{i2k}(0)\right]K_{i1}\right)(K_{i1}-K_{i2})}{(K_{i1}+K_{i2})K_{i1}K_{i2}}\right]
\]
\[
=E\left[\frac{\left(\sum_{k=1}^{K_{i1}}E\left[Y_{i1k}(1)-Y_{i1k}(0)|K_{i1},K_{i2}\right]K_{i2}-\sum_{k=1}^{K_{i2}}E\left[Y_{i2k}(1)-Y_{i2k}(0)|K_{i1},K_{i2}\right]K_{i1}\right)(K_{i1}-K_{i2})}{(K_{i1}+K_{i2})K_{i1}K_{i2}}\right]
\]
For the above equation to equal zero, then (a) and (b) need to be true, or (f) needs to be true.

\subsection{pATE versus cpATE}
We can compare the pATE and cpATE estimands to determine under which general conditions the two estimands coincide.
\[
pATE - cpATE = \frac{1}{2} \sum_{j=1}^{2} \left[ \frac{E\left[ \sum_{k=1}^{K_{ij}} \left[Y_{ijk}(1)-Y_{ijk}(0)\right] \right]}{E\left[K_{ij}\right]} \right] - E\left[ \frac{1}{2} \sum_{j=1}^{2} \frac{ \sum_{k=1}^{K_{ij}} \left[Y_{ijk}(1)-Y_{ijk}(0)\right]}{K_{ij}} \right]
\]
\[
= E\left[\sum_{k=1}^{K_{i1}}[Y_{i1k}(1)-Y_{i1k}(0)]\left(\frac{1}{E[K_{i1}]}-\frac{1}{K_{i1}}\right)\right] + E\left[\sum_{k=1}^{K_{i2}}[Y_{i2k}(1)-Y_{i2k}(0)]\left(\frac{1}{E[K_{i2}]}-\frac{1}{K_{i2}}\right)\right]
\]
\[
= E\left[\sum_{k=1}^{K_{i1}}E[Y_{i1k}(1)-Y_{i1k}(0)|K_{i1}]\left(\frac{1}{E[K_{i1}]}-\frac{1}{K_{i1}}\right)\right] + E\left[\sum_{k=1}^{K_{i2}}E[Y_{i2k}(1)-Y_{i2k}(0)|K_{i2}]\left(\frac{1}{E[K_{i2}]}-\frac{1}{K_{i2}}\right)\right]
\]
For the above equation to equal zero, then (d) needs to be true.

\section{Proof of Proposition \ref{proposition:ICPS}}
\label{sect:appendix_proposition}

Using the sufficient conditions described in Table \ref{tab:estimands_equiv}, as informed by the calculations in Appendix (\ref{sect:appendix_estimands_equiv}), we make the following proposition:

\noindent\textbf{Proposition \ref{proposition:ICPS}}
\begin{enumerate}
    \item \textit{ICPS requires either ICS or IPS to occur.}
    \item \textit{ICPS does not require both ICS and IPS to occur.}
\end{enumerate}

We can prove Proposition \ref{proposition:ICPS}.1 with a proof by contradiction, using the sufficient true conditions for estimands to be equal, as outlined in Table \ref{tab:estimands_equiv}.i. Assume there can be ICPS despite there being no ICS nor IPS, cpATE $\neq$ (iATE = cATE = pATE). For cATE = pATE to be true, conditions (a) $\cap$ (b) must be true, which then indicates that cpATE = cATE. Therefore, we contradict the initial assumption that ICPS can occur despite there being no ICS nor IPS, proving Proposition \ref{proposition:ICPS}.1.

We can prove Proposition \ref{proposition:ICPS}.2 with a proof by induction, using the sufficient false conditions for informative sizes to occur, as outlined in Table \ref{tab:estimands_equiv}.ii. ICPS only requires conditions (d) $\cap$ (f) to be false. However, no ICS (iATE = cATE) can still occur through condition (e) being true, and no IPS (iATE = pATE) can still occur through condition (c) being true, proving Proposition \ref{proposition:ICPS}.2.

It may appear that we can contradict Proposition \ref{proposition:ICPS}.1 by having ICPS occur with conditions (d) $\cap$ (f) being false, while there simultaneously being no ICS nor IPS with conditions (e) $\cap$ (c) being true. Indeed, separately, condition (e) or (c) can be true while conditions (d) $\cap$ (f) are false. However, we cannot have both conditions (e) $\cap$ (c) simultaneously be true while conditions (d) $\cap$ (f) are false, since this only occurs when $K_{ij}=K \, \forall \, i,j$, which contradicts the specified condition (f) being false.

\section{Derivation of the unweighted \& weighted Independence estimating equation estimator (IEE, IEEcpw, IEEcw)}
\label{sect:appendix_IEE_IEEcpw_IEEcw}

\subsection{Derivation of the Independence estimating equation (IEE) estimator}
\label{sect:appendix_IEE}

In a standard 2-period, 2-sequence CRXO trial design (Figure \ref{fig:CRXO_design}), an independence estimating equation (IEE) estimator yields the following treatment effect estimator:
\[\hat{\delta}_{IEE}=(Z'Z)^{-1}Z'Y|_{\delta}\]
where $Z$ is the design matrix of the indicator variables for treatment effect $(\delta)$ and for the two period effects $(\Phi_{1}, \Phi_{2})$, and $Y$ is the vector of outcomes. The Fisher information matrix is:
\begin{align*}
    Z'Z = 
    \begin{bmatrix}
        \sum_{i=1}^{I}(S_{i}K_{i1} + (1-S_{i})K_{i2})) & \sum_{i=1}^{I}S_{i}K_{i1} & \sum_{i=1}^{I}(1-S_{i})K_{i2}\\
        \sum_{i=1}^{I}S_{i}K_{i1} & \sum_{i=1}^{I}K_{i1} & 0\\
        \sum_{i=1}^{I}(1-S_{i})K_{i2} & 0 & \sum_{i=1}^{I}K_{i2}
    \end{bmatrix} \,,
\end{align*}
where notably, $S_{i}$ is the sequence indicator variable for each cluster $i$ and is either $=1$ or $0$. Here, the top-left entry in $Z'Z$ corresponds with the number of individuals in cluster-period cells receiving the treatment.

Accordingly, the first row of the inverse of the Fisher information matrix, corresponding with the treatment effect $(\delta)$, is:
\begin{align*}
    (Z'Z)^{-1}|_{\delta} =
    \begin{bmatrix}
        \frac{ (\sum_{i=1}^{I}K_{i1})(\sum_{i=1}^{I}K_{i2}) }
        { (\sum_{i=1}^{I}S_{i}K_{i1}) (\sum_{i=1}^{I}(1-S_{i})K_{i1}) (\sum_{i=1}^{I}K_{i2}) + (\sum_{i=1}^{I}S_{i}K_{i2}) (\sum_{i=1}^{I}(1-S_{i})K_{i2}) (\sum_{i=1}^{I}K_{i1}) } \\
        \frac{ -(\sum_{i=1}^{I}S_{i}K_{i1})(\sum_{i=1}^{I}K_{i2}) }
        { (\sum_{i=1}^{I}S_{i}K_{i1}) (\sum_{i=1}^{I}(1-S_{i})K_{i1}) (\sum_{i=1}^{I}K_{i2}) + (\sum_{i=1}^{I}S_{i}K_{i2}) (\sum_{i=1}^{I}(1-S_{i})K_{i2}) (\sum_{i=1}^{I}K_{i1}) } \\
        \frac{ -(\sum_{i=1}^{I}K_{i1})(\sum_{i=1}^{I}(1-S_{i})K_{i2}) }
        { (\sum_{i=1}^{I}S_{i}K_{i1}) (\sum_{i=1}^{I}(1-S_{i})K_{i1}) (\sum_{i=1}^{I}K_{i2}) + (\sum_{i=1}^{I}S_{i}K_{i2}) (\sum_{i=1}^{I}(1-S_{i})K_{i2}) (\sum_{i=1}^{I}K_{i1}) }
    \end{bmatrix}' \,,
\end{align*}
and with:
\begin{align*}
    Z'Y = 
    \begin{bmatrix}
        \sum_{i=1}^{I}(S_{i}\sum_{k=1}^{K_{i1}}Y_{i1k} + (1-S_{i})\sum_{k=1}^{K_{i2}}Y_{i2k}) \\
        \sum_{i=1}^{I}\sum_{k=1}^{K_{i1}}Y_{i1k} \\
        \sum_{i=1}^{I}\sum_{k=1}^{K_{i2}}Y_{i2k}
    \end{bmatrix} \,,
\end{align*}
yields the following treatment effect point estimator:
\begin{align*}
    \begin{split}
        (Z'Z)^{-1}Z'Y|_{\delta} = \hat{\delta}_{IEE} =
    \end{split}\\
    \begin{split}
        \left(\frac{ (\sum_{i=1}^{I}S_{i}\sum_{k=1}^{K_{i1}}Y_{i1k}) (\sum_{i=1}^{I}(1-S_{i})K_{i1}) (\sum_{i=1}^{I}K_{i2}) + (\sum_{i=1}^{I}S_{i}K_{i2}) (\sum_{i=1}^{I}(1-S_{i})\sum_{k=1}^{K_{i2}}Y_{i2k}) (\sum_{i=1}^{I}K_{i1})  }
        { (\sum_{i=1}^{I}S_{i}K_{i1}) (\sum_{i=1}^{I}(1-S_{i})K_{i1}) (\sum_{i=1}^{I}K_{i2}) + (\sum_{i=1}^{I}S_{i}K_{i2}) (\sum_{i=1}^{I}(1-S_{i})K_{i2}) (\sum_{i=1}^{I}K_{i1}) }\right)
        \\ -
        \left(\frac{ (\sum_{i=1}^{I}S_{i}K_{i1}) (\sum_{i=1}^{I}(1-S_{i})\sum_{k=1}^{K_{i1}}Y_{i1k}) (\sum_{i=1}^{I}K_{i2}) + (\sum_{i=1}^{I}S_{i}\sum_{k=1}^{K_{i2}}Y_{i2k}) (\sum_{i=1}^{I}(1-S_{i})K_{i2}) (\sum_{i=1}^{I}K_{i1}) }
        { (\sum_{i=1}^{I}S_{i}K_{i1}) (\sum_{i=1}^{I}(1-S_{i})K_{i1}) (\sum_{i=1}^{I}K_{i2}) + (\sum_{i=1}^{I}S_{i}K_{i2}) (\sum_{i=1}^{I}(1-S_{i})K_{i2}) (\sum_{i=1}^{I}K_{i1}) }\right)
    \end{split} \,.
\end{align*}
Connecting this to potential outcomes (eq. \ref{eq:PO}) yields:
\small
\[\hat{\delta}_{IEE} =\]
\[
    \left(\frac{ (\sum_{i=1}^{I}S_{i}\sum_{k=1}^{K_{i1}}Y_{i1k}(1)) (\sum_{i=1}^{I}(1-S_{i})K_{i1}) (\sum_{i=1}^{I}K_{i2}) + (\sum_{i=1}^{I}S_{i}K_{i2}) (\sum_{i=1}^{I}(1-S_{i})\sum_{k=1}^{K_{i2}}Y_{i2k}(1)) (\sum_{i=1}^{I}K_{i1}) }
    { (\sum_{i=1}^{I}S_{i}K_{i1}) (\sum_{i=1}^{I}(1-S_{i})K_{i1}) (\sum_{i=1}^{I}K_{i2}) + (\sum_{i=1}^{I}S_{i}K_{i2}) (\sum_{i=1}^{I}(1-S_{i})K_{i2}) (\sum_{i=1}^{I}K_{i1}) }\right)
\]
\[
    -\left(\frac{ (\sum_{i=1}^{I}S_{i}K_{i1}) (\sum_{i=1}^{I}(1-S_{i})\sum_{k=1}^{K_{i1}}Y_{i1k}(0)) (\sum_{i=1}^{I}K_{i2}) + (\sum_{i=1}^{I}S_{i}\sum_{k=1}^{K_{i2}}Y_{i2k}(0)) (\sum_{i=1}^{I}(1-S_{i})K_{i2}) (\sum_{i=1}^{I}K_{i1})  }
    { (\sum_{i=1}^{I}S_{i}K_{i1}) (\sum_{i=1}^{I}(1-S_{i})K_{i1}) (\sum_{i=1}^{I}K_{i2}) + (\sum_{i=1}^{I}S_{i}K_{i2}) (\sum_{i=1}^{I}(1-S_{i})K_{i2}) (\sum_{i=1}^{I}K_{i1}) }\right) \,.
\]
\normalsize
We can then demonstrate that this unweighted estimator is consistent and asymptotically unbiased for the iATE:
\small
\[\lim_{I\to\infty} \hat{\delta}_{IEE} =\]
\[
    \lim_{I\to\infty} \left(\frac{ (\sum_{i=1}^{I}S_{i}\sum_{k=1}^{K_{i1}}Y_{i1k}(1)) (\sum_{i=1}^{I}(1-S_{i})K_{i1}) (\sum_{i=1}^{I}K_{i2}) + (\sum_{i=1}^{I}S_{i}K_{i2}) (\sum_{i=1}^{I}(1-S_{i})\sum_{k=1}^{K_{i2}}Y_{i2k}(1)) (\sum_{i=1}^{I}K_{i1}) }
    { (\sum_{i=1}^{I}S_{i}K_{i1}) (\sum_{i=1}^{I}(1-S_{i})K_{i1}) (\sum_{i=1}^{I}K_{i2}) + (\sum_{i=1}^{I}S_{i}K_{i2}) (\sum_{i=1}^{I}(1-S_{i})K_{i2}) (\sum_{i=1}^{I}K_{i1}) }\right)
\]
\[
    -\lim_{I\to\infty} \left(\frac{ (\sum_{i=1}^{I}S_{i}K_{i1}) (\sum_{i=1}^{I}(1-S_{i})\sum_{k=1}^{K_{i1}}Y_{i1k}(0)) (\sum_{i=1}^{I}K_{i2}) + (\sum_{i=1}^{I}S_{i}\sum_{k=1}^{K_{i2}}Y_{i2k}(0)) (\sum_{i=1}^{I}(1-S_{i})K_{i2}) (\sum_{i=1}^{I}K_{i1}) }
    { (\sum_{i=1}^{I}S_{i}K_{i1}) (\sum_{i=1}^{I}(1-S_{i})K_{i1}) (\sum_{i=1}^{I}K_{i2}) + (\sum_{i=1}^{I}S_{i}K_{i2}) (\sum_{i=1}^{I}(1-S_{i})K_{i2}) (\sum_{i=1}^{I}K_{i1}) }\right)
\]
\normalsize
and:
\[
    \hat{\delta}_{IEE} \xrightarrow{P} \left(
        \frac{E\left[\sum_{k=1}^{K_{i1}}Y_{i1k}(1)|S_{i}=1\right]E[K_{i1}|S_{i}=0]E[K_{i2}] + E[K_{i2}|S_{i}=1]E\left[\sum_{k=1}^{K_{i2}}Y_{i2k}(1)|S_{i}=0\right]E[K_{i1}]}{E[K_{i1}|S_{i}=1]E[K_{i1}|S_{i}=0]E[K_{i2}] + E[K_{i2}|S_{i}=1]E[K_{i2}|S_{i}=0]E[K_{i1}]}
    \right)
\]
\[
    \indent - \left(
        \frac{E[K_{i1}|S_{i}=1]E\left[\sum_{k=1}^{K_{i1}}Y_{i1k}(0)|S_{i}=0\right]E[K_{i2}] + E\left[\sum_{k=1}^{K_{i2}}Y_{i2k}(0)|S_{i}=1\right]E[K_{i2}|S_{i}=0]E[K_{i1}]}{E[K_{i1}|S_{i}=1]E[K_{i1}|S_{i}=0]E[K_{i2}] + E[K_{i2}|S_{i}=1]E[K_{i2}|S_{i}=0]E[K_{i1}]}
    \right)
\]
where with randomization, the sequence variable $S_{i}$ is independent of the potential outcomes and cluster-period sizes, $S_{i} \indep \Omega$, and $\Omega=\{Y_{ijk}(0), Y_{ijk}(1), K_{ij}\}_{i=1, k=1}^{I, K_{ij}}$, such that:
\[
    \hat{\delta}_{IEE} \xrightarrow{P} \left(
        \frac{E\left[\sum_{k=1}^{K_{i1}}Y_{i1k}(1)\right]E[K_{i1}]E[K_{i2}] + E[K_{i2}]E\left[\sum_{k=1}^{K_{i2}}Y_{i2k}(1)\right]E[K_{i1}]}{E[K_{i1}]E[K_{i1}]E[K_{i2}] + E[K_{i2}]E[K_{i2}]E[K_{i1}]}
    \right)
\]
\[
    \indent - \left(
        \frac{E[K_{i1}]E\left[\sum_{k=1}^{K_{i1}}Y_{i1k}(0)\right]E[K_{i2}] + E\left[\sum_{k=1}^{K_{i2}}Y_{i2k}(0)\right]E[K_{i2}]E[K_{i1}]}{E[K_{i1}]E[K_{i1}]E[K_{i2}] + E[K_{i2}]E[K_{i2}]E[K_{i1}]}
    \right) \,,
\]

Overall, we can prove that the IEE estimator is consistent for iATE:
\begin{equation}
    \hat{\delta}_{IEE} \xrightarrow{P}  \frac{E\left[\sum_{j=1}^{2} \sum_{k=1}^{K_{ij}} \left[Y_{ijk}(1)-Y_{ijk}(0)\right] \right]}{E\left[\sum_{j=1}^{2}K_{ij}\right]} \,.
\end{equation}
We note that no assumptions about non-informative sizes are required for this result to hold.

\subsection{Derivation of the Independence estimating equation with inverse cluster-period size weights (IEEcpw) estimator}
\label{sect:appendix_IEEcpw}

It is then straightforward using the result from Section \ref{sect:appendix_IEE} to derive the treatment effect estimator for the independence estimating equation with inverse cluster-period size weights:
\[\hat{\delta}_{IEEcpw} =\]
\[
    \left(\frac{ (\sum_{i=1}^{I}S_{i}\frac{\sum_{k=1}^{K_{i1}}Y_{i1k}(1)}{K_{i1}}) (\frac{I}{2}) (I)) + (\frac{I}{2}) (\sum_{i=1}^{I}(1-S_{i})\frac{\sum_{k=1}^{K_{i2}}Y_{i2k}(1)}{K_{i2}}) (I) }
    { (\frac{I}{2}) (\frac{I}{2}) (I) + (\frac{I}{2}) (\frac{I}{2}) (I) }\right)
\]
\[
    \indent -\left(\frac{ (\frac{I}{2}) (\sum_{i=1}^{I}(1-S_{i})\frac{\sum_{k=1}^{K_{i1}}Y_{i1k}(0)}{K_{i1}}) (I) + (\sum_{i=1}^{I}S_{i}\frac{\sum_{k=1}^{K_{i2}}Y_{i2k}(0)}{K_{i2}}) (\frac{I}{2}) (I) }
    { (\frac{I}{2}) (\frac{I}{2}) (I) + (\frac{I}{2}) (\frac{I}{2}) (I) }\right)
\]
which simplifies to:
\[
    =\frac{1}{I}\sum_{i=1}^{I} 
    \left(
        S_{i}\left( \frac{\sum_{k=1}^{K_{i1}}Y_{i1k}(1)}{K_{i1}} - \frac{\sum_{k=1}^{K_{i2}}Y_{i2k}(0)}{K_{i2}} \right) + 
        (1-S_{i}) \left(\frac{\sum_{k=1}^{K_{i2}}Y_{i2k}(1)}{K_{i2}} - \frac{\sum_{k=1}^{K_{i1}}Y_{i1k}(0)}{K_{i1}} \right) 
    \right) \,.
\]

It can be easily demonstrated that this IEEcpw estimator is consistent for the cpATE estimand:
\begin{equation}
    \hat{\delta}_{IEEcpw} \xrightarrow{P} E\left[ \frac{1}{2} \sum_{j=1}^{2} \frac{ \sum_{k=1}^{K_{ij}} \left[Y_{ijk}(1)-Y_{ijk}(0)\right]}{K_{ij}} \right] \,.
\end{equation}

Additionally, we can demonstrate that the IEEcpw estimator is unbiased for the cpATE in expectation over the sampling distribution. We can formally define the set of all potential outcomes for all $K_{ij}$ individuals in periods $j=1,2$ in sampled clusters $i=1,...,I$ as $\Omega=\{Y_{ijk}(0), Y_{ijk}(1), K_{ij}\}_{i=1, k=1}^{I, K_{ij}}$.
Formally, we take the expectation $\hat{\delta}_{IEEcpw}$ by treating the potential outcomes of the samples as fixed quantities and the sequence assignment $S_{i}$ as random.
Therefore, conditioning the expectation on the set of sampled potential outcomes and cluster-period sample sizes $\Omega$ yields:
\[E[\hat{\delta}_{IEEcpw}|\Omega] =\]
\[
    \frac{1}{I}\sum_{i=1}^{I} 
    \left(
        E[S_{i}|\Omega]\left( \frac{\sum_{k=1}^{K_{i1}}Y_{i1k}(1)}{K_{i1}} - \frac{\sum_{k=1}^{K_{i2}}Y_{i2k}(0)}{K_{i2}} \right) + 
        E[1-S_{i}|\Omega] \left(\frac{\sum_{k=1}^{K_{i2}}Y_{i2k}(1)}{K_{i2}} - \frac{\sum_{k=1}^{K_{i1}}Y_{i1k}(0)}{K_{i1}} \right) 
    \right) \,.
\]
where with randomization, the sequence variable $S_{i}$ is independent of the potential outcomes and cluster-period sizes, $S_{i} \indep \Omega$, and $\Omega=\{Y_{ijk}(0), Y_{ijk}(1), K_{ij}\}_{i=1, k=1}^{I, K_{ij}}$ and $E[S_{i}|\Omega] = E[1-S_{i}|\Omega] = \frac{1}{2}$. Accordingly:
\[E[\hat{\delta}_{IEEcpw}|\Omega] =
    \frac{1}{I}\sum_{i=1}^{I} 
    \left(
        \frac{1}{2}\left( \frac{\sum_{k=1}^{K_{i1}}Y_{i1k}(1)}{K_{i1}} + \frac{\sum_{k=1}^{K_{i2}}Y_{i2k}(1)}{K_{i2}} - \frac{\sum_{k=1}^{K_{i1}}Y_{i1k}(0)}{K_{i1}} - \frac{\sum_{k=1}^{K_{i2}}Y_{i2k}(0)}{K_{i2}} \right) 
    \right)
\]
\[
    = \frac{1}{I}\sum_{i=1}^{I}
    \left(
        \frac{1}{2} \sum_{j=1}^{2} \left( \frac{\sum_{k=1}^{K_{ij}}Y_{ijk}(1)}{K_{ij}} - \frac{\sum_{k=1}^{K_{ij}}Y_{ijk}(0)}{K_{ij}} \right) 
    \right) \,.
\]

We assume that the sample of clusters is a simple random sample from a superpopulation of clusters. With this superpopulation framework, we have two sources of randomness, random sampling from a superpopulation of clusters and subsequent randomization of treatment assignment. Overall, we can prove that the IEEcpw estimator is unbiased in expectation for the cpATE, where with the law of total expectation:
\[E[E[\hat{\delta}_{IEEcpw}|\Omega]] = E[\hat{\delta}_{IEEcpw}] = E\left[ \frac{1}{2} \sum_{j=1}^{2} \frac{ \sum_{k=1}^{K_{ij}} \left[Y_{ijk}(1)-Y_{ijk}(0)\right]}{K_{ij}} \right] \,.\]

\subsection{Derivation of the Independence estimating equation with inverse cluster size weights (IEEcw) estimator}
\label{sect:appendix_IEEcw}
We can use the result from Section \ref{sect:appendix_IEE} to similarly demonstrate that the independence estimating equation with inverse cluster size weights (IEEcw) estimator is consistent for the cATE estimand. 
We can show that the IEEcw estimator is:
\[\hat{\delta}_{IEEcw} =\]
\[
    \left(\frac{
        \begin{tabular}{l}
            $\left(\sum_{i=1}^{I}S_{i}\frac{\sum_{k=1}^{K_{i1}}Y_{i1k}(1)}{\sum_{j=1}^{2}K_{ij}}\right) \left(\sum_{i=1}^{I}(1-S_{i})\frac{K_{i1}}{\sum_{j=1}^{2}K_{ij}}\right) \left(\sum_{i=1}^{I}\frac{K_{i2}}{\sum_{j=1}^{2}K_{ij}}\right)$ \\
            \indent $+ \left(\sum_{i=1}^{I}S_{i}\frac{K_{i2}}{\sum_{j=1}^{2}K_{ij}}\right) \left(\sum_{i=1}^{I}(1-S_{i})\frac{\sum_{k=1}^{K_{i2}}Y_{i2k}(1)}{\sum_{j=1}^{2}K_{ij}}\right) \left(\sum_{i=1}^{I}\frac{K_{i1}}{\sum_{j=1}^{2}K_{ij}}\right)$
        \end{tabular}
    }{ 
        \begin{tabular}{l}
            $\left(\sum_{i=1}^{I}S_{i}\frac{K_{i1}}{\sum_{j=1}^{2}K_{ij}}\right) \left(\sum_{i=1}^{I}(1-S_{i})\frac{K_{i1}}{\sum_{j=1}^{2}K_{ij}}\right) \left(\sum_{i=1}^{I}\frac{K_{i2}}{\sum_{j=1}^{2}K_{ij}}\right)$ \\
            \indent $+ \left(\sum_{i=1}^{I}S_{i}\frac{K_{i2}}{\sum_{j=1}^{2}K_{ij}}\right) \left(\sum_{i=1}^{I}(1-S_{i})\frac{K_{i2}}{\sum_{j=1}^{2}K_{ij}}\right) \left(\sum_{i=1}^{I}\frac{K_{i1}}{\sum_{j=1}^{2}K_{ij}}\right)$
        \end{tabular} 
    }\right)
\]
\[
    \indent -\left(\frac{ 
        \begin{tabular}{l}
        $\left(\sum_{i=1}^{I}S_{i}\frac{K_{i1}}{\sum_{j=1}^{2}K_{ij}}\right) \left(\sum_{i=1}^{I}(1-S_{i})\frac{\sum_{k=1}^{K_{i1}}Y_{i1k}(0)}{\sum_{j=1}^{2}K_{ij}}\right) \left(\sum_{i=1}^{I}\frac{K_{i2}}{\sum_{j=1}^{2}K_{ij}}\right)$ \\
        \indent $+ \left(\sum_{i=1}^{I}S_{i}\frac{\sum_{k=1}^{K_{i2}}Y_{i2k}(0)}{\sum_{j=1}^{2}K_{ij}}\right) \left(\sum_{i=1}^{I}(1-S_{i})\frac{K_{i2}}{\sum_{j=1}^{2}K_{ij}}\right) \left(\sum_{i=1}^{I}\frac{K_{i1}}{\sum_{j=1}^{2}K_{ij}}\right)$
        \end{tabular}
    }{ 
    \begin{tabular}{l}
        $\left(\sum_{i=1}^{I}S_{i}\frac{K_{i1}}{\sum_{j=1}^{2}K_{ij}}\right) \left(\sum_{i=1}^{I}(1-S_{i})\frac{K_{i1}}{\sum_{j=1}^{2}K_{ij}}\right) \left(\sum_{i=1}^{I}\frac{K_{i2}}{\sum_{j=1}^{2}K_{ij}}\right)$ \\
        \indent $+ \left(\sum_{i=1}^{I}S_{i}\frac{K_{i2}}{\sum_{j=1}^{2}K_{ij}}\right) \left(\sum_{i=1}^{I}(1-S_{i})\frac{K_{i2}}{\sum_{j=1}^{2}K_{ij}}\right) \left(\sum_{i=1}^{I}\frac{K_{i1}}{\sum_{j=1}^{2}K_{ij}}\right)$ 
        \end{tabular}
    }\right) \,.
\]

We can then demonstrate that this weighted estimator is consistent and asymptotically unbiased for the cATE:
\[\lim_{I\to\infty} \hat{\delta}_{IEEcw} =\]
\[
    \lim_{I\to\infty} \left(\frac{
        \begin{tabular}{l}
            $\left(\sum_{i=1}^{I}S_{i}\frac{\sum_{k=1}^{K_{i1}}Y_{i1k}(1)}{\sum_{j=1}^{2}K_{ij}}\right) \left(\sum_{i=1}^{I}(1-S_{i})\frac{K_{i1}}{\sum_{j=1}^{2}K_{ij}}\right) \left(\sum_{i=1}^{I}\frac{K_{i2}}{\sum_{j=1}^{2}K_{ij}}\right)$ \\
            \indent $+ \left(\sum_{i=1}^{I}S_{i}\frac{K_{i2}}{\sum_{j=1}^{2}K_{ij}}\right) \left(\sum_{i=1}^{I}(1-S_{i})\frac{\sum_{k=1}^{K_{i2}}Y_{i2k}(1)}{\sum_{j=1}^{2}K_{ij}}\right) \left(\sum_{i=1}^{I}\frac{K_{i1}}{\sum_{j=1}^{2}K_{ij}}\right)$
        \end{tabular}
    }{ 
        \begin{tabular}{l}
            $\left(\sum_{i=1}^{I}S_{i}\frac{K_{i1}}{\sum_{j=1}^{2}K_{ij}}\right) \left(\sum_{i=1}^{I}(1-S_{i})\frac{K_{i1}}{\sum_{j=1}^{2}K_{ij}}\right) \left(\sum_{i=1}^{I}\frac{K_{i2}}{\sum_{j=1}^{2}K_{ij}}\right)$ \\
            \indent $+ \left(\sum_{i=1}^{I}S_{i}\frac{K_{i2}}{\sum_{j=1}^{2}K_{ij}}\right) \left(\sum_{i=1}^{I}(1-S_{i})\frac{K_{i2}}{\sum_{j=1}^{2}K_{ij}}\right) \left(\sum_{i=1}^{I}\frac{K_{i1}}{\sum_{j=1}^{2}K_{ij}}\right)$
        \end{tabular} 
    }\right)
\]
\[
    \indent -\lim_{I\to\infty} \left(\frac{ 
        \begin{tabular}{l}
            $\left(\sum_{i=1}^{I}S_{i}\frac{K_{i1}}{\sum_{j=1}^{2}K_{ij}}\right) \left(\sum_{i=1}^{I}(1-S_{i})\frac{\sum_{k=1}^{K_{i1}}Y_{i1k}(0)}{\sum_{j=1}^{2}K_{ij}}\right) \left(\sum_{i=1}^{I}\frac{K_{i2}}{\sum_{j=1}^{2}K_{ij}}\right)$ \\
            \indent $+ \left(\sum_{i=1}^{I}S_{i}\frac{\sum_{k=1}^{K_{i2}}Y_{i2k}(0)}{\sum_{j=1}^{2}K_{ij}}\right) \left(\sum_{i=1}^{I}(1-S_{i})\frac{K_{i2}}{\sum_{j=1}^{2}K_{ij}}\right) \left(\sum_{i=1}^{I}\frac{K_{i1}}{\sum_{j=1}^{2}K_{ij}}\right)$ 
        \end{tabular}
    }{ 
        \begin{tabular}{l}
            $\left(\sum_{i=1}^{I}S_{i}\frac{K_{i1}}{\sum_{j=1}^{2}K_{ij}}\right) \left(\sum_{i=1}^{I}(1-S_{i})\frac{K_{i1}}{\sum_{j=1}^{2}K_{ij}}\right) \left(\sum_{i=1}^{I}\frac{K_{i2}}{\sum_{j=1}^{2}K_{ij}}\right)$ \\
            \indent $+ \left(\sum_{i=1}^{I}S_{i}\frac{K_{i2}}{\sum_{j=1}^{2}K_{ij}}\right) \left(\sum_{i=1}^{I}(1-S_{i})\frac{K_{i2}}{\sum_{j=1}^{2}K_{ij}}\right) \left(\sum_{i=1}^{I}\frac{K_{i1}}{\sum_{j=1}^{2}K_{ij}}\right)$ 
        \end{tabular}
    }\right) \,.
\]
and:
\[
    \hat{\delta}_{IEEcw} \xrightarrow{P} \left(
        \frac{
        \begin{tabular}{l}
            $E\left[\frac{\sum_{k=1}^{K_{i1}}Y_{i1k}(1)}{\sum_{j=1}^{2}K_{ij}}|S_{i}=1\right]E\left[\frac{K_{i1}}{\sum_{j=1}^{2}K_{ij}}|S_{i}=0\right]E\left[\frac{K_{i2}}{\sum_{j=1}^{2}K_{ij}}\right]$ \\
            \indent $+ E\left[\frac{K_{i2}}{\sum_{j=1}^{2}K_{ij}}|S_{i}=1\right]E\left[\frac{\sum_{k=1}^{K_{i2}}Y_{i2k}(1)}{\sum_{j=1}^{2}K_{ij}}|S_{i}=0\right]E\left[\frac{K_{i1}}{\sum_{j=1}^{2}K_{ij}}\right]$
        \end{tabular}
        }{
        \begin{tabular}{l}
            $E\left[\frac{K_{i1}}{\sum_{j=1}^{2}K_{ij}}|S_{i}=1\right]E\left[\frac{K_{i1}}{\sum_{j=1}^{2}K_{ij}}|S_{i}=0\right]E\left[\frac{K_{i2}}{\sum_{j=1}^{2}K_{ij}}\right]$ \\
            \indent $+ E\left[\frac{K_{i2}}{\sum_{j=1}^{2}K_{ij}}|S_{i}=1\right]E\left[\frac{K_{i2}}{\sum_{j=1}^{2}K_{ij}}|S_{i}=0\right]E\left[\frac{K_{i1}}{\sum_{j=1}^{2}K_{ij}}\right]$
        \end{tabular}
        }
    \right)
\]
\[
    \indent - \left(
        \frac{
        \begin{tabular}{l}
            $E\left[\frac{K_{i1}}{\sum_{j=1}^{2}K_{ij}}|S_{i}=1\right]E\left[\frac{\sum_{k=1}^{K_{i1}}Y_{i1k}(0)}{\sum_{j=1}^{2}K_{ij}}|S_{i}=0\right]E\left[\frac{K_{i2}}{\sum_{j=1}^{2}K_{ij}}\right]$ \\
            \indent $+ E\left[\frac{\sum_{k=1}^{K_{i2}}Y_{i2k}(0)}{\sum_{j=1}^{2}K_{ij}}|S_{i}=1\right]E\left[\frac{K_{i2}}{\sum_{j=1}^{2}K_{ij}}|S_{i}=0\right]E\left[\frac{K_{i1}}{\sum_{j=1}^{2}K_{ij}}\right]$
        \end{tabular}
        }{
        \begin{tabular}{l}
            $E\left[\frac{K_{i1}}{\sum_{j=1}^{2}K_{ij}}|S_{i}=1\right]E\left[\frac{K_{i1}}{\sum_{j=1}^{2}K_{ij}}|S_{i}=0\right]E\left[\frac{K_{i2}}{\sum_{j=1}^{2}K_{ij}}\right]$ \\
            \indent $+ E\left[\frac{K_{i2}}{\sum_{j=1}^{2}K_{ij}}|S_{i}=1\right]E\left[\frac{K_{i2}}{\sum_{j=1}^{2}K_{ij}}|S_{i}=0\right]E\left[\frac{K_{i1}}{\sum_{j=1}^{2}K_{ij}}\right]$
        \end{tabular}
        }
    \right) \,,
\]

As in Section \ref{sect:appendix_IEE}, with randomization, the sequence variable $S_{i}$ is independent of the potential outcomes and cluster-period sizes, $S_{i} \indep \Omega$, and $\Omega=\{Y_{ijk}(0), Y_{ijk}(1), K_{ij}\}_{i=1, k=1}^{I, K_{ij}}$, such that:
\[
    \hat{\delta}_{IEEcw} \xrightarrow{P} \left(
        \frac{
            E\left[\frac{\sum_{k=1}^{K_{i1}}Y_{i1k}(1)}{\sum_{j=1}^{2}K_{ij}}\right]E\left[\frac{K_{i1}}{\sum_{j=1}^{2}K_{ij}}\right]E\left[\frac{K_{i2}}{\sum_{j=1}^{2}K_{ij}}\right]
            + E\left[\frac{K_{i2}}{\sum_{j=1}^{2}K_{ij}}\right]E\left[\frac{\sum_{k=1}^{K_{i2}}Y_{i2k}(1)}{\sum_{j=1}^{2}K_{ij}}\right]E\left[\frac{K_{i1}}{\sum_{j=1}^{2}K_{ij}}\right]
        }{
            E\left[\frac{K_{i1}}{\sum_{j=1}^{2}K_{ij}}\right]E\left[\frac{K_{i1}}{\sum_{j=1}^{2}K_{ij}}\right]E\left[\frac{K_{i2}}{\sum_{j=1}^{2}K_{ij}}\right]
            + E\left[\frac{K_{i2}}{\sum_{j=1}^{2}K_{ij}}\right]E\left[\frac{K_{i2}}{\sum_{j=1}^{2}K_{ij}}\right]E\left[\frac{K_{i1}}{\sum_{j=1}^{2}K_{ij}}\right]
        }
    \right)
\]
\[
    \indent - \left(
        \frac{
            E\left[\frac{K_{i1}}{\sum_{j=1}^{2}K_{ij}}\right]E\left[\frac{\sum_{k=1}^{K_{i1}}Y_{i1k}(0)}{\sum_{j=1}^{2}K_{ij}}\right]E\left[\frac{K_{i2}}{\sum_{j=1}^{2}K_{ij}}\right]
            + E\left[\frac{\sum_{k=1}^{K_{i2}}Y_{i2k}(0)}{\sum_{j=1}^{2}K_{ij}}\right]E\left[\frac{K_{i2}}{\sum_{j=1}^{2}K_{ij}}\right]E\left[\frac{K_{i1}}{\sum_{j=1}^{2}K_{ij}}\right]
        }{
            E\left[\frac{K_{i1}}{\sum_{j=1}^{2}K_{ij}}\right]E\left[\frac{K_{i1}}{\sum_{j=1}^{2}K_{ij}}\right]E\left[\frac{K_{i2}}{\sum_{j=1}^{2}K_{ij}}\right]
            + E\left[\frac{K_{i2}}{\sum_{j=1}^{2}K_{ij}}\right]E\left[\frac{K_{i2}}{\sum_{j=1}^{2}K_{ij}}\right]E\left[\frac{K_{i1}}{\sum_{j=1}^{2}K_{ij}}\right]
        }
    \right) \,,
\]
\[
= \left(
        \frac{
            E\left[\frac{\sum_{k=1}^{K_{i1}}Y_{i1k}(1)}{\sum_{j=1}^{2}K_{ij}}\right]
            + E\left[\frac{\sum_{k=1}^{K_{i2}}Y_{i2k}(1)}{\sum_{j=1}^{2}K_{ij}}\right]
        }{
            E\left[\frac{K_{i1}}{\sum_{j=1}^{2}K_{ij}}\right]
            + E\left[\frac{K_{i2}}{\sum_{j=1}^{2}K_{ij}}\right]
        }
    \right)
    - \left(
        \frac{
            E\left[\frac{\sum_{k=1}^{K_{i1}}Y_{i1k}(0)}{\sum_{j=1}^{2}K_{ij}}\right]
            + E\left[\frac{\sum_{k=1}^{K_{i2}}Y_{i2k}(0)}{\sum_{j=1}^{2}K_{ij}}\right]
        }{
            E\left[\frac{K_{i1}}{\sum_{j=1}^{2}K_{ij}}\right]
            + E\left[\frac{K_{i2}}{\sum_{j=1}^{2}K_{ij}}\right]
        }
    \right) \,,
\]

Accordingly, we can prove that the IEEcw estimator is consistent for cATE:
\begin{equation}
    \hat{\delta}_{IEEcw} \xrightarrow{P} E\left[ \frac{\sum_{j=1}^{2} \sum_{k=1}^{K_{ij}} \left[Y_{ijk}(1)-Y_{ijk}(0)\right]}{\sum_{j=1}^{2}K_{ij}} \right]  \,.
\end{equation}

\subsection{Derivation of the Independence estimating equation with inverse period size weights (IEEpw) estimator}
\label{sect:appendix_IEEpw}
We can use the result from Section \ref{sect:appendix_IEE} to similarly demonstrate that the independence estimating equation with inverse period size weights (IEEpw) estimator is consistent for the pATE estimand. 
We can show that the IEEpw estimator is:
\[\hat{\delta}_{IEEpw} =\]
\[
    \left(\frac{
        \begin{tabular}{l}
            $\left(\sum_{i=1}^{I}S_{i}\frac{\sum_{k=1}^{K_{i1}}Y_{i1k}(1)}{\sum_{i=1}^{I}K_{i1}}\right) \left(\sum_{i=1}^{I}(1-S_{i})\frac{K_{i1}}{\sum_{i=1}^{I}K_{i1}}\right) \left(\sum_{i=1}^{I}\frac{K_{i2}}{\sum_{i=1}^{I}K_{i2}}\right)$ \\
            \indent $+ \left(\sum_{i=1}^{I}S_{i}\frac{K_{i2}}{\sum_{i=1}^{I}K_{i2}}\right) \left(\sum_{i=1}^{I}(1-S_{i})\frac{\sum_{k=1}^{K_{i2}}Y_{i2k}(1)}{\sum_{i=1}^{I}K_{i2}}\right) \left(\sum_{i=1}^{I}\frac{K_{i1}}{\sum_{i=1}^{I}K_{i1}}\right)$
        \end{tabular}
    }{ 
        \begin{tabular}{l}
            $\left(\sum_{i=1}^{I}S_{i}\frac{K_{i1}}{\sum_{i=1}^{I}K_{i1}}\right) \left(\sum_{i=1}^{I}(1-S_{i})\frac{K_{i1}}{\sum_{i=1}^{I}K_{i1}}\right) \left(\sum_{i=1}^{I}\frac{K_{i2}}{\sum_{i=1}^{I}K_{i2}}\right)$ \\
            \indent $+ \left(\sum_{i=1}^{I}S_{i}\frac{K_{i2}}{\sum_{i=1}^{I}K_{i2}}\right) \left(\sum_{i=1}^{I}(1-S_{i})\frac{K_{i2}}{\sum_{i=1}^{I}K_{i2}}\right) \left(\sum_{i=1}^{I}\frac{K_{i1}}{\sum_{i=1}^{I}K_{i1}}\right)$
        \end{tabular} 
    }\right)
\]
\[
    \indent -\left(\frac{ 
        \begin{tabular}{l}
        $\left(\sum_{i=1}^{I}S_{i}\frac{K_{i1}}{\sum_{i=1}^{I}K_{i1}}\right) \left(\sum_{i=1}^{I}(1-S_{i})\frac{\sum_{k=1}^{K_{i1}}Y_{i1k}(0)}{\sum_{i=1}^{I}K_{i1}}\right) \left(\sum_{i=1}^{I}\frac{K_{i2}}{\sum_{i=1}^{I}K_{i2}}\right)$ \\
        \indent $+ \left(\sum_{i=1}^{I}S_{i}\frac{\sum_{k=1}^{K_{i2}}Y_{i2k}(0)}{\sum_{i=1}^{I}K_{i2}}\right) \left(\sum_{i=1}^{I}(1-S_{i})\frac{K_{i2}}{\sum_{i=1}^{I}K_{i2}}\right) \left(\sum_{i=1}^{I}\frac{K_{i1}}{\sum_{i=1}^{I}K_{i1}}\right)$
        \end{tabular}
    }{ 
    \begin{tabular}{l}
        $\left(\sum_{i=1}^{I}S_{i}\frac{K_{i1}}{\sum_{i=1}^{I}K_{i1}}\right) \left(\sum_{i=1}^{I}(1-S_{i})\frac{K_{i1}}{\sum_{i=1}^{I}K_{i1}}\right) \left(\sum_{i=1}^{I}\frac{K_{i2}}{\sum_{i=1}^{I}K_{i2}}\right)$ \\
        \indent $+ \left(\sum_{i=1}^{I}S_{i}\frac{K_{i2}}{\sum_{i=1}^{I}K_{i2}}\right) \left(\sum_{i=1}^{I}(1-S_{i})\frac{K_{i2}}{\sum_{i=1}^{I}K_{i2}}\right) \left(\sum_{i=1}^{I}\frac{K_{i1}}{\sum_{i=1}^{I}K_{i1}}\right)$ 
        \end{tabular}
    }\right)
\]
\[
    =\left(\frac{              
        \left(\sum_{i=1}^{I}S_{i}\frac{\sum_{k=1}^{K_{i1}}Y_{i1k}(1)}{\sum_{i=1}^{I}K_{i1}}\right) \left(\sum_{i=1}^{I}(1-S_{i})\frac{K_{i1}}{\sum_{i=1}^{I}K_{i1}}\right)
        + \left(\sum_{i=1}^{I}S_{i}\frac{K_{i2}}{\sum_{i=1}^{I}K_{i2}}\right) \left(\sum_{i=1}^{I}(1-S_{i})\frac{\sum_{k=1}^{K_{i2}}Y_{i2k}(1)}{\sum_{i=1}^{I}K_{i2}}\right)
    }{ 
        \left(\sum_{i=1}^{I}S_{i}\frac{K_{i1}}{\sum_{i=1}^{I}K_{i1}}\right) \left(\sum_{i=1}^{I}(1-S_{i})\frac{K_{i1}}{\sum_{i=1}^{I}K_{i1}}\right)
        + \left(\sum_{i=1}^{I}S_{i}\frac{K_{i2}}{\sum_{i=1}^{I}K_{i2}}\right) \left(\sum_{i=1}^{I}(1-S_{i})\frac{K_{i2}}{\sum_{i=1}^{I}K_{i2}}\right)
    }\right)
\]
\[
    \indent -\left(\frac{ 
        \left(\sum_{i=1}^{I}S_{i}\frac{K_{i1}}{\sum_{i=1}^{I}K_{i1}}\right) \left(\sum_{i=1}^{I}(1-S_{i})\frac{\sum_{k=1}^{K_{i1}}Y_{i1k}(0)}{\sum_{i=1}^{I}K_{i1}}\right)
        + \left(\sum_{i=1}^{I}S_{i}\frac{\sum_{k=1}^{K_{i2}}Y_{i2k}(0)}{\sum_{i=1}^{I}K_{i2}}\right) \left(\sum_{i=1}^{I}(1-S_{i})\frac{K_{i2}}{\sum_{i=1}^{I}K_{i2}}\right)
    }{ 
        \left(\sum_{i=1}^{I}S_{i}\frac{K_{i1}}{\sum_{i=1}^{I}K_{i1}}\right) \left(\sum_{i=1}^{I}(1-S_{i})\frac{K_{i1}}{\sum_{i=1}^{I}K_{i1}}\right)
        + \left(\sum_{i=1}^{I}S_{i}\frac{K_{i2}}{\sum_{i=1}^{I}K_{i2}}\right) \left(\sum_{i=1}^{I}(1-S_{i})\frac{K_{i2}}{\sum_{i=1}^{I}K_{i2}}\right)
    }\right) \,.
\]

We can then demonstrate that this weighted estimator is consistent and asymptotically unbiased for the pATE:
\[\lim_{I\to\infty} \hat{\delta}_{IEEpw} =\]
\[
    \lim_{I\to\infty} \left(\frac{              
        \left(\sum_{i=1}^{I}S_{i}\frac{\sum_{k=1}^{K_{i1}}Y_{i1k}(1)}{\sum_{i=1}^{I}K_{i1}}\right) \left(\sum_{i=1}^{I}(1-S_{i})\frac{K_{i1}}{\sum_{i=1}^{I}K_{i1}}\right)
        + \left(\sum_{i=1}^{I}S_{i}\frac{K_{i2}}{\sum_{i=1}^{I}K_{i2}}\right) \left(\sum_{i=1}^{I}(1-S_{i})\frac{\sum_{k=1}^{K_{i2}}Y_{i2k}(1)}{\sum_{i=1}^{I}K_{i2}}\right)
    }{ 
        \left(\sum_{i=1}^{I}S_{i}\frac{K_{i1}}{\sum_{i=1}^{I}K_{i1}}\right) \left(\sum_{i=1}^{I}(1-S_{i})\frac{K_{i1}}{\sum_{i=1}^{I}K_{i1}}\right)
        + \left(\sum_{i=1}^{I}S_{i}\frac{K_{i2}}{\sum_{i=1}^{I}K_{i2}}\right) \left(\sum_{i=1}^{I}(1-S_{i})\frac{K_{i2}}{\sum_{i=1}^{I}K_{i2}}\right)
    }\right)
\]
\[
    \indent -\lim_{I\to\infty} \left(\frac{ 
        \left(\sum_{i=1}^{I}S_{i}\frac{K_{i1}}{\sum_{i=1}^{I}K_{i1}}\right) \left(\sum_{i=1}^{I}(1-S_{i})\frac{\sum_{k=1}^{K_{i1}}Y_{i1k}(0)}{\sum_{i=1}^{I}K_{i1}}\right)
        + \left(\sum_{i=1}^{I}S_{i}\frac{\sum_{k=1}^{K_{i2}}Y_{i2k}(0)}{\sum_{i=1}^{I}K_{i2}}\right) \left(\sum_{i=1}^{I}(1-S_{i})\frac{K_{i2}}{\sum_{i=1}^{I}K_{i2}}\right)
    }{ 
        \left(\sum_{i=1}^{I}S_{i}\frac{K_{i1}}{\sum_{i=1}^{I}K_{i1}}\right) \left(\sum_{i=1}^{I}(1-S_{i})\frac{K_{i1}}{\sum_{i=1}^{I}K_{i1}}\right)
        + \left(\sum_{i=1}^{I}S_{i}\frac{K_{i2}}{\sum_{i=1}^{I}K_{i2}}\right) \left(\sum_{i=1}^{I}(1-S_{i})\frac{K_{i2}}{\sum_{i=1}^{I}K_{i2}}\right)
    }\right) \,.
\]
and:
\[
    \hat{\delta}_{IEEpw} \xrightarrow{P} \left(\frac{              
        \frac{E\left[\sum_{k=1}^{K_{i1}}Y_{i1k}(1) | S_{i}=1\right]}{E[K_{i1}]} \frac{E\left[K_{i1} | S_i=0\right]}{E[K_{i1}]}
        + \frac{E\left[K_{i2} | S_i=1\right]}{E[K_{i2}]} \frac{E\left[\sum_{k=1}^{K_{i2}}Y_{i2k}(1) | S_i=0\right]}{E[K_{i2}]}
    }{ 
        \frac{E\left[K_{i1} | S_i=1\right]}{E[K_{i1}]} \frac{E\left[K_{i1} | S_i=0\right]}{E[K_{i1}]}
        + \frac{E\left[K_{i2} | S_i=1\right]}{E[K_{i2}]} \frac{E\left[K_{i2} | S_i=0\right]}{E[K_{i2}]}
    }\right)
\]
\[
    \indent - \left(\frac{ 
        \frac{E\left[K_{i1} | S_i=1\right]}{E[K_{i1}]} \frac{E\left[\sum_{k=1}^{K_{i1}}Y_{i1k}(0) | S_i=0\right]}{E[K_{i1}]}
        + \frac{E\left[\sum_{k=1}^{K_{i2}}Y_{i2k}(0) | S_i=1\right]}{E[K_{i2}]} \frac{E\left[K_{i2} | S_i=0\right]}{E[K_{i2}]}
    }{ 
        \frac{E\left[K_{i1} | S_i=1\right]}{E[K_{i1}]} \frac{E\left[K_{i1} | S_i=0\right]}{E[K_{i1}]}
        + \frac{E\left[K_{i2} | S_i=1\right]}{E[K_{i2}]} \frac{E\left[K_{i2} | S_i=0\right]}{E[K_{i2}]}
    }\right) \,,
\]

As in Section \ref{sect:appendix_IEE}, with randomization, the sequence variable $S_{i}$ is independent of the potential outcomes and cluster-period sizes, $S_{i} \indep \Omega$, and $\Omega=\{Y_{ijk}(0), Y_{ijk}(1), K_{ij}\}_{i=1, k=1}^{I, K_{ij}}$, such that:
\[
    \hat{\delta}_{IEEpw} \xrightarrow{P} \left(\frac{              
        \frac{E\left[\sum_{k=1}^{K_{i1}}Y_{i1k}(1)\right]}{E[K_{i1}]} \frac{E\left[K_{i1}\right]}{E[K_{i1}]}
        + \frac{E\left[K_{i2}\right]}{E[K_{i2}]} \frac{E\left[\sum_{k=1}^{K_{i2}}Y_{i2k}(1)\right]}{E[K_{i2}]}
    }{ 
        \frac{E\left[K_{i1}\right]}{E[K_{i1}]} \frac{E\left[K_{i1}\right]}{E[K_{i1}]}
        + \frac{E\left[K_{i2}\right]}{E[K_{i2}]} \frac{E\left[K_{i2}\right]}{E[K_{i2}]}
    }\right)
    - \left(\frac{ 
        \frac{E\left[K_{i1}\right]}{E[K_{i1}]} \frac{E\left[\sum_{k=1}^{K_{i1}}Y_{i1k}(0)\right]}{E[K_{i1}]}
        + \frac{E\left[\sum_{k=1}^{K_{i2}}Y_{i2k}(0)\right]}{E[K_{i2}]} \frac{E\left[K_{i2}\right]}{E[K_{i2}]}
    }{ 
        \frac{E\left[K_{i1}\right]}{E[K_{i1}]} \frac{E\left[K_{i1}\right]}{E[K_{i1}]}
        + \frac{E\left[K_{i2}\right]}{E[K_{i2}]} \frac{E\left[K_{i2}\right]}{E[K_{i2}]}
    }\right)
\]
\[
    = \frac{1}{2}\left[\left(
        \frac{E\left[\sum_{k=1}^{K_{i1}}Y_{i1k}(1)\right]}{E[K_{i1}]} + \frac{E\left[\sum_{k=1}^{K_{i2}}Y_{i2k}(1)\right]}{E[K_{i2}]}
    \right)
    - \left(
        \frac{E\left[\sum_{k=1}^{K_{i1}}Y_{i1k}(0)\right]}{E[K_{i1}]} + \frac{E\left[\sum_{k=1}^{K_{i2}}Y_{i2k}(0)\right]}{E[K_{i2}]}
    \right)
    \right]
\]

Accordingly, we can prove that the IEEpw estimator is consistent for pATE:
\begin{equation}
    \hat{\delta}_{IEEpw} \xrightarrow{P} \frac{1}{2} \sum_{j=1}^{2} \left[ \frac{E\left[ \sum_{k=1}^{K_{ij}} \left[Y_{ijk}(1)-Y_{ijk}(0)\right] \right]}{E\left[K_{ij}\right]} \right]  \,.
\end{equation}

\section{Derivation of the unweighted and weighted Exchangeable mixed effects model (EME, EMEcpw, EMEcw) estimator}
\label{sect:appendix_EME_EMEcw_EMEcw}

\subsection{Derivation of the Exchangeable mixed effects model (EME) estimator}
\label{sect:appendix_EME}

In a standard 2-period, 2-sequence CRXO trial design (Figure \ref{fig:CRXO_design}), an exchangeable mixed effects (EME) estimating equation estimator yields the following treatment effect estimator:
\[\hat{\delta}_{EME}=(Z'\ddot{V}^{-1}Z)^{-1}Z'\ddot{V}^{-1}Y|_{\delta}\]
where $Z$ is the design matrix, $\ddot{V}$ is the variance-covariance matrix of the individual individual outcomes, and $Y$ is the vector of outcomes.

With an exchangeable correlation structure, we can write:
\[Z'\ddot{V}^{-1}Z =
    \begin{bmatrix}
        \sum_{i=1}^{I} (S_{i}A_{i} + (1-S_{i})B_{i}) & \sum_{i=1}^{I} (S_{i}A_{i} + (1-S_{i})C_{i}) & \sum_{i=1}^{I} (S_{i}C_{i} + (1-S_{i})B_{i})\\
        \sum_{i=1}^{I} (S_{i}A_{i} + (1-S_{i})C_{i}) & \sum_{i=1}^{I} A_{i} & \sum_{i=1}^{I} C_{i}\\
        \sum_{i=1}^{I} (S_{i}C_{i} + (1-S_{i})B_{i}) & \sum_{i=1}^{I} C_{i} & \sum_{i=1}^{I} B_{i} 
    \end{bmatrix} \,,
\]
with the top-left entry in $Z'\ddot{V}^{-1}Z$ corresponding with the treatment. We define:
\begin{align*}
    \begin{split}
        A_{i} = K_{i1}(D_{i}+(K_{i1}-1)F_{i})
    \end{split} \\
    \begin{split}
        B_{i} =  K_{i2}(D_{i}+(K_{i2}-1)F_{i})
    \end{split} \\
    \begin{split}
        C_{i} = K_{11}K_{i2}F_{i}
    \end{split}
\end{align*}
where 
$D_{i} = \frac{1}{\sigma^{2}_{w}} \left( \frac{\sigma^{2}_{w} + ((\sum_{j=1}^{2}K_{ij})-1)\tau^{2}_{\alpha})}{\sigma^{2}_{w} + (\sum_{j=1}^{2}K_{ij})\tau^{2}_{\alpha})} \right)$
and
$F_{i} = -\frac{1}{\sigma^{2}_{w}} \left( \frac{\tau^{2}_{\alpha}}{\sigma^{2}_{w} + (\sum_{j=1}^{2}K_{ij})\tau^{2}_{\alpha})} \right)$ are the diagonal and off-diagonal terms, respectively, of the inverse variance-covariance matrix of the individual-level outcomes $\ddot{V}^{-1}$. 
We define $\sigma^{2}_{w}$ and $\tau^{2}_{\alpha}$ as the variances of the residual errors and cluster random intercepts, respectively.

Accordingly, the first row of $(Z'\ddot{V}^{-1}Z)^{-1}$ corresponding with the treatment effect $(\delta)$ is:
\[(Z'\ddot{V}^{-1}Z)^{-1}|_{\delta} =\]
\[
    \left(\frac{1}{|Z'\ddot{V}^{-1}Z|}\right)
    \begin{bmatrix}
        (\sum_{i=1}^{I} A_{i})(\sum_{i=1}^{I} B_{i}) - (\sum_{i=1}^{I} C_{i})^2 \\
         -\left((\sum_{i=1}^{I} [S_{i}A_{i} + (1-S_{i})C_{i}])(\sum_{i=1}^{I} B_{i}) - (\sum_{i=1}^{I} [(1-S_{i})B_{i} + S_{i}C_{i}])(\sum_{i=1}^{I} C_{i})\right)\\
         (\sum_{i=1}^{I} [S_{i}A_{i} + (1-S_{i})C_{i}])(\sum_{i=1}^{I} C_{i}) - (\sum_{i=1}^{I} [(1-S_{i})B_{i} + S_{i}C_{i}])(\sum_{i=1}^{I} A_{i})
    \end{bmatrix}'
\]
with the determinant:
\[
    |Z'\ddot{V}^{-1}Z| = 
\]
\[
    \begin{tabular}{l}
        $\left( \sum_{i=1}^{I} (S_{i}A_{i} + (1-S_{i})B_{i}) \right)
        \left( (\sum_{i=1}^{I} A_{i})(\sum_{i=1}^{I} B_{i}) - (\sum_{i=1}^{I} C_{i})^2 \right)$ \\
        $- \left( (\sum_{i=1}^{I} [S_{i}A_{i} + (1-S_{i})C_{i}]) \right)
        \left((\sum_{i=1}^{I} [S_{i}A_{i} + (1-S_{i})C_{i}])(\sum_{i=1}^{I} B_{i}) - (\sum_{i=1}^{I} [(1-S_{i})B_{i} + S_{i}C_{i}])(\sum_{i=1}^{I} C_{i})\right)$ \\
        $+ \left( (\sum_{i=1}^{I} [(1-S_{i})B_{i} + S_{i}C_{i}]) \right)
        \left((\sum_{i=1}^{I} [S_{i}A_{i} + (1-S_{i})C_{i}])(\sum_{i=1}^{I} C_{i}) - (\sum_{i=1}^{I} [(1-S_{i})B_{i} + S_{i}C_{i}])(\sum_{i=1}^{I} A_{i})\right) \,.$
    \end{tabular}
\]
Alongside:
\[Z'\ddot{V}^{-1}Y =
    \begin{bmatrix}
        Z'\ddot{V}^{-1}Y|_{\delta}\\
        Z'\ddot{V}^{-1}Y|_{\Phi_{1}}\\
        Z'\ddot{V}^{-1}Y|_{\Phi_{2}}
    \end{bmatrix}
\]
where:
\[
\begin{tabular}{l} 
        $Z'\ddot{V}^{-1}Y|_{\delta} = \sum_{i=1}^{I}S_{i}\left(\sum_{k=1}^{K_{i1}}Y_{i1k}(1)\right)\left(\frac{A_{i}}{K_{i1}}\right) + \sum_{i=1}^{I}S_{i}\left(\sum_{k=1}^{K_{i2}}Y_{i2k}(0)\right)\left(\frac{C_{i}}{K_{i2}}\right) +$ \\ 
        \indent\indent $\sum_{i=1}^{I}(1-S_{i})\left(\sum_{k=1}^{K_{i1}}Y_{i1k}(0)\right)\left(\frac{C_{i}}{K_{i1}}\right) + \sum_{i=1}^{I}(1-S_{i})\left(\sum_{k=1}^{K_{i2}}Y_{i2k}(1)\right)\left(\frac{B_{i}}{K_{i2}}\right)$ \,, \\
        $Z'\ddot{V}^{-1}Y|_{\Phi_{1}} = \sum_{i=1}^{I}S_{i}\left(\sum_{k=1}^{K_{i1}}Y_{i1k}(1)\right)\left(\frac{A_{i}}{K_{i1}}\right) + \sum_{i=1}^{I}S_{i}\left(\sum_{k=1}^{K_{i2}}Y_{i2k}(0)\right)\left(\frac{C_{i}}{K_{i2}}\right) +$ \\ 
        \indent\indent $\sum_{i=1}^{I}(1-S_{i})\left(\sum_{k=1}^{K_{i1}}Y_{i1k}(0)\right)\left(\frac{A_{i}}{K_{i1}}\right) + \sum_{i=1}^{I}(1-S_{i})\left(\sum_{k=1}^{K_{i2}}Y_{i2k}(1)\right)\left(\frac{C_{i}}{K_{i2}}\right)$ \,, \\
        $Z'\ddot{V}^{-1}Y|_{\Phi_{2}} = \sum_{i=1}^{I}S_{i}\left(\sum_{k=1}^{K_{i1}}Y_{i1k}(1)\right)\left(\frac{C_{i}}{K_{i1}}\right) + \sum_{i=1}^{I}S_{i}\left(\sum_{k=1}^{K_{i2}}Y_{i2k}(0)\right)\left(\frac{B_{i}}{K_{i2}}\right) +$ \\ 
        \indent\indent $\sum_{i=1}^{I}(1-S_{i})\left(\sum_{k=1}^{K_{i1}}Y_{i1k}(0)]\right)\left(\frac{C_{i}}{K_{i1}}\right) + \sum_{i=1}^{I}(1-S_{i})\left(\sum_{k=1}^{K_{i2}}Y_{i2k}(1)\right)\left(\frac{B_{i}}{K_{i2}}\right)$ \,.
\end{tabular}
\]
Overall, yields the following treatment effect point estimator:
\begin{align*}
    \begin{split}
        \hat{\delta}_{EME} = (Z'\ddot{V}^{-1}Z)^{-1}Z'\ddot{V}^{-1}Y|_{\delta} =
    \end{split}\\
    \begin{split}
        \indent \left(\frac{1}{|Z'\ddot{V}^{-1}Z|}\right) \times
    \end{split}\\
    \begin{split}
        \indent\indent \left(
        \begin{tabular}{l}
            $\left((\sum_{i=1}^{I} A_{i})(\sum_{i=1}^{I} B_{i}) - (\sum_{i=1}^{I} C_{i})^2\right)\left(Z'\ddot{V}^{-1}Y|_{\delta}\right)$ \\
            \indent $-\left((\sum_{i=1}^{I} [S_{i}A_{i} + (1-S_{i})C_{i}])(\sum_{i=1}^{I} B_{i}) - (\sum_{i=1}^{I} [(1-S_{i})B_{i} + S_{i}C_{i}])(\sum_{i=1}^{I} C_{i})\right)\left(Z'\ddot{V}^{-1}Y_{\Phi_{1}}\right)$\\
            \indent $+\left((\sum_{i=1}^{I} [S_{i}A_{i} + (1-S_{i})C_{i}])(\sum_{i=1}^{I} C_{i}) - (\sum_{i=1}^{I} [(1-S_{i})B_{i} + S_{i}C_{i}])(\sum_{i=1}^{I} A_{i})\right)\left(Z'\ddot{V}^{-1}Y_{\Phi_{2}}\right)$
        \end{tabular}
        \right) \,.
    \end{split}
\end{align*}

\begin{align*}
    \begin{split}
        = \left(\frac{1}{ \left(\frac{1}{I}\right)\left(\frac{1}{I}\right)\left(\frac{1}{I/2}\right) |Z'\ddot{V}^{-1}Z|}\right) \times
    \end{split}\\
    \begin{split}
        \left(
        \begin{tabular}{l}
            $\left(
            \left(\frac{1}{I^2}\right)(\sum_{i=1}^{I} A_{i})(\sum_{i=1}^{I} B_{i}) -\left(\frac{1}{I^2}\right) (\sum_{i=1}^{I} C_{i})^2\right)
            \left( 
                \begin{tabular}{l}       $\left(\frac{1}{I/2}\right) \sum_{i=1}^{I}S_{i}\left(\sum_{k=1}^{K_{i1}}Y_{i1k}(1)\right)\left(\frac{A_{i}}{K_{i1}}\right) $\\
                $+\left(\frac{1}{I/2}\right) \sum_{i=1}^{I}S_{i}\left(\sum_{k=1}^{K_{i2}}Y_{i2k}(0)\right)\left(\frac{C_{i}}{K_{i2}}\right)$\\ 
                $+\left(\frac{1}{I/2}\right) \sum_{i=1}^{I}(1-S_{i})\left(\sum_{k=1}^{K_{i1}}Y_{i1k}(0)\right)\left(\frac{C_{i}}{K_{i1}}\right)$\\ 
                $+\left(\frac{1}{I/2}\right) \sum_{i=1}^{I}(1-S_{i})\left(\sum_{k=1}^{K_{i2}}Y_{i2k}(1)\right)\left(\frac{B_{i}}{K_{i2}}\right)$
                \end{tabular}
            \right)$ \\
            $-\left(
                \begin{tabular}{l}
                $\left(\frac{1}{2(I/2)I}\right) (\sum_{i=1}^{I} [S_{i}A_{i} + (1-S_{i})C_{i}])(\sum_{i=1}^{I} B_{i})$\\
                $- \left(\frac{1}{2(I/2)I}\right) (\sum_{i=1}^{I} [(1-S_{i})B_{i} + S_{i}C_{i}])(\sum_{i=1}^{I} C_{i})$
                \end{tabular}
            \right)$
            $\left(
                \begin{tabular}{l}
                $\left(\frac{1}{I/2}\right)\sum_{i=1}^{I}S_{i}\left(\sum_{k=1}^{K_{i1}}Y_{i1k}(1)\right)\left(\frac{A_{i}}{K_{i1}}\right)$\\
                $+\left(\frac{1}{I/2}\right) \sum_{i=1}^{I}S_{i}\left(\sum_{k=1}^{K_{i2}}Y_{i2k}(0)\right)\left(\frac{C_{i}}{K_{i2}}\right)$\\ 
                $+\left(\frac{1}{I/2}\right) \sum_{i=1}^{I}(1-S_{i})\left(\sum_{k=1}^{K_{i1}}Y_{i1k}(0)\right)\left(\frac{A_{i}}{K_{i1}}\right)$\\
                $+\left(\frac{1}{I/2}\right) \sum_{i=1}^{I}(1-S_{i})\left(\sum_{k=1}^{K_{i2}}Y_{i2k}(1)\right)\left(\frac{C_{i}}{K_{i2}}\right)$
                \end{tabular}
            \right)$\\
            $+\left(
                \begin{tabular}{l}
                $\left(\frac{1}{2(I/2)I}\right) (\sum_{i=1}^{I} [S_{i}A_{i} + (1-S_{i})C_{i}])(\sum_{i=1}^{I} C_{i})$\\ 
                $- \left(\frac{1}{2(I/2)I}\right)(\sum_{i=1}^{I} [(1-S_{i})B_{i} + S_{i}C_{i}])(\sum_{i=1}^{I} A_{i})$
                \end{tabular}
            \right)
            \left(
                \begin{tabular}{l}
                $\left(\frac{1}{I/2}\right) \sum_{i=1}^{I}S_{i}\left(\sum_{k=1}^{K_{i1}}Y_{i1k}(1)\right)\left(\frac{C_{i}}{K_{i1}}\right)$\\
                $+\left(\frac{1}{I/2}\right) \sum_{i=1}^{I}S_{i}\left(\sum_{k=1}^{K_{i2}}Y_{i2k}(0)\right)\left(\frac{B_{i}}{K_{i2}}\right)$ \\ 
                $+\left(\frac{1}{I/2}\right) \sum_{i=1}^{I}(1-S_{i})\left(\sum_{k=1}^{K_{i1}}Y_{i1k}(0)]\right)\left(\frac{C_{i}}{K_{i1}}\right)$\\
                $+\left(\frac{1}{I/2}\right) \sum_{i=1}^{I}(1-S_{i})\left(\sum_{k=1}^{K_{i2}}Y_{i2k}(1)\right)\left(\frac{B_{i}}{K_{i2}}\right)$
                \end{tabular}
            \right)$
        \end{tabular}
        \right) \,.
    \end{split}
\end{align*}
This converges in probability to:
\[\lim_{I\to\infty} \hat{\delta}_{EME} =\]
\begin{align*}
    \begin{split}
        \left(
        \begin{tabular}{l}
            $\left( E[A_{i}|S_{i}] + E[B_{i}|1-S_{i}]\right)
            \left( E[A_{i}]E[B_{i}] - E[C_{i}]^2 \right)$ \\
            $- (\frac{1}{2})\left( E[A_{i}|S_{i}] + E[C_{i}|1-S_{i}] \right)
            \left((E[A_{i}|S_{i}] + E[C_{i}|1-S_{i}])E[B_{i}] - (E[B_{i}|1-S_{i}] + E[C_{i}|S_{i}])E[C_{i}]\right)$ \\
            $+ (\frac{1}{2})\left(E[B_{i}|1-S_{i}] + E[C_{i}|S_{i}] \right)
            \left((E[A_{i}|S_{i}] + E[C_{i}|1-S_{i}])E[C_{i}] - (E[B_{i}|1-S_{i}] + E[C_{i}|S_{i}])E[A_{i}]\right) \,.$
        \end{tabular}
        \right)^{-1} \times
    \end{split}\\
    \begin{split}
        \left(
        \begin{tabular}{l}
            $\left(
            E[A_{i}]E[B_{i}] - E[C_{i}]^2\right)
            \left( 
                \begin{tabular}{l}       $E\left[\left(\sum_{k=1}^{K_{i1}}Y_{i1k}(1)\right)\left(\frac{A_{i}}{K_{i1}}\right)|S_{i}\right] $\\
                $+E\left[\left(\sum_{k=1}^{K_{i2}}Y_{i2k}(0)\right)\left(\frac{C_{i}}{K_{i2}}\right)|S_{i}\right]$\\ 
                $+E\left[\left(\sum_{k=1}^{K_{i1}}Y_{i1k}(0)\right)\left(\frac{C_{i}}{K_{i1}}\right)|1-S_{i}\right]$\\ 
                $+E\left[\left(\sum_{k=1}^{K_{i2}}Y_{i2k}(1)\right)\left(\frac{B_{i}}{K_{i2}}\right)|1-S_{i}\right]$
                \end{tabular}
            \right)$ \\
            $-(\frac{1}{2})\left(
                \begin{tabular}{l}
                $(E[A_{i}|S_{i}] + E[C_{i}|1-S_{i}])E[B_{i}]$\\
                $- (E[B_{i}|1-S_{i}] + E[C_{i}|S_{i}]) E[C_{i}]$
                \end{tabular}
            \right)$
            $\left(
                \begin{tabular}{l}
                $E\left[\left(\sum_{k=1}^{K_{i1}}Y_{i1k}(1)\right)\left(\frac{A_{i}}{K_{i1}}\right)|S_{i}\right]$\\
                $+E\left[\left(\sum_{k=1}^{K_{i2}}Y_{i2k}(0)\right)\left(\frac{C_{i}}{K_{i2}}\right)|S_{i}\right]$\\ 
                $+E\left[\left(\sum_{k=1}^{K_{i1}}Y_{i1k}(0)\right)\left(\frac{A_{i}}{K_{i1}}\right)|1-S_{i}\right]$\\
                $+E\left[\left(\sum_{k=1}^{K_{i2}}Y_{i2k}(1)\right)\left(\frac{C_{i}}{K_{i2}}\right)|1-S_{i}\right]$
                \end{tabular}
            \right)$\\
            $+(\frac{1}{2})\left(
                \begin{tabular}{l}
                $(E[A_{i}|S_{i}] + E[C_{i}|1-S_{i}])E[C_{i}]$\\ 
                $- (E[B_{i}|1-S_{i}] + E[C_{i}|S_{i}])E[A_{i}]$
                \end{tabular}
            \right)
            \left(
                \begin{tabular}{l}
                $E\left[\left(\sum_{k=1}^{K_{i1}}Y_{i1k}(1)\right)\left(\frac{C_{i}}{K_{i1}}\right)|S_{i}\right]$\\
                $+ E\left[\left(\sum_{k=1}^{K_{i2}}Y_{i2k}(0)\right)\left(\frac{B_{i}}{K_{i2}}\right)|S_{i}\right]$ \\ 
                $+E\left[\left(\sum_{k=1}^{K_{i1}}Y_{i1k}(0)]\right)\left(\frac{C_{i}}{K_{i1}}\right)|1-S_{i}\right]$\\
                $+E\left[\left(\sum_{k=1}^{K_{i2}}Y_{i2k}(1)\right)\left(\frac{B_{i}}{K_{i2}}\right)|1-S_{i}\right]$
                \end{tabular}
            \right)$
        \end{tabular}
        \right) \,.
    \end{split}
\end{align*}

Overall, with randomization, we can demonstrate that the EME estimator then converges in probability to:
\begin{equation}
    \hat{\delta}_{EME} \xrightarrow{P} 
    \frac{ E\left[ \frac{1}{K_{i1}}(A_{i}-C_{i})\left(\sum_{k=1}^{K_{i1}}Y_{i1k}(1) - \sum_{k=1}^{K_{i1}}Y_{i1k}(0)\right) + \frac{1}{K_{i2}}(B_{i}-C_{i})\left(\sum_{k=1}^{K_{i2}}Y_{i2k}(1) - \sum_{k=1}^{K_{i2}}Y_{i2k}(0)\right) \right] }
    { E\left[ (A_{i}-C_{i})+(B_{i}-C_{i}) \right] }  \,.
\end{equation}
Generally, this estimator does not coincide with any of the iATE, cpATE, cATE, and pATE estimands.

Notably, when cluster-period sizes are equal between-periods, within-clusters, $K_{i1}=K_{i2}=K_{i-} \, \forall \, i$. Furthermore, $A_{i}=B_{i}=K_{i-}(D_{i}+(K_{i-}-1)F_{i})$ and $C_{i}=K_{i-}^{2}F_{i}$, where recall 
$D_{i} = \frac{1}{\sigma^{2}_{w}} \left( \frac{\sigma^{2}_{w} + ((\sum_{j=1}^{2}K_{ij})-1)\tau^{2}_{\alpha})}{\sigma^{2}_{w} + (\sum_{j=1}^{2}K_{ij})\tau^{2}_{\alpha})} \right)$
and
$F_{i} = -\frac{1}{\sigma^{2}_{w}} \left( \frac{\tau^{2}_{\alpha}}{\sigma^{2}_{w} + (\sum_{j=1}^{2}K_{ij})\tau^{2}_{\alpha})} \right)$ are the diagonal and off-diagonal terms, respectively, of the inverse variance-covariance matrix of the individual-level outcomes. 
Accordingly, the above equation simplifies to:
\[
    \hat{\delta}_{EME} \xrightarrow{P} 
    \frac{ E\left[ \frac{1}{K_{i-}}(A_{i}-C_{i})\left(\sum_{j=1}^{2}\sum_{k=1}^{K_{i-}}\left[Y_{ijk}(1)-Y_{ijk}(0)\right]\right)\right] }
    { E\left[ 2(A_{i}-C_{i}) \right] }
\]
\begin{equation}
    =  \frac{ E\left[ (D_{i}-F_{i})\left(\sum_{j=1}^{2}\sum_{k=1}^{K_{i-}}\left[Y_{ijk}(1)-Y_{ijk}(0)\right]\right)\right] }
    { E\left[ 2(D_{i}-F_{i})K_{i-} \right] }  \,.
\end{equation}
Additionally, $D_{i}-F_{i}=\frac{1}{\sigma^{2}_{w}}$, the inverse residual error variance, when cluster-period sizes are equal between-periods and within-clusters. As a result, we can demonstrate that under such conditions the EME estimator is consistent for the iATE and pATE estimands:
\begin{equation}
    \hat{\delta}_{EME} \xrightarrow{P} \frac{ E\left[ \sum_{j=1}^{2}\sum_{k=1}^{K_{i-}}\left[Y_{ijk}(1)-Y_{ijk}(0)\right]\right] }
    { E\left[2K_{i-}\right] } \,.
\end{equation}

\subsection{Derivation of the weighted Exchangeable mixed effects model (EMEcpw, EMEcw) estimators}
\label{sect:appendix_EMEcpw_EMEcw}
In general, we can demonstrate that the unweighted and weighted IEE, EME, and NEME estimators all asymptotically converge to the same general form of:
\[
    \hat{\delta} \xrightarrow{P} \frac{ E\left[ \frac{1}{K_{i-}}(A_{i}-C_{i})\left(\sum_{j=1}^{2}\sum_{k=1}^{K_{i-}}\left[Y_{ijk}(1)-Y_{ijk}(0)\right]\right)\right] }
    { E\left[ 2(A_{i}-C_{i}) \right] }
\]
with model-specific and cluster-specific values of $A_{i}$ and $C_{i}$, when cluster-period sizes are equal between-periods and within-clusters.

Notably, cpATE and cATE are equivalent when cluster period sizes are equal between-periods, within-clusters. It is then straight-forward to demonstrate that the EMEcpw and EMEcw estimators are consistent for the cpATE and cATE:
\begin{equation}
    \hat{\delta}_{EMEcpw} \xrightarrow{P}  E\left[ \frac{1}{2} \sum_{j=1}^{2} \frac{ \sum_{k=1}^{K_{i-}} \left[Y_{ijk}(1)-Y_{ijk}(0)\right]}{K_{i-}} \right]
\end{equation}
and:
\begin{equation}
    \hat{\delta}_{EMEcw} \xrightarrow{P} E\left[ \frac{\sum_{j=1}^{2} \sum_{k=1}^{K_{i-}} \left[Y_{ijk}(1)-Y_{ijk}(0)\right]}{2K_{i-}} \right] \,.
\end{equation}
given $K_{i1}=K_{i2}=K_{i-} \, \forall \, i$.

\section{Derivation of the unweighted and weighted nested exchangeable mixed effects model (NEME, NEMEcpw, NEMEcw) estimator}
\label{sect:appendix_NEME}
As pointed out in Section \ref{sect:appendix_EME_EMEcw_EMEcw}, the unweighted and weighted NEME estimators all asymptotically converge to the same general form of:
\[
    \hat{\delta} \xrightarrow{P} \frac{ E\left[ \frac{1}{K_{i-}}(A_{i}-C_{i})\left(\sum_{j=1}^{2}\sum_{k=1}^{K_{i-}}\left[Y_{ijk}(1)-Y_{ijk}(0)\right]\right)\right] }
    { E\left[ 2(A_{i}-C_{i}) \right] }
\]
with model-specific and cluster-specific values of $A_{i}$ and $C_{i}$, given $K_{i1}=K_{i2}=K_{i-} \, \forall \, i$.

Accordingly, with a nested exchangeable mixed effects (NEME) model treatment effect point estimator $\hat{\delta}_{NEME}$, the model-specific values of $A_{i}$ and $C_{i}$ are:
\[A_{i}=(K_{i-})\left(\frac{\sigma^{2}_{w}+(K_{i-})(\tau^{2}_{\alpha}+\tau^{2}_{\gamma})}{(\sigma^{2}_{w}+(K_{i-})(\tau^{2}_{\alpha}+\tau^{2}_{\gamma}))^{2}-((K_{i-})(\tau^{2}_{\alpha}))^{2}}\right)\]
and:
\[C_{i}=-(K_{i-})\left(\frac{(K_{i-})(\tau^{2}_{\alpha})}{(\sigma^{2}_{w}+(K_{i-})(\tau^{2}_{\alpha}+\tau^{2}_{\gamma}))^{2}-((K_{i-})(\tau^{2}_{\alpha}))^{2}}\right) \,.\]
We define $\sigma^{2}_{w}$,  $\tau^{2}_{\alpha}$, and $\tau^{2}_{\gamma}$ as the variances of the residual errors, cluster random intercepts, and cluster-period random interaction terms, respectively.
In the NEME estimator:
\[A_{i}-C_{i}=(K_{i-})(\sigma^{2}_{w}+\tau^{2}_{\alpha}+\tau^{2}_{\gamma})\left(\frac{1}{1+(K_{i-}-1)\rho_{wp}-(K_{i-})\rho_{bp}}\right)\]
where $\rho_{wp}=\frac{\tau^{2}_{\alpha}+\tau^{2}_{\gamma}}{\sigma^{2}_{w}+\tau^{2}_{\alpha}+\tau^{2}_{\gamma}}$ and $\rho_{bp}=\frac{\tau^{2}_{\alpha}}{\sigma^{2}_{w}+\tau^{2}_{\alpha}+\tau^{2}_{\gamma}}$ are the within-period and between-period intracluster correlation (ICC) respectively.

Altogether,the NEME estimator converges in probability to:
\begin{equation}
    \hat{\delta}_{NEME} \xrightarrow{P} \frac{ E\left[ \left(\frac{1}{1+(K_{i-}-1)\rho_{wp}-(K_{i-})\rho_{bp}}\right)\left(\sum_{j=1}^{2}\sum_{k=1}^{K_{i-}}\left[Y_{ijk}(1)-Y_{ijk}(0)\right]\right)\right] }
    { E\left[ 2\left(\frac{1}{1+(K_{i-}-1)\rho_{wp}-(K_{i-})\rho_{bp}}\right)K_{i-} \right] } \,.
\end{equation}
given $K_{i1}=K_{i2}=K_{i-} \, \forall \, i$. 

Subsequently, the NEMEcpw and NEMEcw estimators converge to:
\begin{equation}
     \hat{\delta}_{NEMEcpw} \xrightarrow{P}  E\left[ \left(\frac{\left(\frac{1}{1+(K_{i-}-1)\rho_{wp}-(K_{i-})\rho_{bp}}\right)}{E\left[\frac{1}{1+(K_{i-}-1)\rho_{wp}-(K_{i-})\rho_{bp}}\right]}\right) \frac{ \sum_{j=1}^{2} \sum_{k=1}^{K_{i-}} \left[Y_{ijk}(1)-Y_{ijk}(0)\right]}{2K_{i-}} \right]
\end{equation}
and:
\begin{equation}
    \hat{\delta}_{NEMEcw} \xrightarrow{P} E\left[ \left(\frac{\left(\frac{1}{1+(K_{i-}-1)\rho_{wp}-(K_{i-})\rho_{bp}}\right)}{E\left[\frac{1}{1+(K_{i-}-1)\rho_{wp}-(K_{i-})\rho_{bp}}\right]}\right) \frac{\sum_{j=1}^{2} \sum_{k=1}^{K_{i-}} \left[Y_{ijk}(1)-Y_{ijk}(0)\right]}{2K_{i-}} \right]
\end{equation}
given $K_{i1}=K_{i2}=K_{i-} \, \forall \, i$.

Therefore, while $A_{i}-C_{i}$ remains cluster-specific, $\hat{\delta}_{NEME}$, $\hat{\delta}_{NEMEcpw}$, and $\hat{\delta}_{NEMEcw}$ converge to weighted average treatment effect estimands with data-dependent model-specific weights that are difficult to interpret and are not consistent for the iATE, cpATE, nor cATE estimands.

\section{Derivation of the unweighted \& weighted Fixed Effects model estimator (FE, FEcpw, FEcw)}
\label{sect:appendix_FE_FEcpw_FEcw}

\subsection{Derivation of the Fixed effects (FE) estimator}
\label{sect:appendix_FE}

Similar to the IEE estimator (Section \ref{sect:appendix_IEE_IEEcpw_IEEcw}), the FE estimator yields a similar OLS estimator:
\[\hat{\delta}_{FE} = (\Tilde{Z}'\Tilde{Z})^{-1}\Tilde{Z}'Y|_{\delta}\]
where $\Tilde{Z}$ is the design matrix of the indicator variables for treatment effect $(\delta)$, the period two effect $(\Phi_{2})$, and the cluster fixed effects $(\alpha_{i})$, and $Y$ is the vector of outcomes. With equal allocation of clusters to sequences $S_i = 1$ and $S_i = 0$, we denote that clusters $i \in \left[1,\frac{I}{2}\right]$ are randomized into sequence $S_i = 1$, and clusters $i \in \left[\frac{I}{2}+1,I\right]$ are randomized into sequence sequence $S_i = 0$ (such that $1-S_i=1$).

Accordingly, the first row of the inverse of the Fisher information matrix, corresponding with the treatment effect $(\delta)$ is:
\small
\begin{align*}
    (\Tilde{Z}'\Tilde{Z})^{-1}|_{\delta} =
    \left(\frac{1}{4}\right)
    \begin{bmatrix}
        L+M
        \\  
        L-M
        \\
        -------------------------------------
        \\
        -(L+M)
        \left(
            \frac{S_1 K_{11}}{S_1 (K_{11}+K_{12})}
        \right)
        -
        (L-M)
        \left(
            \frac{S_1  K_{12}}{S_1 (K_{11}+K_{12})}
        \right)
        \\
        \vdots 
        \\
        -(L+M)
        \left(
            \frac{S_{I/2} K_{I/2,1}}{S_{I/2} (K_{I/2,1}+K_{I/2,2})}
        \right)
        -
        (L-M)
        \left(
            \frac{S_{I/2} K_{I/2,2}}{S_{I/2} (K_{I/2,1}+K_{I/2,2})}
        \right)
        \\
        -------------------------------------
        \\
        -(L+M)
        \left(
            \frac{(1-S_{(I/2)+1}) K_{(I/2)+1,1}}{(1-S_{(I/2)+1}) (K_{(I/2)+1,1}+K_{(I/2)+1,2})}
        \right)
        -
        (L-M)
        \left(
            \frac{(1-S_{(I/2)+1}) K_{(I/2)+1,2}}{(1-S_{(I/2)+1}) (K_{(I/2)+1,1}+K_{(I/2)+1,2})}
        \right)
        \\
        \vdots 
        \\
        -(L+M)
        \left(
            \frac{(1-S_I) K_{I1}}{(1-S_I) (K_{I1}+K_{I2})}
        \right)
        -
        (L-M)
        \left(
            \frac{(1-S_I) K_{I2}}{(1-S_I) (K_{I1}+K_{I2})}
        \right)
    \end{bmatrix}' \,,
\end{align*}
\normalsize
where in this section, we define:
\[
L=
    \frac{1}
    {\left( 
        \sum_{i=1}^{I} (1-S_i) \frac{K_{i1}K_{i2}}{K_{i1}+K_{i2}}
    \right)} \,,
\]
\[
M=
    \frac{1}
    {\left(
        \sum_{i=1}^{I} S_i \frac{K_{i1}K_{i2}}{K_{i1}+K_{i2}} 
    \right)} \,.
\]

Subsequently:
\begin{align*}
    \Tilde{Z}'Y =
    \begin{bmatrix}
        \sum_{i=1}^{I}S_{i}\sum_{k=1}^{K_{i1}}Y_{i1k}
        +
        \sum_{i=1}^{I}(1-S_{i})\sum_{k=1}^{K_{i2}}Y_{i2k}
        \\
        \sum_{i=1}^{I} \sum_{k=1}^{K_{i2}} Y_{i2k}
        \\
        -------------------
        \\
        \sum_{j=1}^{2} \sum_{k=1}^{K_{1j}} Y_{1jk}
        \\
        \vdots
        \\
        \sum_{j=1}^{2} \sum_{k=1}^{K_{I/2,j}} Y_{I/2,jk}
        \\
        -------------------
        \\
        \sum_{j=1}^{2} \sum_{k=1}^{K_{(I/2)+1,j}} Y_{(I/2)+1,jk}
        \\
        \vdots
        \\
        \sum_{j=1}^{2} \sum_{k=1}^{K_{Ij}} Y_{Ijk}        
    \end{bmatrix}' \,.
\end{align*}

Accordingly, altogether:
\begin{align*}
    \begin{split}
       \hat{\delta}_{FE} = (\Tilde{Z}'\Tilde{Z})^{-1}\Tilde{Z}'Y|_{\delta} = 
    \end{split}\\
    \begin{split}
        \indent\indent \left(\frac{1}{4}\right) \left(
        \begin{tabular}{l}
            $(L+M) \left(\sum_{i=1}^{I} S_{i} \sum_{k=1}^{K_{i1}} Y_{i1k} \right) + (L+M) \left(\sum_{i=1}^{I} (1-S_{i}) \sum_{k=1}^{K_{i2}} Y_{i2k} \right)$ 
            \\
            \indent $+ (L-M) \left(\sum_{i=1}^{I} S_{i} \sum_{k=1}^{K_{i2}} Y_{i2k} \right) + (L-M) \left(\sum_{i=1}^{I} (1-S_{i}) \sum_{k=1}^{K_{i2}} Y_{i2k} \right)$ 
            \\
            \indent $- L \left(\sum_{i=1}^{I} S_{i} \sum_{j=1}^{2} \sum_{k=1}^{K_{ij}} Y_{ijk} \right) - 
            M \left( \sum_{i=1}^{I} S_{i} \left( \frac{K_{i1}-K_{i2}}{K_{i1}+K_{i2}} \right) \sum_{j=1}^{2} \sum_{k=1}^{K_{ij}} Y_{ijk} \right)$
            \\
            \indent $- 2 L \left( \sum_{i=1}^{I} (1-S_i) \left( \frac{K_{i2}}{K_{i1}+K_{i2}} \right) \sum_{j=1}^{2} \sum_{k=1}^{K_{ij}} Y_{ijk} \right)$
        \end{tabular}
        \right)
    \end{split}\\
    \begin{split}
    \indent\indent = \left(\frac{1}{4}\right) \left(
        \begin{tabular}{l}
            $M \left(\sum_{i=1}^{I} S_{i} (\sum_{k=1}^{K_{i1}} Y_{i1k} - \sum_{k=1}^{K_{i2}} Y_{i2k}) \right)$
            \\
            \indent $+ L \left( \sum_{i=1}^{I} (1-S_i) \sum_{k=1}^{K_{i2}} Y_{i2k} - 2 \sum_{i=1}^{I} (1-S_i) \left( \frac{K_{i2}}{K_{i1}+K_{i2}} \right) \sum_{k=1}^{K_{i1}} Y_{i1k} \right)$ 
            \\
            \indent $- M \left( \sum_{i=1}^{I} S_i \left( \frac{K_{i1}-K_{i2}}{K_{i1}+K_{i2}} \right) \sum_{j=1}^{2} \sum_{k=1}^{K_{ij}} Y_{ijk} \right)$
            \\
            \indent $+ L \left( \sum_{i=1}^{I} (1-S_i) \left( \frac{K_{i1}-K_{i2}}{K_{i1}+K_{i2}} \right) \sum_{k=1}^{K_{i2}} Y_{i2k} \right)$
        \end{tabular}
        \right) \,.
    \end{split}
\end{align*}
Using the previously mentioned definitions for $L$ and $M$:
\small
\begin{align*}
    \begin{split}
    \hat{\delta}_{FE} = \left(\frac{1}{4}\right) \left(
        \begin{tabular}{l}
            $\frac{1}{\left(\sum_{i=1}^{I} S_i \frac{K_{i1}K_{i2}}{K_{i1}+K_{i2}} \right)} \left(\sum_{i=1}^{I} S_{i} (\sum_{k=1}^{K_{i1}} Y_{i1k} - \sum_{k=1}^{K_{i2}} Y_{i2k}) \right)$
            \\
            \indent $+ \frac{1}{\left( \sum_{i=1}^{I} (1-S_i) \frac{K_{i1}K_{i2}}{K_{i1}+K_{i2}}\right)}
            \left( \sum_{i=1}^{I} (1-S_i) \sum_{k=1}^{K_{i2}} Y_{i2k} - 2 \sum_{i=1}^{I} (1-S_i) \left( \frac{K_{i2}}{K_{i1}+K_{i2}} \right) \sum_{k=1}^{K_{i1}} Y_{i1k} \right)$ 
            \\
            \indent $- \frac{1}{\left(\sum_{i=1}^{I} S_i \frac{K_{i1}K_{i2}}{K_{i1}+K_{i2}} \right)} \left( \sum_{i=1}^{I} S_i \left( \frac{K_{i1}-K_{i2}}{K_{i1}+K_{i2}} \right) \sum_{j=1}^{2} \sum_{k=1}^{K_{ij}} Y_{ijk} \right)$
            \\
            \indent $+ \frac{1}{\left( \sum_{i=1}^{I} (1-S_i) \frac{K_{i1}K_{i2}}{K_{i1}+K_{i2}}\right)} \left( \sum_{i=1}^{I} (1-S_i) \left( \frac{K_{i1}-K_{i2}}{K_{i1}+K_{i2}} \right) \sum_{k=1}^{K_{i2}} Y_{i2k} \right)$
        \end{tabular}
        \right)
    \end{split}
\end{align*}
\begin{equation}
    \indent = \left(\frac{1}{4}\right) \left(
        \begin{tabular}{l}
            $\frac{1}{\left( \frac{1}{I/2} \right) \left(\sum_{i=1}^{I} S_i \frac{K_{i1}K_{i2}}{K_{i1}+K_{i2}} \right)} \left( \left( \frac{1}{I/2} \right) \sum_{i=1}^{I} S_{i} \left(\sum_{k=1}^{K_{i1}} Y_{i1k}(1) - \sum_{k=1}^{K_{i2}} Y_{i2k}(0) \right) \right)$
            \\
            \indent $+ \frac{1}{\left( \frac{1}{I/2} \right) \left( \sum_{i=1}^{I} (1-S_i) \frac{K_{i1}K_{i2}}{K_{i1}+K_{i2}}\right)}
            \left( \left( \frac{1}{I/2} \right) \sum_{i=1}^{I} (1-S_i) \sum_{k=1}^{K_{i2}} Y_{i2k}(1) \right)$
            \\ 
            \indent $- \frac{1}{\left( \frac{1}{I/2} \right) \left( \sum_{i=1}^{I} (1-S_i) \frac{K_{i1}K_{i2}}{K_{i1}+K_{i2}}\right)} \left(2 \left( \frac{1}{I/2} \right) \sum_{i=1}^{I} (1-S_i) \left( \frac{K_{i2}}{K_{i1}+K_{i2}} \right) \sum_{k=1}^{K_{i1}} Y_{i1k}(0) \right)$ 
            \\
            \indent $- \frac{1}{\left( \frac{1}{I/2} \right) \left(\sum_{i=1}^{I} S_i \frac{K_{i1}K_{i2}}{K_{i1}+K_{i2}} \right)} \left( \left( \frac{1}{I/2} \right) \sum_{i=1}^{I} S_i \left( \frac{K_{i1}-K_{i2}}{K_{i1}+K_{i2}} \right) \left(\sum_{k=1}^{K_{i1}} Y_{i1k}(1) + \sum_{k=1}^{K_{i2}} Y_{i2k}(0) \right) \right)$
            \\
            \indent $+ \frac{1}{ \left( \frac{1}{I/2} \right) \left( \sum_{i=1}^{I} (1-S_i) \frac{K_{i1}K_{i2}}{K_{i1}+K_{i2}}\right)} \left( \left( \frac{1}{I/2} \right) \sum_{i=1}^{I} (1-S_i) \left( \frac{K_{i1}-K_{i2}}{K_{i1}+K_{i2}} \right) \sum_{k=1}^{K_{i2}} Y_{i2k}(1) \right)$
        \end{tabular} 
        \right)
\end{equation}
\normalsize
when connected to the potential outcomes with equation \ref{eq:PO}.

The FE estimator then converges in probability to:
\begin{align*}
    \hat{\delta}_{FE} \xrightarrow{P} \left(\frac{1}{4}\right) \left(
        \begin{tabular}{l}
           $\frac{1}{ E\left[ \frac{K_{i1}K_{i2}}{K_{i1}+K_{i2}} | S_i \right]} E\left[ \sum_{k=1}^{K_{i1}} Y_{i1k}(1) - \sum_{k=1}^{K_{i2}} Y_{i2k}(0) | S_{i} \right]$
            \\
            \indent $+ \frac{1}{ E\left[\frac{K_{i1}K_{i2}}{K_{i1}+K_{i2}} | 1-S_i \right]}
            E\left[ \sum_{k=1}^{K_{i2}} Y_{i2k}(1) | 1-S_i \right]$
            \\ 
            \indent $- \frac{1}{ E\left[ \frac{K_{i1}K_{i2}}{K_{i1}+K_{i2}} | 1-S_i \right]} 2 E\left[ \left( \frac{K_{i2}}{K_{i1}+K_{i2}} \right) \sum_{k=1}^{K_{i1}} Y_{i1k}(0) | 1-S_i \right]$ 
            \\
            \indent $- \frac{1}{ E\left[ \frac{K_{i1}K_{i2}}{K_{i1}+K_{i2}} | S_i \right]} E\left[ \left( \frac{K_{i1}-K_{i2}}{K_{i1}+K_{i2}} \right) \left(\sum_{k=1}^{K_{i1}} Y_{i1k}(1) + \sum_{k=1}^{K_{i2}} Y_{i2k}(0) \right) | S_i \right]$
            \\
            \indent $+ \frac{1}{ E\left[ \frac{K_{i1}K_{i2}}{K_{i1}+K_{i2}} | 1-S_i \right]} E \left[ \left( \frac{K_{i1}-K_{i2}}{K_{i1}+K_{i2}} \right) \sum_{k=1}^{K_{i2}} Y_{i2k}(1) | 1-S_i \right]$
        \end{tabular} 
        \right)
    \,.
\end{align*}

As in Section \ref{sect:appendix_IEE}, with randomization, the sequence variable $S_{i}$ is independent of the potential outcomes and cluster-period sizes, $S_{i} \indep \Omega$, and $\Omega=\{Y_{ijk}(0), Y_{ijk}(1), K_{ij}\}_{i=1, k=1}^{I, K_{ij}}$.

\small
\[
\hat{\delta}_{FE} \xrightarrow{P} 
\left(
    \frac{
    \left(
        E\left[\left(\frac{K_{i2}}{K_{i1}+K_{i2}}\right)\left(\sum_{k=1}^{K_{i1}} Y_{i1k}(1) - \sum_{k=1}^{K_{i1}} Y_{i1k}(0)\right)\right]
        +
        E\left[\left(\frac{K_{i1}}{K_{i1}+K_{i2}}\right)\left(\sum_{k=1}^{K_{i2}} Y_{i2k}(1) - \sum_{k=1}^{K_{i2}} Y_{i2k}(0)\right)\right]
    \right)
    }
    {2E\left[\frac{K_{i1}K_{i2}}{K_{i1}+K_{i2}}\right]}
\right)
\]
\normalsize

In a CRXO trial with IPS, we can surprisingly demonstrate that the FE estimator can be consistent for the pATE estimand. Where $\lambda_{i} =K_{i2}/K_{i1}$, we can re-write:
\[
\frac{\left(\frac{K_{i2}}{K_{i1}+K_{i2}}\right)}{E\left[\frac{K_{i1}K_{i2}}{K_{i1}+K_{i2}}\right]} = \frac{\left(\frac{\lambda_{i}}{1+\lambda_{i}}\right)}{E\left[\frac{\lambda_{i}K_{i1}}{1+\lambda_{i}}\right]} \,,
\]
\[
\frac{\left(\frac{K_{i1}}{K_{i1}+K_{i2}}\right)}{E\left[\frac{K_{i1}K_{i2}}{K_{i1}+K_{i2}}\right]} = \frac{\left(\frac{1}{1+\lambda_{i}}\right)}{E\left[\frac{\lambda_{i}K_{i1}}{1+\lambda_{i}}\right]} \,.
\]
Importantly, if $\lambda_i=\lambda \, \forall \, i$, then:
\[
\frac{\left(\frac{K_{i2}}{K_{i1}+K_{i2}}\right)}{E\left[\frac{K_{i1}K_{i2}}{K_{i1}+K_{i2}}\right]} = \frac{1}{E\left[K_{i1}\right]} \,,
\]
\[
\frac{\left(\frac{K_{i1}}{K_{i1}+K_{i2}}\right)}{E\left[\frac{K_{i1}K_{i2}}{K_{i1}+K_{i2}}\right]} = \frac{1}{E\left[K_{i1}\right]} \,,
\]
and:
\begin{equation}
    \hat{\delta}_{FE} \xrightarrow{P} \frac{1}{2} \sum_{j=1}^{2} \left[ \frac{E\left[ \sum_{k=1}^{K_{ij}} \left[Y_{ijk}(1)-Y_{ijk}(0)\right] \right]}{E\left[K_{ij}\right]} \right]  \,.
\end{equation}

To reiterate, if the sample sizes in period $j=2$, relative to period $j=1$, are inflated by a fixed ratio of $\lambda$ for all clusters $i$ ($\lambda_i = \lambda \, \forall \, i$), the FE estimator is surprisingly a consistent estimator for the pATE estimand in a CRXO design. A special case of this condition includes when cluster-period cell sizes are fixed between periods, within clusters ($K_{i1} = K_{i2}\, \forall \, i$), which also represent a special case of no IPS. In scenarios without IPS, the fixed effects model estimator is still consistent for the pATE which happens to coincide with the iATE estimand.

\subsection{Derivation of the Fixed effects model with inverse cluster-period size weights (FEcpw) estimator}
\label{sect:appendix_FEcpw}

It is then straightforward using the result from Section \ref{sect:appendix_FE} to derive the treatment effect estimator for the Fixed effects model with inverse cluster-period size weights:
\begin{align*}
    \begin{split}
    \hat{\delta}_{FEcpw} = \left(\frac{1}{2}\right) \left(
        \begin{tabular}{l}
            $\frac{1}{\left( I/2 \right)} \left(\sum_{i=1}^{I} S_{i} (\frac{\sum_{k=1}^{K_{i1}} Y_{i1k}(1)}{K_{i1}} - \frac{\sum_{k=1}^{K_{i2}} Y_{i2k}(0)}{K_{i2}}) \right)$
            \\
            \indent $+ \frac{1}{\left( I/2\right)}
            \left( \sum_{i=1}^{I} (1-S_i) \frac{\sum_{k=1}^{K_{i2}} Y_{i2k}(1)}{K_{i2}} - \sum_{i=1}^{I} (1-S_i) \frac{\sum_{k=1}^{K_{i1}} Y_{i1k}(0)}{K_{i1}} \right)$ 
        \end{tabular}
        \right)
    \end{split}
    \\
    \begin{split}
    =\frac{1}{I}\sum_{i=1}^{I} 
    \left(
        S_{i}\left( \frac{\sum_{k=1}^{K_{i1}}Y_{i1k}(1)}{K_{i1}} - \frac{\sum_{k=1}^{K_{i2}}Y_{i2k}(0)}{K_{i2}} \right) + 
        (1-S_{i}) \left(\frac{\sum_{k=1}^{K_{i2}}Y_{i2k}(1)}{K_{i2}} - \frac{\sum_{k=1}^{K_{i1}}Y_{i1k}(0)}{K_{i1}} \right) 
    \right) \,.
    \end{split}
\end{align*}

It can be easily demonstrated that this FEcpw estimator is consistent for the cpATE estimand:
\begin{equation}
    \hat{\delta}_{FEcpw} \xrightarrow{P}  E\left[ \frac{1}{2} \sum_{j=1}^{2} \frac{ \sum_{k=1}^{K_{ij}} \left[Y_{ijk}(1)-Y_{ijk}(0)\right]}{K_{ij}} \right] \,.
\end{equation}

Additionally, like the IEEcpw estimator described in Section \ref{sect:appendix_IEEcpw}, we can demonstrate that the FEcpw estimator is unbiased for the cpATE in expectation over the sampling distribution. We can formally define the set of all potential outcomes for all $K_{ij}$ individuals in periods $j=1,2$ in sampled clusters $i=1,...,I$ as $\Omega=\{Y_{ijk}(0), Y_{ijk}(1), K_{ij}\}_{i=1, k=1}^{I, K_{ij}}$.
Formally, we take the expectation $\hat{\delta}_{FEcpw}$ by treating the potential outcomes of the samples as fixed quantities and the sequence assignment $S_{i}$ as random.
Therefore, conditioning the expectation on the set of sampled potential outcomes and cluster-period sample sizes $\Omega$ yields:
\[E[\hat{\delta}_{FEcpw}|\Omega] =\]
\[
    \frac{1}{I}\sum_{i=1}^{I} 
    \left(
        E[S_{i}|\Omega]\left( \frac{\sum_{k=1}^{K_{i1}}Y_{i1k}(1)}{K_{i1}} - \frac{\sum_{k=1}^{K_{i2}}Y_{i2k}(0)}{K_{i2}} \right) + 
        E[1-S_{i}|\Omega] \left(\frac{\sum_{k=1}^{K_{i2}}Y_{i2k}(1)}{K_{i2}} - \frac{\sum_{k=1}^{K_{i1}}Y_{i1k}(0)}{K_{i1}} \right) 
    \right) \,.
\]
where with randomization, the sequence variable $S_{i}$ is independent of the potential outcomes and cluster-period sizes, $S_{i} \indep \Omega$, and $\Omega=\{Y_{ijk}(0), Y_{ijk}(1), K_{ij}\}_{i=1, k=1}^{I, K_{ij}}$ and $E[S_{i}|\Omega] = E[1-S_{i}|\Omega] = \frac{1}{2}$. Accordingly:
\[E[\hat{\delta}_{FEcpw}|\Omega] =
    \frac{1}{I}\sum_{i=1}^{I} 
    \left(
        \frac{1}{2}\left( \frac{\sum_{k=1}^{K_{i1}}Y_{i1k}(1)}{K_{i1}} + \frac{\sum_{k=1}^{K_{i2}}Y_{i2k}(1)}{K_{i2}} - \frac{\sum_{k=1}^{K_{i1}}Y_{i1k}(0)}{K_{i1}} - \frac{\sum_{k=1}^{K_{i2}}Y_{i2k}(0)}{K_{i2}} \right) 
    \right)
\]
\[
    = \frac{1}{I}\sum_{i=1}^{I}
    \left(
        \frac{1}{2} \sum_{j=1}^{2}  \left( \frac{\sum_{k=1}^{K_{ij}}Y_{ijk}(1)}{K_{ij}} - \frac{\sum_{k=1}^{K_{ij}}Y_{ijk}(0)}{K_{ij}} \right) 
    \right) \,.
\]

We assume that the sample of clusters is a simple random sample from a superpopulation of clusters. With this superpopulation framework, we have two sources of randomness, random sampling from a superpopulation of clusters and subsequent randomization of treatment assignment. Overall, we can prove that the IEEcpw estimator is unbiased in expectation for the cpATE, where with the law of total expectation:
\[E[E[\hat{\delta}_{FEcpw}|\Omega]] = E[\hat{\delta}_{FEcpw}] = E\left[ \frac{1}{2} \sum_{j=1}^{2} \frac{ \sum_{k=1}^{K_{ij}} \left[Y_{ijk}(1)-Y_{ijk}(0)\right]}{K_{ij}} \right] \,.\]

\subsection{Derivation of the Fixed effects model with inverse cluster size weights (FEcw) estimator}
\label{sect:appendix_FEcw}

We can use the result from Section \ref{sect:appendix_FE} to similarly demonstrate that the fixed effects model with inverse cluster size weights (FEcw) estimator is consistent for the cATE estimand. We can show that the FEcw estimator is:
\small
\begin{equation}
    \hat{\delta}_{FEcw} = 
\end{equation}
\[
\left(\frac{1}{4}\right) \left(
        \begin{tabular}{l}
            $\frac{1}{\left(\sum_{i=1}^{I} S_i \frac{K_{i1}K_{i2}}{(K_{i1}+K_{i2})^2} \right)} \left(\sum_{i=1}^{I} S_{i} \left( \frac{\sum_{k=1}^{K_{i1}} Y_{i1k}}{\sum_{j=1}^2 K_{ij}} - \frac{\sum_{k=1}^{K_{i2}} Y_{i2k}}{\sum_{j=1}^2 K_{ij}} \right) \right)$
            \\
            \indent $+ \frac{1}{\left( \sum_{i=1}^{I} (1-S_i) \frac{K_{i1}K_{i2}}{(K_{i1}+K_{i2})^2}\right)}
            \left( \sum_{i=1}^{I} (1-S_i) \left(\frac{\sum_{k=1}^{K_{i2}} Y_{i2k}}{\sum_{j=1}^2 K_{ij}}\right) 
            - 2 \sum_{i=1}^{I} (1-S_i) \left( \frac{K_{i2}}{K_{i1}+K_{i2}} \right) \left(\frac{\sum_{k=1}^{K_{i1}} Y_{i1k}}{\sum_{j=1}^2 K_{ij}}\right) \right)$ 
            \\
            \indent $- \frac{1}{\left(\sum_{i=1}^{I} S_i \frac{K_{i1}K_{i2}}{(K_{i1}+K_{i2})^2} \right)} \left( \sum_{i=1}^{I} S_i \left( \frac{K_{i1}-K_{i2}}{K_{i1}+K_{i2}} \right) \sum_{j=1}^{2} \left(\frac{\sum_{k=1}^{K_{ij}} Y_{ijk}}{\sum_{j=1}^2 K_{ij}}\right) \right)$
            \\
            \indent $+ \frac{1}{\left( \sum_{i=1}^{I} (1-S_i) \frac{K_{i1}K_{i2}}{(K_{i1}+K_{i2})^2}\right)} \left( \sum_{i=1}^{I} (1-S_i) \left( \frac{K_{i1}-K_{i2}}{K_{i1}+K_{i2}} \right) \left(\frac{\sum_{k=1}^{K_{i2}} Y_{i2k}}{\sum_{j=1}^2 K_{ij}}\right) \right)$
        \end{tabular}
        \right)
\]
\normalsize

The FEcw estimator connected to potential outcomes then converges in probability to:
\begin{align*}
    \hat{\delta}_{FEcw} \xrightarrow{P} \left(\frac{1}{4}\right) \left(
        \begin{tabular}{l}
           $\frac{1}{ E\left[ \frac{K_{i1}K_{i2}}{(K_{i1}+K_{i2})^2} | S_i \right]} E\left[ \frac{\sum_{k=1}^{K_{i1}} Y_{i1k}(1)}{\sum_{j=1}^2 K_{ij}} - \frac{\sum_{k=1}^{K_{i2}} Y_{i2k}(0)}{\sum_{j=1}^2 K_{ij}} | S_{i} \right]$
            \\
            \indent $+ \frac{1}{ E\left[\frac{K_{i1}K_{i2}}{(K_{i1}+K_{i2})^2} | 1-S_i \right]}
            E\left[ \frac{\sum_{k=1}^{K_{i2}} Y_{i2k}(1)}{\sum_{j=1}^2 K_{ij}} | 1-S_i \right]$
            \\ 
            \indent $- \frac{1}{ E\left[ \frac{K_{i1}K_{i2}}{(K_{i1}+K_{i2})^2} | 1-S_i \right]} 2 E\left[ \left( \frac{K_{i2}}{K_{i1}+K_{i2}} \right) \frac{\sum_{k=1}^{K_{i1}} Y_{i1k}(0)}{\sum_{j=1}^2 K_{ij}} | 1-S_i \right]$ 
            \\
            \indent $- \frac{1}{ E\left[ \frac{K_{i1}K_{i2}}{(K_{i1}+K_{i2})^2} | S_i \right]} E\left[ \left( \frac{K_{i1}-K_{i2}}{K_{i1}+K_{i2}} \right) \left(\frac{\sum_{k=1}^{K_{i1}} Y_{i1k}(1)}{\sum_{j=1}^2 K_{ij}} + \frac{\sum_{k=1}^{K_{i2}} Y_{i2k}(0)}{\sum_{j=1}^2 K_{ij}} \right) | S_i \right]$
            \\
            \indent $+ \frac{1}{ E\left[ \frac{K_{i1}K_{i2}}{(K_{i1}+K_{i2})^2} | 1-S_i \right]} E \left[ \left( \frac{K_{i1}-K_{i2}}{K_{i1}+K_{i2}} \right) \frac{\sum_{k=1}^{K_{i2}} Y_{i2k}(1)}{\sum_{j=1}^2 K_{ij}} | 1-S_i \right]$
        \end{tabular} 
        \right)
    \,.
\end{align*}

As in Section \ref{sect:appendix_IEE}, with randomization, the sequence variable $S_{i}$ is independent of the potential outcomes and cluster-period sizes, $S_{i} \indep \Omega$, and $\Omega=\{Y_{ijk}(0), Y_{ijk}(1), K_{ij}\}_{i=1, k=1}^{I, K_{ij}}$.
\begin{align*}
    \hat{\delta}_{FEcw} \xrightarrow{P} \left(\frac{1}{4}\right) \left(\frac{1}{ E\left[ \frac{K_{i1}K_{i2}}{(K_{i1}+K_{i2})^2} \right]}\right) \left(
        \begin{tabular}{l}
           $ E\left[ \frac{\sum_{k=1}^{K_{i1}} Y_{i1k}(1)}{\sum_{j=1}^2 K_{ij}} - \frac{\sum_{k=1}^{K_{i2}} Y_{i2k}(0)}{\sum_{j=1}^2 K_{ij}} \right]$
            \\
            \indent $+ 
            E\left[ \frac{\sum_{k=1}^{K_{i2}} Y_{i2k}(1)}{\sum_{j=1}^2 K_{ij}} \right]$
            \\ 
            \indent $-  2 E\left[ \left( \frac{K_{i2}}{K_{i1}+K_{i2}} \right) \frac{\sum_{k=1}^{K_{i1}} Y_{i1k}(0)}{\sum_{j=1}^2 K_{ij}} \right]$ 
            \\
            \indent $- E\left[ \left( \frac{K_{i1}-K_{i2}}{K_{i1}+K_{i2}} \right) \left(\frac{\sum_{k=1}^{K_{i1}} Y_{i1k}(1)}{\sum_{j=1}^2 K_{ij}} + \frac{\sum_{k=1}^{K_{i2}} Y_{i2k}(0)}{\sum_{j=1}^2 K_{ij}} \right) \right]$
            \\
            \indent $+ E \left[ \left( \frac{K_{i1}-K_{i2}}{K_{i1}+K_{i2}} \right) \frac{\sum_{k=1}^{K_{i2}} Y_{i2k}(1)}{\sum_{j=1}^2 K_{ij}} \right]$
        \end{tabular} 
        \right)
    \,.
\end{align*}
\begin{small}
\[ 
 =  \left(\frac{1}{2E\left[ \frac{K_{i1}K_{i2}}{(\sum_{j=1}^2 K_{ij})^2} \right]}\right) \left(
    E\left[ 
        \frac{K_{i2}}{\sum_{j=1}^2 K_{ij}}
        \left( \frac{\sum_{k=1}^{K_{i1}} Y_{i1k}(1)}{\sum_{j=1}^2 K_{ij}} - \frac{\sum_{k=1}^{K_{i1}} Y_{i1k}(0)}{\sum_{j=1}^2 K_{ij}} \right) 
        +
        \frac{K_{i1}}{\sum_{j=1}^2 K_{ij}}
        \left( \frac{\sum_{k=1}^{K_{i2}} Y_{i2k}(1)}{\sum_{j=1}^2 K_{ij}} - \frac{\sum_{k=1}^{K_{i2}} Y_{i2k}(0)}{\sum_{j=1}^2 K_{ij}} \right) 
    \right]
 \right)
 \,.
\]
\end{small}
Where $\lambda_i=K_{i2}/K_{i1}$, we can re-write:
\[
 =  \left(\frac{1}{2E\left[ \frac{\lambda_i}{(1+\lambda_i)^2} \right]}\right) \left(
    E\left[ 
        \frac{\lambda_i}{1+\lambda_i}
        \left( \frac{\sum_{k=1}^{K_{i1}} Y_{i1k}(1)}{\sum_{j=1}^2 K_{ij}} - \frac{\sum_{k=1}^{K_{i1}} Y_{i1k}(0)}{\sum_{j=1}^2 K_{ij}} \right) 
        +
        \frac{1}{1+\lambda_i}
        \left( \frac{\sum_{k=1}^{K_{i2}} Y_{i2k}(1)}{\sum_{j=1}^2 K_{ij}} - \frac{\sum_{k=1}^{K_{i2}} Y_{i2k}(0)}{\sum_{j=1}^2 K_{ij}} \right) 
    \right]
 \right)
 \,.
\]
If $\lambda_i=\lambda \, \forall \, i$, then:
\[
 \hat{\delta}_{FEcw} \xrightarrow{P} \left(\frac{1}{2\frac{\lambda}{(1+\lambda)^2}}\right) \left(
        \frac{\lambda}{1+\lambda}
        E\left[ \frac{\sum_{k=1}^{K_{i1}} Y_{i1k}(1)}{\sum_{j=1}^2 K_{ij}} - \frac{\sum_{k=1}^{K_{i1}} Y_{i1k}(0)}{\sum_{j=1}^2 K_{ij}} \right]
        +
        \frac{1}{1+\lambda}
        E\left[ \frac{\sum_{k=1}^{K_{i2}} Y_{i2k}(1)}{\sum_{j=1}^2 K_{ij}} - \frac{\sum_{k=1}^{K_{i2}} Y_{i2k}(0)}{\sum_{j=1}^2 K_{ij}} \right] 
 \right)
\]
\[
= \left(\frac{1+\lambda}{2\lambda}\right) \left(
        \lambda
        E\left[ \frac{\sum_{k=1}^{K_{i1}} Y_{i1k}(1)}{\sum_{j=1}^2 K_{ij}} - \frac{\sum_{k=1}^{K_{i1}} Y_{i1k}(0)}{\sum_{j=1}^2 K_{ij}} \right]
        +
        E\left[ \frac{\sum_{k=1}^{K_{i2}} Y_{i2k}(1)}{\sum_{j=1}^2 K_{ij}} - \frac{\sum_{k=1}^{K_{i2}} Y_{i2k}(0)}{\sum_{j=1}^2 K_{ij}} \right] 
 \right)
\]
\[
= \left(\frac{1+\lambda}{2\lambda}\right) \left(
        \lambda
        E\left[ \frac{\sum_{k=1}^{K_{i1}} Y_{i1k}(1)}{K_{i1}(1+\lambda)} - \frac{\sum_{k=1}^{K_{i1}} Y_{i1k}(0)}{K_{i1}(1+\lambda)} \right]
        +
        E\left[ \frac{\sum_{k=1}^{K_{i2}} Y_{i2k}(1)}{K_{i1}(1+\lambda)} - \frac{\sum_{k=1}^{K_{i2}} Y_{i2k}(0)}{K_{i1}(1+\lambda)} \right] 
 \right)
\]
\[
= \left(\frac{1}{2\lambda}\right) \left(
        \lambda
        E\left[ \frac{\sum_{k=1}^{K_{i1}} Y_{i1k}(1)}{K_{i1}} - \frac{\sum_{k=1}^{K_{i1}} Y_{i1k}(0)}{K_{i1}} \right]
        +
        E\left[ \frac{\sum_{k=1}^{K_{i2}} Y_{i2k}(1)}{K_{i1}} - \frac{\sum_{k=1}^{K_{i2}} Y_{i2k}(0)}{K_{i1}} \right] 
 \right)
\]
\[
= \left(\frac{1}{2}\right) \left(
        E\left[ \frac{\sum_{k=1}^{K_{i1}} Y_{i1k}(1)}{K_{i1}} - \frac{\sum_{k=1}^{K_{i1}} Y_{i1k}(0)}{K_{i1}} \right]
        +
        E\left[ \frac{\sum_{k=1}^{K_{i2}} Y_{i2k}(1)}{K_{i2}} - \frac{\sum_{k=1}^{K_{i2}} Y_{i2k}(0)}{K_{i2}} \right] 
 \right)
\]
and:
\begin{equation}
    \hat{\delta}_{FEcw} \xrightarrow{P}  E\left[ \frac{1}{2} \sum_{j=1}^{2} \frac{ \sum_{k=1}^{K_{ij}} \left[Y_{ijk}(1)-Y_{ijk}(0)\right]}{K_{ij}} \right] \,.
\end{equation}
To reiterate, if the sample sizes in period $j=2$, relative to period $j=1$, are inflated by a fixed ratio of $\lambda$ for all clusters $i$ ($\lambda_i = \lambda \, \forall \, i$), the FEcw estimator is surprisingly a consistent estimator for the cpATE estimand in a CRXO design.

\subsection{Derivation of the Fixed effects model with inverse period size weights (FEpw) estimator}
\label{sect:appendix_FEpw}

We can use the result from Section \ref{sect:appendix_FE} to similarly demonstrate that the fixed effects model with inverse period size weights (FEpw) estimator is consistent for the pATE estimand. We can show that the FEpw estimator is:
\small
\begin{equation}
    \hat{\delta}_{FEpw} = 
\end{equation}
\[
\left(\frac{1}{4}\right) \left(
        \begin{tabular}{l}
            $\frac{1}{\left(\sum_{i=1}^{I} S_i \textbf{A}_i \right)} \left(\sum_{i=1}^{I} S_{i} \left( \frac{\sum_{k=1}^{K_{i1}} Y_{i1k}}{\sum_{i=1}^I K_{i1}} - \frac{\sum_{k=1}^{K_{i2}} Y_{i2k}}{\sum_{i=1}^I K_{i2}} \right) \right)$
            \\
            \indent $+ \frac{1}{\left( \sum_{i=1}^{I} (1-S_i) \textbf{A}_i\right)} \left( \sum_{i=1}^{I} (1-S_i) \left(\frac{\sum_{k=1}^{K_{i2}} Y_{i2k}}{\sum_{i=1}^I K_{i2}}\right) 
            - 2 \sum_{i=1}^{I} (1-S_i) \left(\textbf{C}_i\right) \left(\frac{\sum_{k=1}^{K_{i1}} Y_{i1k}}{\sum_{i=1}^I K_{i1}}\right) \right)$ 
            \\
            \indent $- \frac{1}{\left(\sum_{i=1}^{I} S_i \textbf{A}_i \right)} \left( \sum_{i=1}^{I} S_i \left( \textbf{B}_i-\textbf{C}_i \right) \sum_{j=1}^{2} \left(\frac{\sum_{k=1}^{K_{ij}} Y_{ijk}}{\sum_{i=1}^I K_{ij}}\right) \right)$
            \\
            \indent $+ \frac{1}{\left( \sum_{i=1}^{I} (1-S_i) \textbf{A}_i\right)} \left( \sum_{i=1}^{I} (1-S_i) \left( \textbf{B}_i-\textbf{C}_i \right) \left(\frac{\sum_{k=1}^{K_{i2}} Y_{i2k}}{\sum_{i=1}^I K_{i2}}\right) \right)$
        \end{tabular}
        \right)
\]
\normalsize
where we define:
\[
    \textbf{A}_i = \frac{\left(\frac{K_{i1}}{\sum_{i=1}^I K_{i1}}\right)\left(\frac{K_{i2}}{\sum_{i=1}^I K_{i2}}\right)}{\left(\frac{K_{i1}}{\sum_{i=1}^I K_{i1}}+\frac{K_{i2}}{\sum_{i=1}^I K_{i2}}\right)} = \frac{K_{i1}K_{i2}}{(\sum_{i=1}^I K_{i2})K_{i1}+(\sum_{i=1}^I K_{i1})K_{i2}} \,,
\]
\[
    \textbf{B}_i = \frac{\left(\frac{K_{i1}}{\sum_{i=1}^I K_{i1}}\right)}{\left(\frac{K_{i1}}{\sum_{i=1}^I K_{i1}}+\frac{K_{i2}}{\sum_{i=1}^I K_{i2}}\right)} = \frac{(\sum_{i=1}^I K_{i2})K_{i1}}{(\sum_{i=1}^I K_{i2})K_{i1}+(\sum_{i=1}^I K_{i1})K_{i2}} \,,
\]
\[
    \textbf{C}_i = \frac{\left(\frac{K_{i2}}{\sum_{i=1}^I K_{i2}}\right)}{\left(\frac{K_{i1}}{\sum_{i=1}^I K_{i1}}+\frac{K_{i2}}{\sum_{i=1}^I K_{i2}}\right)} = \frac{(\sum_{i=1}^I K_{i1})K_{i2}}{(\sum_{i=1}^I K_{i2})K_{i1}+(\sum_{i=1}^I K_{i1})K_{i2}} \,.
\]

Accordingly, the FEpw estimator connected to potential outcomes converges in probability to:
\begin{align*}
    \hat{\delta}_{FEpw} \xrightarrow{P} \left(\frac{1}{4}\right) \left(
        \begin{tabular}{l}
           $\frac{1}{ E\left[ \textbf{A}_i | S_i \right]} E\left[ \frac{\sum_{k=1}^{K_{i1}} Y_{i1k}(1)}{E[K_{i1}]} - \frac{\sum_{k=1}^{K_{i2}} Y_{i2k}(0)}{E[K_{i2}]} | S_{i} \right]$
            \\
            \indent $+ \frac{1}{ E\left[\textbf{A}_i | 1-S_i \right]}
            E\left[ \frac{\sum_{k=1}^{K_{i2}} Y_{i2k}(1)}{E[K_{i2}]} | 1-S_i \right]$
            \\ 
            \indent $- \frac{1}{ E\left[ \textbf{A}_i | 1-S_i \right]} 2 E\left[ \left( \textbf{C}_i \right) \frac{\sum_{k=1}^{K_{i1}} Y_{i1k}(0)}{E[K_{i1}]} | 1-S_i \right]$ 
            \\
            \indent $- \frac{1}{ E\left[ \textbf{A}_i | S_i \right]} E\left[ \left( \textbf{B}_i-\textbf{C}_i \right) \left(\frac{\sum_{k=1}^{K_{i1}} Y_{i1k}(1)}{E[K_{i1}]} + \frac{\sum_{k=1}^{K_{i2}} Y_{i2k}(0)}{E[K_{i2}]} \right) | S_i \right]$
            \\
            \indent $+ \frac{1}{ E\left[ \textbf{A}_i | 1-S_i \right]} E \left[ \left( \textbf{B}_i-\textbf{C}_i \right) \frac{\sum_{k=1}^{K_{i2}} Y_{i2k}(1)}{E[K_{i2}]} | 1-S_i \right]$
        \end{tabular} 
        \right)
    \,.
\end{align*}

As in Section \ref{sect:appendix_IEE}, with randomization, the sequence variable $S_{i}$ is independent of the potential outcomes and cluster-period sizes, $S_{i} \indep \Omega$, and $\Omega=\{Y_{ijk}(0), Y_{ijk}(1), K_{ij}\}_{i=1, k=1}^{I, K_{ij}}$.
\begin{align*}
    \hat{\delta}_{FEpw} \xrightarrow{P} \left(\frac{1}{4}\right) \left(\frac{1}{ E\left[ \textbf{A}_i \right]}\right) \left(
        \begin{tabular}{l}
           $ E\left[ \frac{\sum_{k=1}^{K_{i1}} Y_{i1k}(1)}{E[K_{i1}]} - \frac{\sum_{k=1}^{K_{i2}} Y_{i2k}(0)}{E[K_{i2}]} \right]$
            \\
            \indent $+ 
            E\left[ \frac{\sum_{k=1}^{K_{i2}} Y_{i2k}(1)}{E[K_{i2}]} \right]$
            \\ 
            \indent $-  2 E\left[ \left( \textbf{C}_i \right) \frac{\sum_{k=1}^{K_{i1}} Y_{i1k}(0)}{E[K_{i1}]} \right]$ 
            \\
            \indent $- E\left[ \left( \textbf{B}_i-\textbf{C}_i \right) \left(\frac{\sum_{k=1}^{K_{i1}} Y_{i1k}(1)}{E[K_{i1}]} + \frac{\sum_{k=1}^{K_{i2}} Y_{i2k}(0)}{E[K_{i2}]} \right) \right]$
            \\
            \indent $+ E \left[ \left( \textbf{B}_i-\textbf{C}_i \right) \frac{\sum_{k=1}^{K_{i2}} Y_{i2k}(1)}{E[K_{i2}]} \right]$
        \end{tabular} 
        \right)
\end{align*}
\begin{align*}
    = \left(\frac{1}{4}\right) \left(\frac{1}{ E\left[ \textbf{A}_i \right]}\right) \left(
        \begin{tabular}{l}
           $E\left[ \frac{\sum_{k=1}^{K_{i1}} Y_{i1k}(1)}{E[K_{i1}]} - 2\left( \textbf{C}_i \right) \frac{\sum_{k=1}^{K_{i1}} Y_{i1k}(0)}{E[K_{i1}]} \right]$
            \\
            \indent $+ 
            E\left[ \frac{\sum_{k=1}^{K_{i2}} Y_{i2k}(1)}{E[K_{i2}]}- \frac{\sum_{k=1}^{K_{i2}} Y_{i2k}(0)}{E[K_{i2}]} \right]$
            \\ 
            \indent $- E\left[ \left( \textbf{B}_i-\textbf{C}_i \right) \left(\frac{\sum_{k=1}^{K_{i1}} Y_{i1k}(1)}{E[K_{i1}]} \right) \right]$
            \\
            \indent $+ E \left[ \left( \textbf{B}_i-\textbf{C}_i \right) \left( \frac{\sum_{k=1}^{K_{i2}} Y_{i2k}(1)}{E[K_{i2}]} - \frac{\sum_{k=1}^{K_{i2}} Y_{i2k}(0)}{E[K_{i2}]} \right) \right]$
        \end{tabular} 
        \right)
\end{align*}
and:
\[
    \textbf{A}_i \xrightarrow{P} \frac{\left(\frac{K_{i1}}{E[K_{i1}]}\right)\left(\frac{K_{i2}}{E[K_{i2}]}\right)}{\left(\frac{K_{i1}}{E[K_{i1}]}+\frac{K_{i2}}{E[K_{i2}]}\right)} = \frac{K_{i1}K_{i2}}{E[K_{i2}]K_{i1}+E[K_{i1}]K_{i2}} \,,
\]
\[
    \textbf{B}_i \xrightarrow{P} \frac{\left(\frac{K_{i1}}{E[K_{i1}]}\right)}{\left(\frac{K_{i1}}{E[K_{i1}]}+\frac{K_{i2}}{E[K_{i2}]}\right)} = \frac{E[K_{i2}]K_{i1}}{E[K_{i2}]K_{i1}+E[K_{i1}]K_{i2}} \,,
\]
\[
    \textbf{C}_i \xrightarrow{P} \frac{\left(\frac{K_{i2}}{E[K_{i2}]}\right)}{\left(\frac{K_{i1}}{E[K_{i1}]}+\frac{K_{i2}}{E[K_{i2}]}\right)} = \frac{E[K_{i1}]K_{i2}}{E[K_{i2}]K_{i1}+E[K_{i1}]K_{i2}} \,.
\]
Accordingly, where we specify $\lambda_i=K_{i2}/K_{i1}$:
\begin{align*}
 \hat{\delta}_{FEpw} \xrightarrow{P} \left(\frac{1}{E\left[\frac{2K_{i1}\lambda_i}{(E[K_{i1}\lambda_i]+E[K_{i1}]\lambda_i)}\right]}\right) \left(
    \begin{tabular}{l}
        $\frac{E[K_{i1}]\lambda_i}{E[K_{i1}\lambda_i]+E[K_{i1}]\lambda_i}
        E\left[ \frac{\sum_{k=1}^{K_{i1}} Y_{i1k}(1)}{\sum_{j=1}^2 K_{ij}} - \frac{\sum_{k=1}^{K_{i1}} Y_{i1k}(0)}{\sum_{j=1}^2 K_{ij}} \right]$ \\
        \indent $+ \frac{E[K_{i1}\lambda_i]}{E[K_{i1}\lambda_i]+E[K_{i1}]\lambda_i}
        E\left[ \frac{\sum_{k=1}^{K_{i2}} Y_{i2k}(1)}{\sum_{j=1}^2 K_{ij}} - \frac{\sum_{k=1}^{K_{i2}} Y_{i2k}(0)}{\sum_{j=1}^2 K_{ij}} \right]$
    \end{tabular}
 \right) \,.
\end{align*}

Assuming IPS, but in the specific scenarios where $\lambda_i=\lambda \, \forall \, i$, we can demonstrate that the FEpw estimator can be consistent for the pATE estimand in a CRXO trial with IPS:
\begin{equation}
    \hat{\delta}_{FEpw} \xrightarrow{P} \frac{1}{2} \sum_{j=1}^{2} \left[ \frac{E\left[ \sum_{k=1}^{K_{ij}} \left[Y_{ijk}(1)-Y_{ijk}(0)\right] \right]}{E\left[K_{ij}\right]} \right]  \,.
\end{equation}

\newpage
\section{Additional simulation results with non-informative cluster sizes}
\label{sect:appendix_noICS_efficiency}

We include additional efficiency results in simulation scenarios with non-informative cluster sizes.
The efficiency of the different unweighted and weighted models as captured by the average of the model-based, leave-one-cluster jackknife, bias-reduced linearization, and ``CR0" robust variance estimates over the 1000 simulation replicates are graphed alongside the corresponding empirical variances.
Notably, the biased reduced linearization (BRL) robust variance estimates are not included for the EMEw and NEMEw estimators, due to the unclear implementation of this robust variance estimator with analyses conducted with WeMix in R.

\begin{figure}[htp]
    \centering
    \includegraphics[width=15cm]{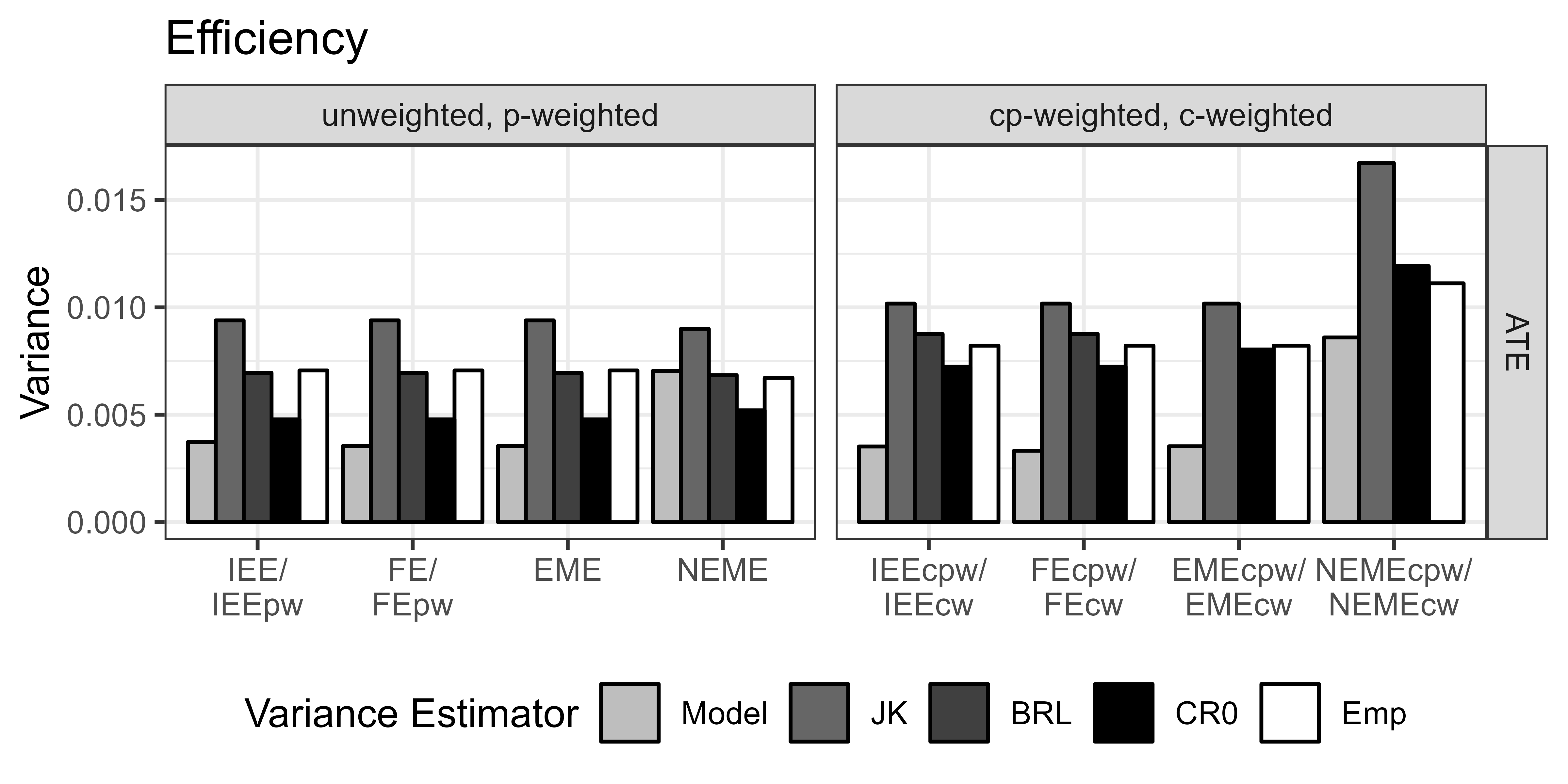}
    \caption{The efficiency of the different unweighted and weighted models as captured by the average of the model-based ``Model", leave-one-cluster jackknife ``JK", bias-reduced linearization ``BRL", and ``CR0" robust variance estimates over the 1000 simulation replicates, graphed alongside the corresponding empirical ``Emp" variances.}
\end{figure}

Furthermore, we present the coverage probability and power of the different unweighted and weighted models using the model-based, leave-one-cluster jackknife, bias-reduced linearization, and ``CR0" robust variance estimates over the 1000 simulation replicates are graphed alongside the corresponding empirical variances, in simulation scenarios with non-informative cluster sizes

\begin{figure}[htp]
    \centering
    \includegraphics[width=15cm]{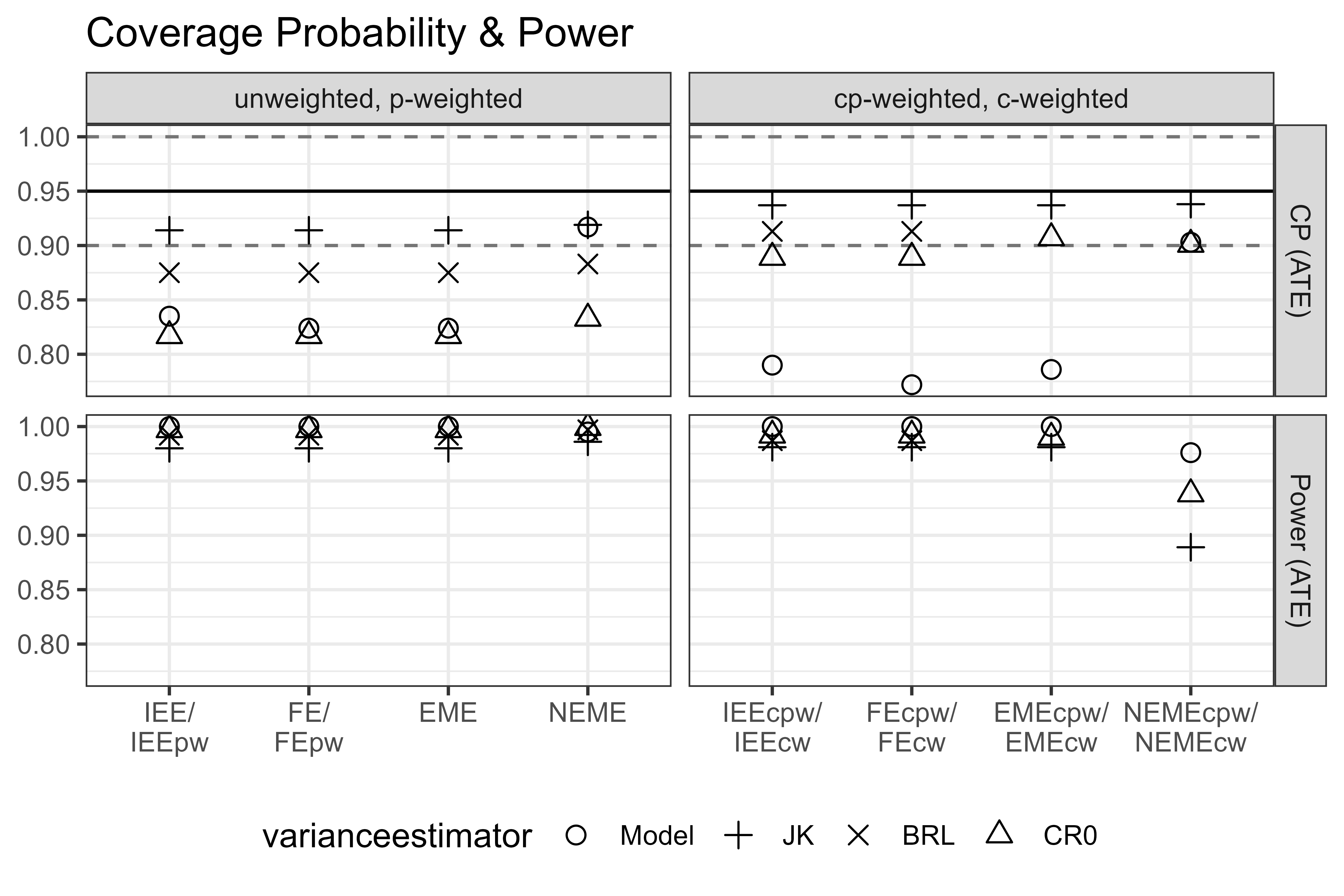}
    \caption{The coverage probability of the 95\% confidence interval and the power of the different unweighted and weighted models using the model-based, leave-one-cluster jackknife ``JK", bias-reduced linearization ``BRL", and ``CR0" variance estimators in scenarios with a homogeneous treatment effect (non-informative cluster sizes). The solid lines lines show a coverage probability of 0.95, with the dashed lines denoting the range from 0.90 to 1.}
\end{figure}

\newpage

\section{Additional simulation results with unequal cluster-period sizes and informative cluster sizes}
\label{sect:appendix_ICPS}

We simulated data with informative cluster sizes but now additionally allowing cluster-period sizes to vary between-periods, within-clusters.
The average cluster-period sizes from subpopulations $u=1$ and $2$ were generated with $E[K_{ij,1}] \sim Poisson(20)$ and $E[K_{ij,2}] \sim Poisson(100)$, respectively. The observed cluster-period sizes were then further subsampled from $K_{ij,u} \sim E[K_{ij,u}]$.

Since we simulate with informative cluster sizes, but not informative cluster-period sizes, the cpATE and cATE are again equivalent.
In this scenario, we again simulated heterogeneous treatment effects $\delta_1 = 0.2$ and $\delta_2 = 0.6$, yielding a iATE of $0.5\bar{3}$ and a cpATE and cATE of $0.4$.

With the described DGP, the true iATE estimand is:
\[
    iATE = E\left[ \frac{\sum_{k=1}^{K_{ij}} [Y_{ijk}(1)-Y_{ijk}(0)]}{E[K_{ij}]} \right]
\]
\[
    = \frac{E[E[K_{ij}\delta_u|u]]}{E[E[K_{ij}|u]]} = \frac{P(u=1)E[K_{ij}\delta_u|u=1] + P(u=2)E[K_{ij}\delta_u|u=2]}{P(u=1)E[K_{ij}|u=1] + P(u=2)E[K_{ij}|u=2]}
\]
\[
    = \frac{0.5(20\delta_1) + 0.5(100\delta_2)}{0.5(20) + 0.5(100)} = \frac{10\delta_1 + 50\delta_2}{60}
\]
which is equal to $0.5\overline{3}$ in scenarios with informative cluster sizes (where $\delta_1 = 0.2$ and $\delta_2 = 0.5$). Subsequently, the true cpATE is then:
\[
    cpATE = E\left[ \frac{\sum_{k=1}^{K_{ij}}[Y_{ijk}(1)-Y_{ijk}(0)]}{K_{ij}} \right]
\]
\[
    = E\left[E\left[\frac{K_{ij,u}\delta_u}{K_{ij,u}}|u]\right]\right] = E[E[\delta_u|u]]
\]
\[
     = P(u=1)E[\delta_u|u=1] + P(u=2)E[\delta_u|u=2] = 0.5(\delta_1) + 0.5(\delta_2)
\]
which is equal to 0.4 in simulation scenarios with informative cluster sizes. Finally, the true cATE is:
 \[
    cATE=E\left[ \frac{\sum_{j=1}^{2} \sum_{k=1}^{K_{ij}} \left[Y_{ijk}(1)-Y_{ijk}(0)\right]}{\sum_{j=1}^{2}K_{ij}} \right]
 \]
 \[
    = E\left[E\left[\frac{(K_{i1,u}+K_{i2,u})\delta_u}{(K_{i1,u}+K_{i2,u})}|u]\right]\right] = E[E[\delta_u|u]]
\]
\[
     = P(u=1)E[\delta_u|u=1] + P(u=2)E[\delta_u|u=2] = 0.5\delta_1 + 0.5\delta_2
\]
which is equal to 0.4 in simulation scenarios with informative cluster sizes.

\begin{figure}[htp]
    \centering
    \includegraphics[width=15cm]{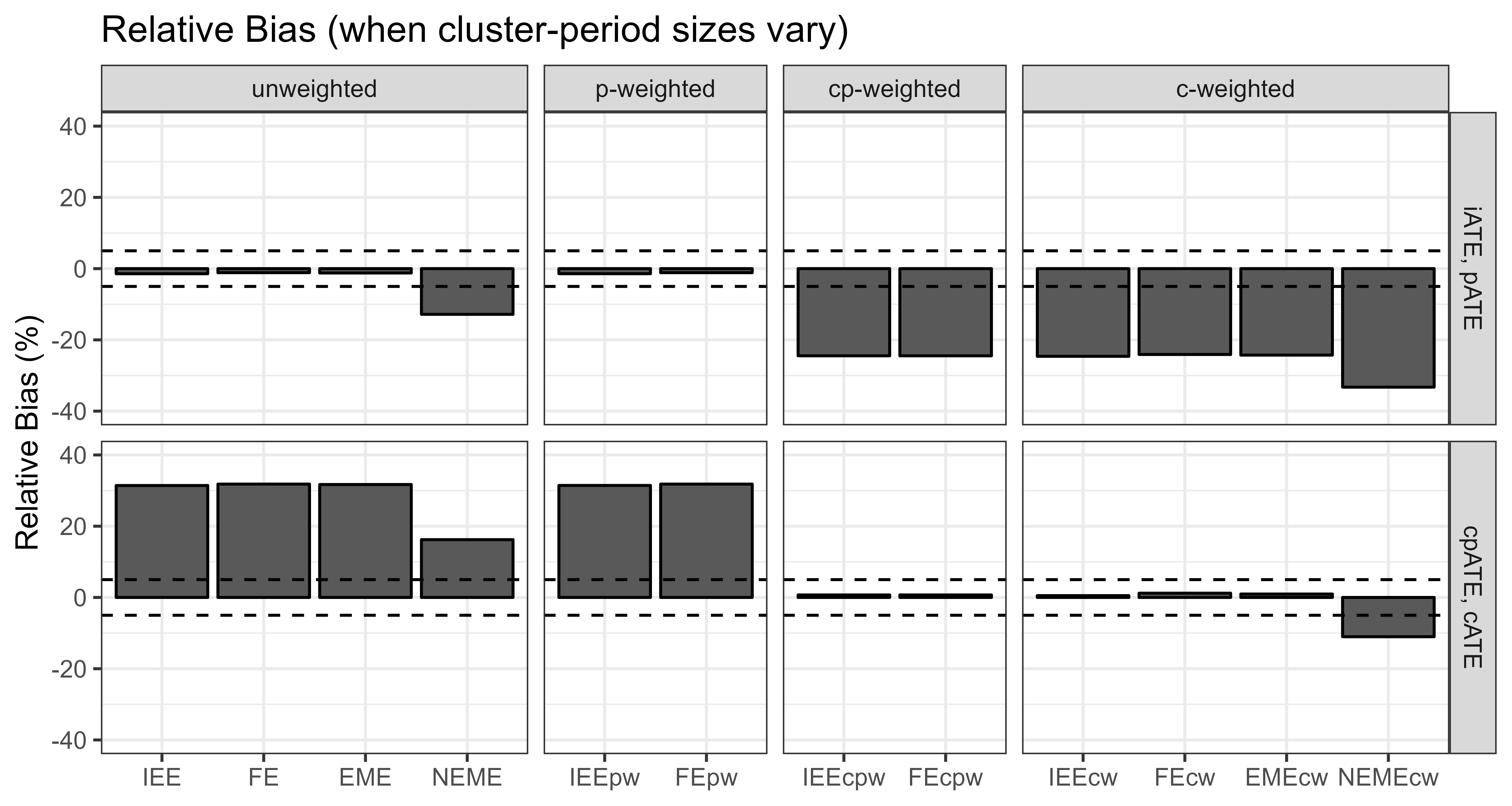}
    \caption{Simulation bias results in scenarios with heterogeneous treatment effects (informative cluster sizes) and cluster-period sizes that vary between-periods, within-clusters. Dashed lines show a relative bias ($\%$) of $5\%$ and $-5\%$.}
\end{figure}

We observe in these scenarios with informative cluster sizes that the NEME and NEMEcw estimators can yield biased results for the iATE, pATE, cpATE, and cATE estimands.
In contrast, the IEE, FE, and EME yielded empirically unbiased results for the coinciding iATE and pATE estimands, and the IEEcpw, FEcpw, IEEcw, FEcw, EMEcw estimators for the coinciding cpATE and cATE estimands, despite the variable cluster-period sizes between-periods, within-clusters.

Notably, the EMEcpw and NEMEcpw estimators are not included in these analyses.
To our knowledge, the extension of inverse cluster-period size weighting to analyses with variable cluster-period size and correlated errors (EMEcpw and NEMEcpw) is relatively unclear. Notably, specifying such weights deviate from the weighted estimating equation described by Williamson et al. \cite{williamson_marginal_2003}, which specifies weights on the cluster-level rather than the cluster-period-level.
However, unless there are informative cluster-period sizes, the cpATE and cATE estimands are expected to coincide. As a result, the EMEcw estimator may be appropriate to target the cpATE estimand in such a scenario.

In contrast, it is straightforward to specify variable inverse cluster-period size weights in analyses with an independence correlation structure, as is the case with the IEEcpw and FEcpw estimators which we have demonstrated are theoretically consistent estimators for the cpATE estimand. In R, such weights can be easily be specified with the “weights” option in lm(). We observe that the FEcpw estimator has similar properties to the EMEcpw estimator (Figures \ref{fig:noICS_bias_var} - \ref{fig:ICS_bias}), is a theoretically consistent estimator for the cpATE, and can be easily implemented in most statistical software. In general, if cluster-period sizes vary between-periods, within-clusters, the IEEcpw and FEcpw estimators can be used to estimate the cpATE in the analysis of CRXO trials with informative cluster sizes.

\newpage
\section{Additional Simulation results with 50 total clusters}
\label{sect:appendix_sim50}

\begin{figure}[htp]
    \centering
    \includegraphics[width=12cm]{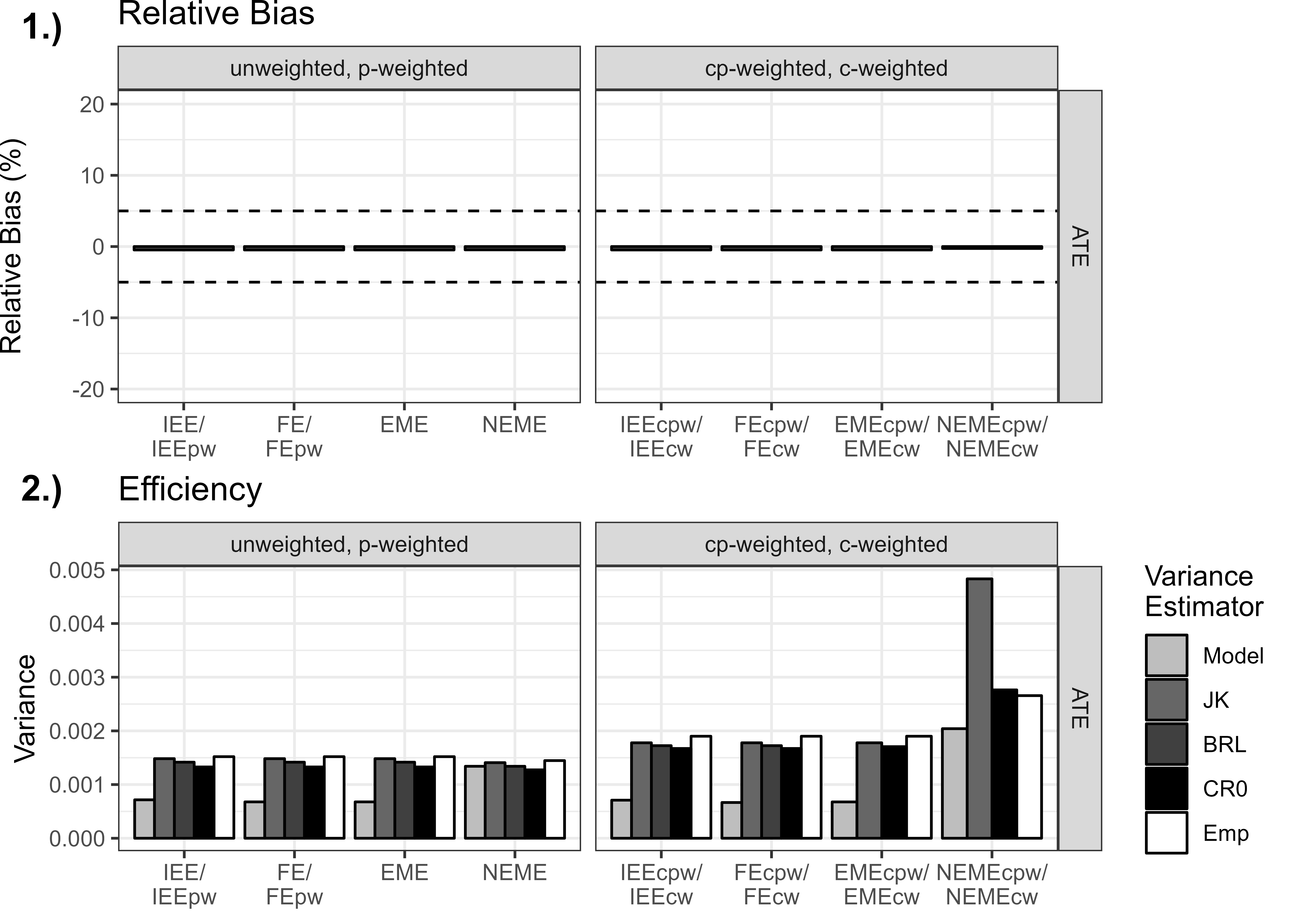}
    \caption{1.) Simulation relative bias (\%) results in scenarios with homogeneous treatment effects (non-informative sizes). Dashed lines show a relative bias of $5\%$ and $-5\%$. 2.) The efficiency of the different unweighted and weighted models as captured by the average of the model-based (``Model"), leave-one-cluster jackknife (``JK"), bias-reduced linearization (``BRL"), and ``CRO" robust variance estimates over the 1000 simulation replicates, graphed alongside the corresponding empirical (``Emp") variances.}
    \label{fig:appendix_sim50_bias_efficiency}
\end{figure}

\begin{figure}[htp]
    \centering
    \includegraphics[width=12cm]{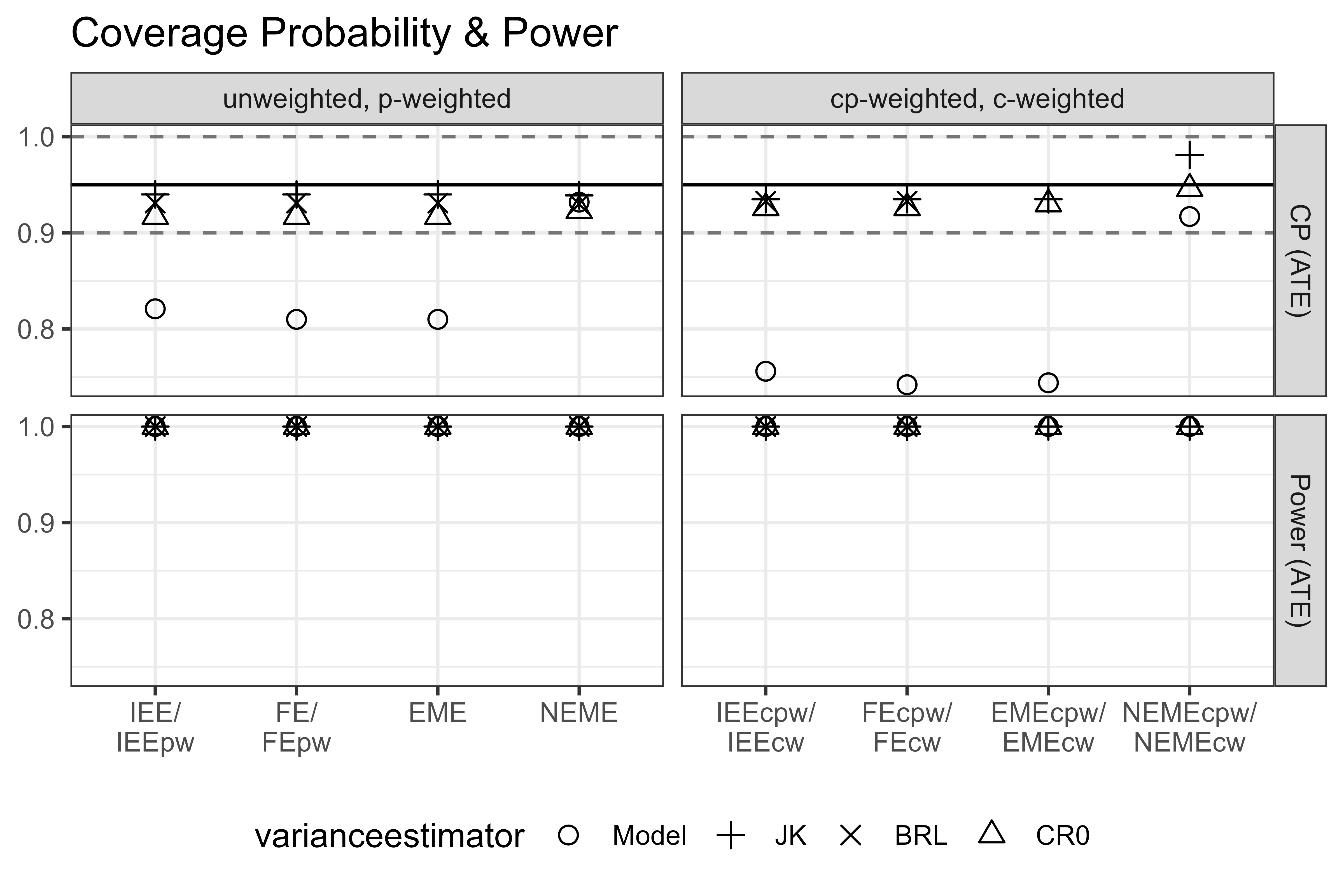}
    \caption{The coverage probability of the 95\% confidence interval and the power of the different unweighted and weighted models using the model-based (``Model"), leave-one-cluster jackknife (``JK"), bias-reduced linearization (``BRL"), and ``CRO" robust variance estimators in scenarios with a homogeneous treatment effect (non-informative sizes). The solid lines lines show a coverage probability of 0.95, with the dashed lines denoting the range from 0.90 to 1.}
    \label{fig:appendix_sim50_CP}
\end{figure}

We first simulated data without informative sizes, using the data generating process described in Section \ref{sect:simulation_no_ics}, but with 50 clusters (25 clusters/sequence). The results are included in Figures \ref{fig:appendix_sim50_bias_efficiency}-\ref{fig:appendix_sim50_CP}.
The bias results here (with 50 clusters) (Figure \ref{fig:appendix_sim50_bias_efficiency}.1) correspond with the results described in Section \ref{sect:simulation_no_ics} with 10 clusters (5 clusters/sequence).
Overall, we observe that the leave-one-cluster jackknife variance estimator closely approximates the empirical standard errors (Figure \ref{fig:appendix_sim50_bias_efficiency}.2) and yielded close to nominal coverage probabilities of the 95\% confidence intervals (Figure \ref{fig:appendix_sim50_CP}).

\begin{figure}[htp]
    \centering
    \includegraphics[width=12cm]{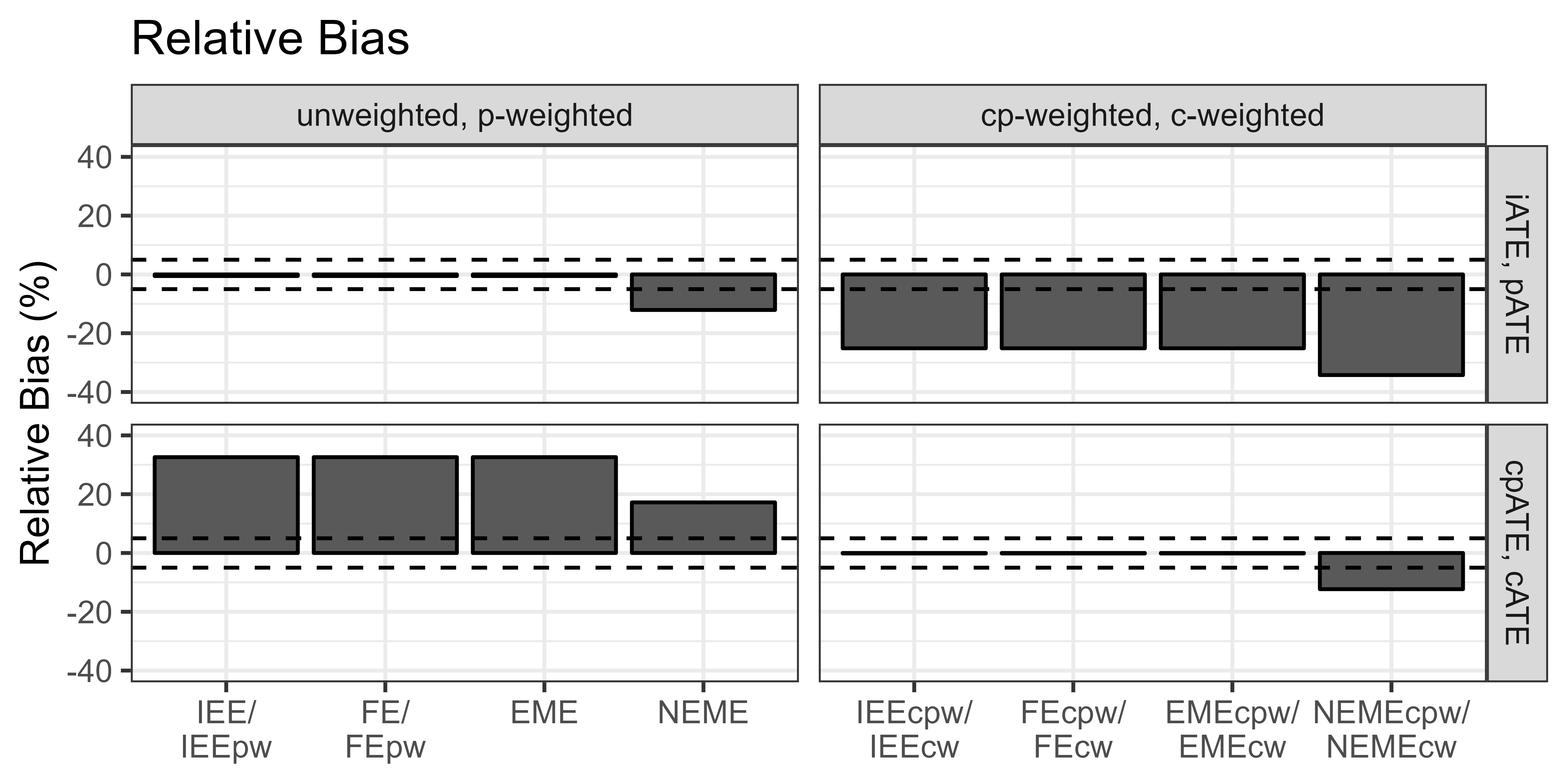}
    \caption{Simulation relative bias (\%) results in scenarios with informative cluster sizes. Dashed lines show a relative bias of $5\%$ and $-5\%$.}
    \label{fig:appendix_sim50_ICS_bias}
\end{figure}

We also simulated data with informative cluster sizes, using the data generating process described in Section \ref{sect:simulation_ics}, but with 50 clusters (25 clusters/sequence) in Figure \ref{fig:appendix_sim50_ICS_bias}. The bias results here (with 50 clusters) (Figure \ref{fig:appendix_sim50_ICS_bias}) correspond with the results described in Section \ref{sect:simulation_ics} with 10 clusters (5 clusters/sequence).

\begin{figure}[htp]
    \centering
    \includegraphics[width=15cm]{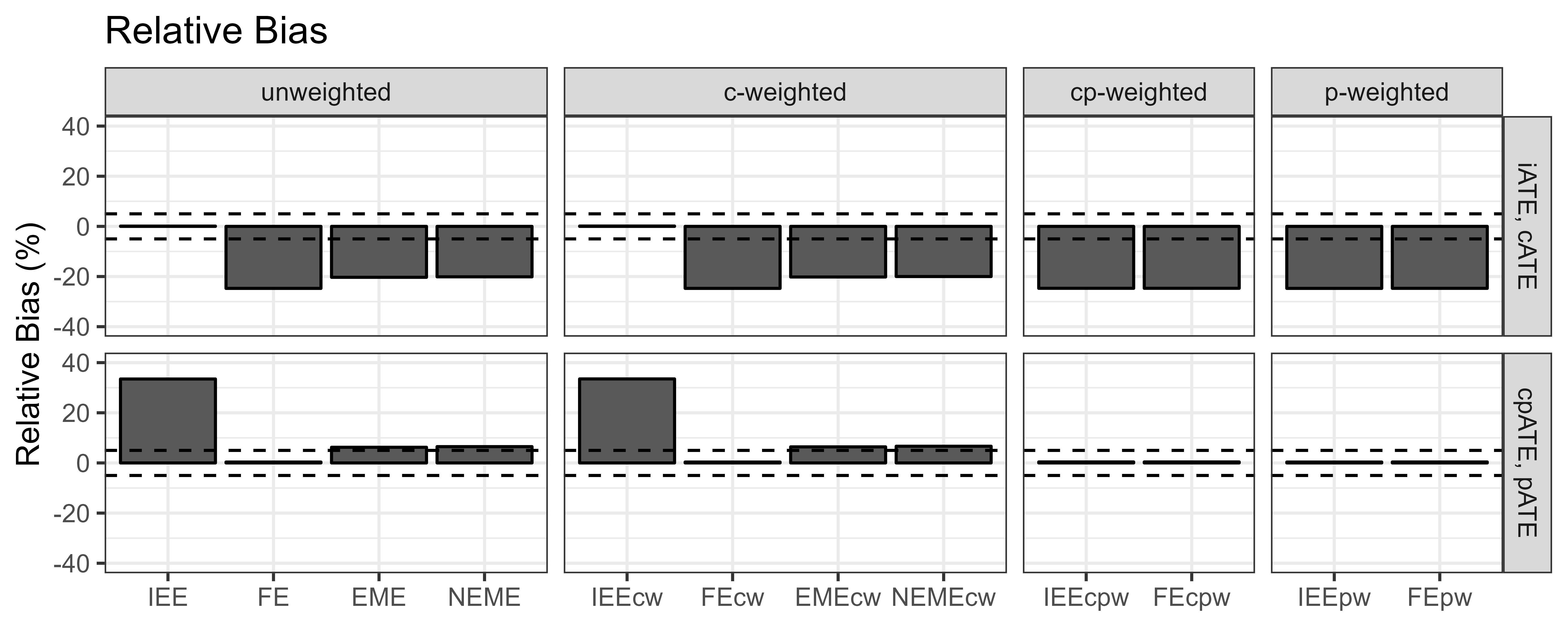}
    \caption{Simulation relative bias (\%) results in scenarios with informative period sizes. Dashed lines show a relative bias of $5\%$ and $-5\%$.}
   \label{fig:appendix_sim50_IPS_bias}
\end{figure}

We also simulated data with informative period sizes, using the data generating process described in Section \ref{sect:simulation_ips}, but with 50 clusters (25 clusters/sequence) in Figure \ref{fig:appendix_sim50_IPS_bias}. The bias results here (with 50 clusters) (Figure \ref{fig:appendix_sim50_IPS_bias}) correspond with the results described in Section \ref{sect:simulation_ips} with 10 clusters (5 clusters/sequence).

\newpage
\section{Case Study design}
\label{sect:appendix_casestudy}

The design of a 2-period, 49 cluster cross-sectional CRXO design exploring the effect of stress ulcer prophylaxis with proton pump inhibitors (PPIs; treatment) versus histamine-2 receptor blockers (H$_{2}$RBs; control) on log-length of stay (log-LOS) among patients receiving invasive mechanical ventilation \cite{the_peptic_investigators_for_the_australian_and_new_zealand_intensive_care_society_clinical_trials_group_alberta_health_services_critical_care_strategic_clinical_network_and_the_irish_critical_care_trials_group_effect_2020}.
Clusters (hospital ICU's) receiving the treatment (PPIs) or control (H$_2$RBs) during period $j$ are marked in gray or white, respectively.

\begin{figure}[htp]
    \centering
    \includegraphics[width=6cm]{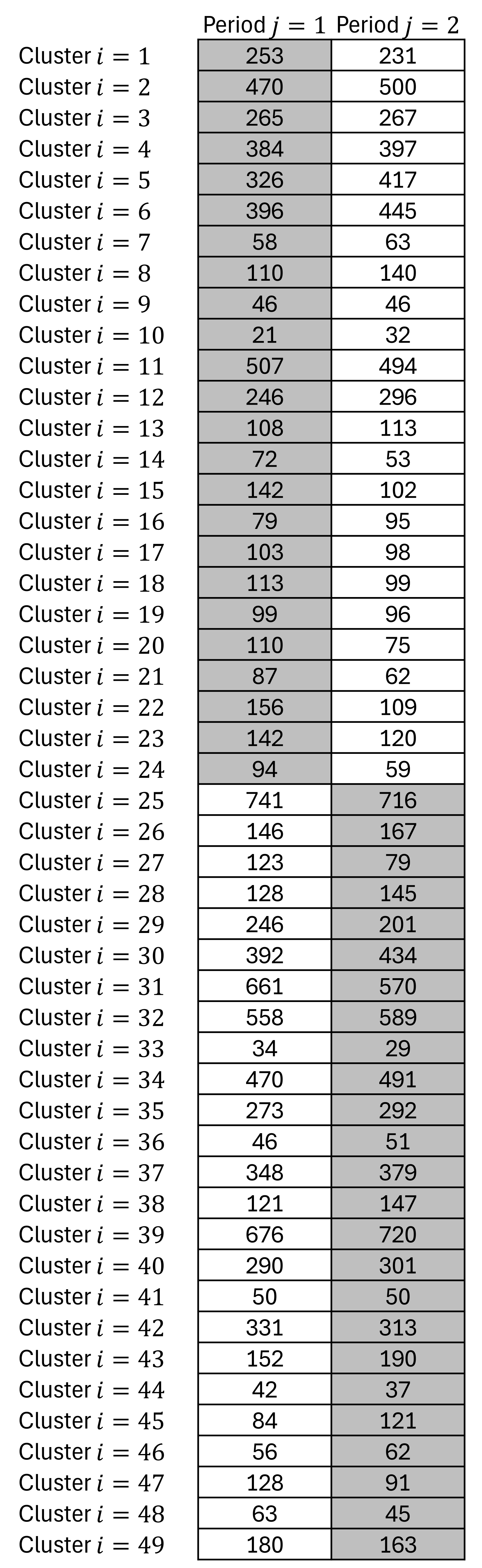}
\end{figure}
\end{document}